\documentclass[acmsmall,screen]{acmart}
\usepackage{amsmath,amsfonts}
\usepackage{multirow}
\usepackage{multicol}
\usepackage{amsthm}
\usepackage{listings}
\usepackage{algorithm}
\usepackage{algpseudocode}
\usepackage{graphicx}
\usepackage{textcomp}
\usepackage{xcolor}
\usepackage{cleveref}
\usepackage{booktabs}
\usepackage{comment}
\usepackage{tabularx}
\usepackage{multirow}
\usepackage{bm}
\usepackage{url}
\usepackage{xspace}
\usepackage{pifont}
\usepackage{verbatim}
\usepackage[referable]{threeparttablex}
\usepackage{calc} 
\usepackage{soul}
\usepackage{color}
\usepackage{float}
\usepackage{subcaption}
\usepackage{tikz}
\usetikzlibrary{positioning,calc,fit,decorations.pathmorphing,shapes.geometric, shapes.gates.logic.US, calc}
\usetikzlibrary{arrows,arrows.meta,decorations.markings,shapes,shapes.arrows}
\usetikzlibrary{backgrounds}
\usepackage{pgfplots}
\usepackage{pgfplotstable}
\usepackage{scalefnt}
\usepgfplotslibrary{groupplots}
\pgfplotsset{compat=newest}
\usepackage{caption}
\Crefformat{figure}{Fig.~#2#1#3}                           % "Fig.", instead of "Figure"
\Crefname{subfigure}{Fig.}{Figs.}
\Crefname{figure}{Fig.}{Figs.}
\Crefformat{table}{TABLE~#2#1#3}                           % "TABLE", instead of "Table"
\usepackage[figuresright]{rotating}
\definecolor{USTgold}{RGB}{153,102,0}
\definecolor{USTyellow}{RGB}{204,153,0}
\definecolor{USTyellowlight}{RGB}{255,212,0}
\definecolor{USTorange}{RGB}{255,166,26}
\definecolor{USTpink}{RGB}{255,157,157}
\definecolor{USTblue}{RGB}{0,51,102}
\definecolor{USTmiddle}{RGB}{0,116,188}
\definecolor{USTlight}{RGB}{99,202,225}
\definecolor{USTgray}{RGB}{204,204,204}
\definecolor{USTred}{RGB}{237,27,47}
\definecolor{USTdarkred}{RGB}{124,35,72}

\definecolor{CUHKorange}{RGB}{244,106,18} %F47012
\definecolor{CUHKblue}{RGB}{0,111,190}    %006FBE
\definecolor{CUHKgreen}{RGB}{0,127,128}   %007F80
\definecolor{CUHKred}{RGB}{228,46,36}     %E42E24
\definecolor{CUHKyellow}{RGB}{198,148,34} %C69422
\definecolor{CUHKdark}{RGB}{114,44,114}   %722C72
\definecolor{CUHKmiddle}{RGB}{144,44,144} %902C90
\definecolor{CUHKlight}{RGB}{167,44,167} 
\definecolor{CUHKpurple}{RGB}{117,15,109}
\definecolor{CUHKgold}{RGB}{221,163,0}
\definecolor{CUHKribbon}{RGB}{244,223,176}
\definecolor{CUHKblack}{RGB}{34,24,21}

\renewcommand{\vec}[1]{\boldsymbol{#1}}    % re-define vec command

\newcommand*\bcircled[1]{                  % black circled
    \tikz[baseline=(char.base)]{
        \node[shape=circle,fill,inner sep=1pt] (char) {\textcolor{white}{#1}};
    }
}

\usepackage{tcolorbox}
\tcbuselibrary{skins,breakable}
    {\endtcolorbox}
    {\endtcolorbox}

\crefname{mytheorem}{Theorem}{Theorems}
\crefname{mylemma}{Lemma}{Lemmas}
\crefname{myclaim}{Claim}{Claims}
\crefname{myproperty}{Property}{Properties}
\crefname{mycorollary}{Corollary}{Corollaries}

\algrenewcommand\textproc{\texttt}

\makeatletter
\let\OldStatex\Statex
\renewcommand{\Statex}[1][3]{%
  \setlength\@tempdima{\algorithmicindent}%
  \OldStatex\hskip\dimexpr#1\@tempdima\relax
}
\makeatother

\algblockdefx[Foreach]{Foreach}{EndForeach}[1]{\textbf{for each} #1 \textbf{do}}{\textbf{end for}}
\makeatletter
\ifthenelse{\equal{\ALG@noend}{t}}{\algtext*{EndForeach}}{}
\makeatother

\RequirePackage[normalem]{ulem} %DIF PREAMBLE
\RequirePackage{color}\definecolor{RED}{rgb}{1,0,0}\definecolor{BLUE}{rgb}{0,0,1} %DIF PREAMBLE
\setcopyright{acmcopyright}
\acmJournal{TACO}
\newcommand{\ie}{\textit{i}.\textit{e}., }
\newcommand{\eg}{\textit{e}.\textit{g}. }
\DeclareMathAlphabet\mathbfcal{OMS}{cmsy}{b}{n}

\begin{document}

\title{GPU-Accelerated Gate-Level Time-Based Power Analysis via Event-Density-Aware Partitioning and Kernel Fusion}

\author{Weihao Wang}
\orcid{0009-0005-3469-5618}
\affiliation{%
    \institution{The Hong Kong University of Science and Technology - Guangzhou Campus}
    \department{Microelectronics Thrust, Function Hub}
    \city{Guangzhou}
    \state{Guangdong}
    \postcode{510530}
    \country{China}
}
\email{wwang103@connect.hkust-gz.edu.cn}

\author{Yikang Ouyang}
\orcid{0009-0007-0714-9501}
\affiliation{%
    \institution{The Hong Kong University of Science and Technology - Guangzhou Campus}
    \department{the Microelectronics Thrust}
    \city{Guangzhou}
    \state{Guangdong}
    \country{China}
}
\email{youyang929@connect.hkust-gz.edu.cn}

\author{Hongyuan Liu}
\orcid{0000-0002-6961-6394}
\affiliation{%
    \institution{Stevens Institute of Technology}
    \department{the Department of Computer Science}
    \city{Hoboken}
    \state{NJ}
    \country{USA}
}
\email{hliu96@stevens.edu}

\author{Yuzhe Ma}
\orcid{0000-0002-3612-4182}
\affiliation{%
    \institution{The Hong Kong University of Science and Technology - Guangzhou Campus}
    \department{the Microelectronics Thrust}
    \city{Guangzhou}
    \state{Guangdong}
    \country{China}
}
\email{yuzhema@hkust-gz.edu.cn}

\begin{abstract}
Power analysis is crucial in modern chip design flow.
Particularly, the time-based power analysis can provide fine-grained power consumption information to facilitate the diagnosis of power issues and guide power optimization accordingly.
However, it may take tens of hours to conduct time-based power analysis on modern large-scale circuits, which greatly slows down the power optimization flow.
In this paper, we present the first GPU-accelerated gate-level time-based power analysis framework.
% We propose an event-equalized chunking method to mitigate the potential for excessive memory usage on GPUs when working with large designs, which additionally enables the CPU-GPU pipelining to hide event preparation latency.
% We introduce a spatial-temporal parallelization strategy that optimizes workload distribution among GPU threads.
We propose a novel data structure to enable efficient state-dependent power retrieval.
To accommodate the imbalanced event distribution across gates, we propose an event-density-aware partitioning strategy that allocates GPU threads based on gate event density.
Finally, we fuse the power computation into a single kernel invocation to reduce redundant work in separate kernels.
Experimental results show that our proposed framework achieves high accuracy while delivering up to 37.63$\times$ end-to-end speedup compared to multi-threaded Synopsys PrimeTime PX.
\end{abstract}



\maketitle

\section{Introduction}
\renewcommand{\thefootnote}{}
\footnotetext{New Paper, Not an Extension of a Conference Paper.}
\renewcommand{\thefootnote}{\arabic{footnote}}
%Low-power IC designs are very essential for modern digital devices. 
%For example, low-power designs for portable devices can increase the battery life and also allow for more features at the same time~\cite{weste2015cmos}. 
% Power is becoming increasingly critical in modern VLSI design. 
%The functionality of chips is increasingly limited by power, and energy efficiency is becoming as important as high performance in IC designs~\cite{weste2015cmos}. 
% Thus, power analysis is essential in the design flow. 
%it is important to do a power analysis of the IC design and reduce the power dissipated in the circuit according to the power analysis result. 
%Power has become a primary design constraint in modern very-large-scale integration (VLSI) systems, affecting performance, thermal management, and reliability.
%For extremely accurate, high-temporal-resolution power reporting, time-based power analysis is paramount using switching activity obtained from gate-level simulation. 
Power has become a primary design constraint in modern very-large-scale integration (VLSI) systems, critically affecting performance, thermal management, and reliability. 
Consequently, accurate power analysis is indispensable throughout the design flow.
While time-based power analysis using gate-level simulation offers the high temporal resolution required for accurate power reporting, it is computationally intensive. 
%The power analysis result can be obtained by time-based power analysis.
% Notably, the time-based power analysis can also provide various important power composition analysis results, such as glitch power and peak power. 
% \Cref{fig:overview_of_flow} illustrates a typical workflow of gate-level time-based power analysis.
% The register-transfer level (RTL) design is first synthesized into a gate-level netlist with all timing information stored in a standard delay format (SDF) file.
% Next, the netlist is simulated with a certain testbench.
% Finally, the netlist and gate-level activity file are fed into the power analysis tool for time-based power analysis. 
% The tool processes all gate events, each defined by a value-timestamp pair representing a signal transition to value at that timestamp, to compute per-gate power and generate the power waveform~\cite{ptpx_user_guide}.
%One major challenge in time-based power analysis is the computational demand, particularly with the increasing complexity of modern designs and long simulation traces.
%For instance, conducting a comprehensive analysis of a design encompassing 288K gates using a commercial tool requires several days~\cite{aspdac_2025_smart_gpo}, which presents substantial challenges for power optimization.
The increasing complexity of modern designs and long simulation traces exacerbate this challenge.
For instance, analyzing a 288k-gate design with commercial tools can take several days~\cite{aspdac_2025_smart_gpo}. 
This computational bottleneck hinders effective optimization, necessitating the acceleration of gate-level time-based power analysis.
% The time-consuming power analysis hinders power optimization and thus impedes the whole IC design flow. 
%Consequently, there is a need to expedite the gate-level time-based power analysis process.

Prior research on power analysis can be broadly categorized into machine learning (ML)-based approaches and simulation-based approaches. 
Existing ML-based approaches~\cite{dac_2019_primal, dac_2020_grannite, tcad_2025_gript, iccad_2023_master_rtl, micro_2021_apollo} typically formulate power analysis as an \emph{estimation} problem using register-transfer-level (RTL) activity proxies, and thus are not designed to explicitly perform event-driven, per-gate accumulation required for precise time-based power computation. 
Simulation-based approaches~\cite{isca_2016_strober, aspdac_2025_smart_gpo} are more physically grounded, yet they often rely on \emph{sampling or limited temporal coverage} to reduce overhead, which can miss short-duration glitches and transient peak power behaviors. 
In contrast, commercial tools such as Synopsys PrimeTime PX~\cite{synopsys_primetime} support gate-level time-based power analysis and can generate complete and precise power reports (e.g., component breakdowns, waveforms, and peak values), but their runtime becomes prohibitive for large-scale designs. 
These limitations motivate accelerating gate-level time-based power analysis with high precision.

% Although GPU acceleration has been investigated in these domains, the acceleration of gate-level time-based power analysis using GPUs remains underexplored.
%Considering that gate-level time-based power analysis is time-consuming and exhibits both spatial (gate) and temporal (event/cycle) parallelism, leveraging GPU acceleration for power analysis has great potential to boost the entire design flow further.
% Given the inherent spatial and temporal parallelism of gate-level time-based power analysis, GPU acceleration is well suited to exploit this parallelism, alleviating computational bottlenecks and improving the overall design flow.
Given its inherent spatial and temporal parallelism, gate-level time-based power analysis is well suited for GPU acceleration, thereby alleviating computational bottlenecks and improving the overall design flow.
However, fully exploiting this potential is non-trivial.
We identify three key challenges to effectively leveraging GPUs for this workload:
(1) \emph{Inefficient state-dependent power retrieval.}
Standard state-dependent power retrieval methods require two iterative list traversals per event. 
This incurs a substantial per-event cost, necessitating a more optimized data structure to handle high-frequency lookups.
% This process requires two iterative list traversals per event, constituting a computationally intensive operation that significantly constrains the overall system throughput.
%Consequently, designing an efficient data structure for state-dependent power retrieval becomes essential.
(2) \emph{Imbalanced event distribution across gates.}
The distribution of events per gate is highly skewed, and a globally fixed partitioning strategy is ineffective.  
%A fine-grained global partitioning over-allocates threads to sparse gates, causing many threads to do little work before exiting and incurring scheduling overhead.
Fine-grained partitioning incurs scheduling overhead for sparse gates with few events, whereas coarse-grained partitioning limits parallelism for dense gates with many events.
Consequently, a workload-aware, per-gate granularity control is required.
%Conversely, a coarse-grained global partitioning under-allocates threads to dense gates, limiting workload distribution and increasing the per-thread work depth.  
%Therefore, we need per-gate, workload-aware control of parallel granularity.  
(3) \emph{Redundant work in separate kernels.}
%Computing dynamic power and leakage power with two separate GPU kernels introduces substantial redundant work, because both kernels require per-thread gate-state reconstruction for state-dependent power retrieval.
Calculating dynamic and leakage power in separate kernels leads to significant redundancy, as both processes must independently perform costly gate-state reconstruction.
Moreover, cycle-level information needed for glitch detection in dynamic power computation is already exposed during leakage power computation through cycle-level traversal.
Therefore, a unified formulation is desirable to share gate-state reconstruction and reuse cycle-level information.

To address the above challenges, we propose a GPU-accelerated gate-level time-based power analysis framework. 
% Specifically, we introduce a novel data structure to enable efficient state-dependent power retrieval.
% We further incorporate an event-density-aware, per-gate cycle partitioning scheme that allocates threads to gates according to their event density, accommodating the inherently unbalanced event distribution across gates.
% Moreover, we fuse the power computation into a single kernel invocation to reduce redundant work arising from separate kernels.
This framework supports efficient state-dependent power retrieval, uses adaptive per-gate partitioning to handle imbalanced event distribution across gates, and reduces kernel overhead through fused power computation.
% The implementation of our framework is built upon OpenSTA~\cite{opensta}.
% Note that OpenSTA supports only averaged rather than time-based power analysis. 
% OpenSTA functions solely as an static timing analysis (STA) engine in our framework, and we leverage the timing information from it for power computation. 
% We emphasize that the generality of our framework allows for easy integration into other STA tools without loss of functionality.
To the best of our knowledge, it is the \emph{first} work that explores time-based power analysis on the GPU.  
This work\footnote{The source code is publicly available at \url{https://github.com/whwang1996/GTPower}.} makes the following contributions:

\begin{itemize}
% \item We propose an event-equalized chunking method to mitigate the limitations imposed by constrained GPU memory. 
% Additionally, we introduce a pipelined CPU-GPU execution technique to reduce runtime overhead caused by large file sizes. %(\Cref{sec: event_chunking_and_pipelined_cpu_gpu_exe})

% \item We present the first end-to-end GPU-accelerated gate-level time-based power analysis framework.

\item We identify three key challenges in effectively leveraging GPUs for time-based power computation.~(\Cref{sec:characterization_and_motivation})

\item We propose a novel data structure that encodes state patterns into binary indices, enabling direct and efficient state-dependent power retrieval.~(\Cref{sec:efficient_power_retrieval_with_bsim})

\item We propose an event-density-aware, per-gate cycle partitioning strategy that efficiently distributes workload across GPU threads by leveraging each gate's event density and assigning threads accordingly.~(\Cref{sec:adaptive_temporal_part})

\item We propose a kernel fusion strategy that combines dynamic and leakage power computation within a single kernel invocation, thereby reducing redundant gate-state reconstruction and eliminating repeated cycle-level traversal of the same event streams.~(\Cref{sec:kernel_fusion})

\item The proposed GPU-accelerated time-based power analysis framework leverages both spatial (gate-level) and temporal (cycle-level) parallelism to fully exploit the capabilities of GPUs. This framework achieves high accuracy and delivers an end-to-end speedup of up to 37.63$\times$ compared to multi-threaded PrimeTime PX.~(\Cref{sec:exp})

\end{itemize}

% The remainder of this paper is organized as follows: 
% \Cref{sec:preliminaries} introduces the background of power analysis and presents the naive implementation.
% \Cref{sec:characterization_and_motivation} discusses three challenges associated with the naive implementation.
% % our CPU-GPU pipelined execution strategy, and an overview of the proposed framework.
% \Cref{sec:multi_level_gpu_opt_framework} presents the methodologies for addressing the three challenges.
% \Cref{sec:end_to_end_framework} introduces the end-to-end framework.
% \Cref{sec:exp} presents the experimental results.
% \Cref{sec:related_work} summarizes the related work.
% \Cref{sec:conclusion_and_future} concludes this paper.

\section{Preliminaries}\label{sec:preliminaries}

\subsection{Basic Concepts and Terminology}

\begin{figure}[!t]
  \centering
  \begin{minipage}{1.0\linewidth}
    \centering
    \subfloat[]{\includegraphics[width=0.38\linewidth]{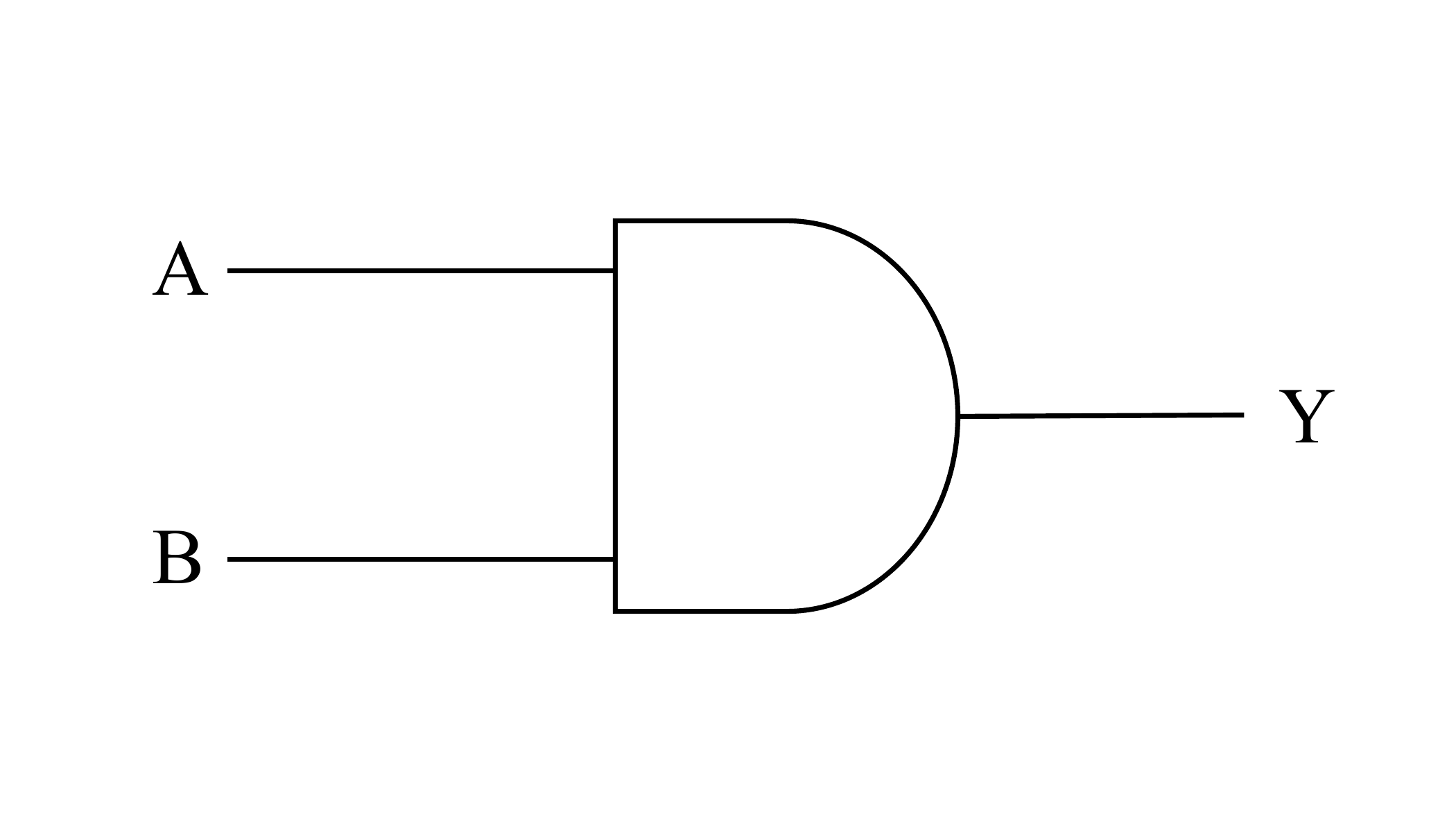}\label{fig:gate_example}}
    \hspace{0.04\linewidth}
    \subfloat[]{\includegraphics[width=0.53\linewidth]{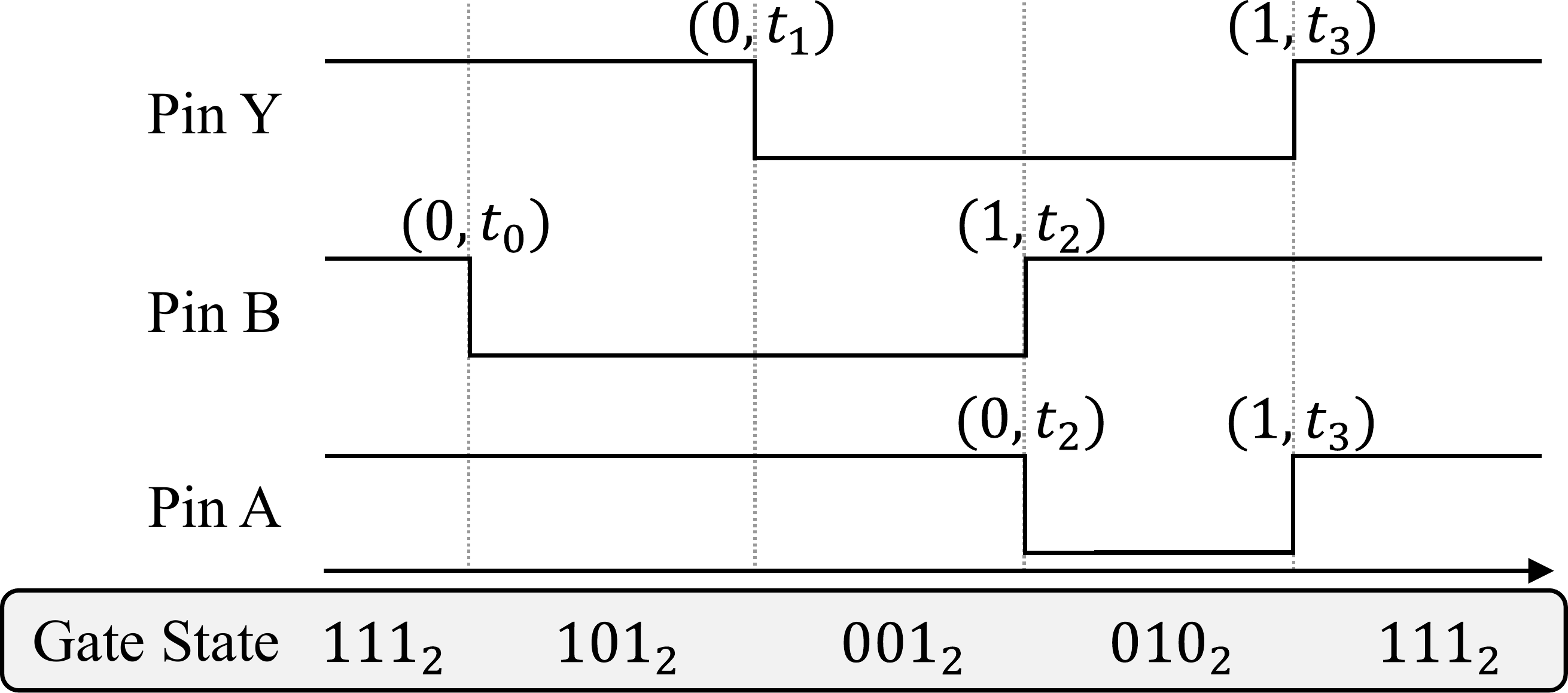}\label{fig:event_and_gate_state_example}}
  \end{minipage}
  \caption{Concept illustration using an \texttt{AND} gate. (a) Gate and its pins. (b) Events and gate states.}
  \label{fig:gate_event_and_gate_state_example}
\end{figure}

\Cref{fig:gate_example} illustrates an example gate with input pins \texttt{A} and \texttt{B} and output pin \texttt{Y}.
%Each pin is associated with a logic signal, represented as a sequence of events in chronological order.
Each pin carries a logic signal represented by a chronologically ordered event sequence.
An \emph{event} is a value--timestamp pair, indicating that the signal transitions to the corresponding logic value (0 or 1) at the associated time.
For example, $(0, t_{1})$ in \Cref{fig:event_and_gate_state_example} indicates the signal of pin \texttt{Y} transitions to 0 at $t_{1}$.
%As depicted in \Cref{fig:event_and_gate_state_example}.
The \emph{gate state} is the vector of logic values on all pins of the gate at a given time, and it changes whenever an event updates one of these pin values.

\subsection{Power Models} \label{sec: power_models}

\subsubsection{Dynamic Power and Leakage Power} \label{sec:pre_leak_power_and_dyn_power}
%The power dissipated in the circuit can be categorized into two types: dynamic power and leakage power~\cite{weste2015cmos}.
Circuit power is commonly decomposed into dynamic power and leakage power~\cite{weste2015cmos}.
%Dynamic power is the power dissipated when an event occurs, which contains two parts: internal power and switching power~\cite{ptpx_user_guide, weste2015cmos}.
Dynamic power is dissipated when an event occurs and consists of internal power and switching power~\cite{ptpx_user_guide, weste2015cmos}.
%Internal power refers to the power dissipated within the boundary of a gate. 
%The internal energy corresponding to an event is determined by interpolating a power lookup table (LUT) in the nonlinear delay model (NLDM)~\cite{bhasker2009STA_for_nm_design}.
Internal power is dissipated within a gate, and the corresponding event energy is obtained by interpolating a power lookup table (LUT) in the nonlinear delay model (NLDM)~\cite{bhasker2009STA_for_nm_design}.
Note that the internal power can be state-dependent~\cite{us_patent_state_dependent_power_modeling, ptpx_user_guide}.
%The LUT corresponding to the current gate state is retrieved from a pre-characterized array, where each entry pairs a gate state with its associated LUT.
As shown in \Cref{fig:overview_of_naive_1t1e_dynamic_kernel}~\bcircled{\bfseries\scriptsize 2}, different gate states are associated with different internal power LUTs.
%The gate state pattern is specified in the \texttt{when} clause, where a pin name without the \texttt{!} prefix denotes a logic value of 1, while a pin name with the \texttt{!} prefix indicates a logic value of 0. 
% The corresponding leakage power value for the specified state pattern is provided in the \texttt{value} field.
% The same pattern also works for internal power, but the retrieved item is a LUT instead of a power value. 
Switching power is the power dissipation when the load capacitance at the gate output pin is charged or discharged.
% The calculation of switching energy is straightforward.
% Note that dynamic power is dissipated only when events occur, and no dynamic power dissipation takes place in the absence of events.
Leakage power is caused by intrinsic transistor leakage current, and its value persists between state-changing events.
%The leakage power value $p_{\mathrm{leak}}^{(g)}$ of gate $g$ represents a constant power dissipation level maintained over the time interval between the current event and the previous one.
%Like internal power, leakage power value also demonstrates state-dependent~\cite{us_patent_state_dependent_power_modeling, ptpx_user_guide} behavior, and $p_{\mathrm{leak}}^{(g)}$ is retrieved from a pre-characterized array~(\Cref{fig:overview_of_naive_leakage_kernel}~\bcircled{\bfseries\scriptsize 1}), where each element stores a gate state together with the corresponding leakage power value.
For gate $g$, the leakage value $p_{\mathrm{leak}}^{(g)}$ is state-dependent~\cite{us_patent_state_dependent_power_modeling, ptpx_user_guide} and is retrieved from a pre-characterized array that maps gate states to leakage values~(\Cref{fig:overview_of_naive_leakage_kernel}~\bcircled{\bfseries\scriptsize 1}).
Note that leakage power is always present; events only change the gate state and therefore modify $p_{\mathrm{leak}}^{(g)}$.
%Events do not create leakage power; they update the gate state and therefore change $p_{\mathrm{leak}}^{(g)}$.

\subsubsection{Glitch Power} \label{sec:pre_glitch_power}
Beyond these fundamental power components, glitch power emerges as another critical consideration in power analysis.
A glitch is an undesirable phenomenon caused by misaligned event timing, which leads to unnecessary events and consequently results in additional dynamic power consumption.
% \Cref{fig: glitch_example} illustrates the mechanism of glitch generation in digital circuits.
% In the zero-delay scenario (\Cref{fig: glitch_zero_delay_and_gate}), the rising edge of pin \texttt{A} coincides precisely with the falling edge of pin \texttt{B}, resulting in no transition at the output pin~\texttt{Y}.
% When considering SDF annotated delays (\Cref{fig: glitch_sdf_delay_and_gate}), the rising edge of pin \texttt{A} precedes the falling edge of pin \texttt{B} by a small time interval. 
% This temporal misalignment generates a narrow high-level pulse at the output pin \texttt{Y}, which constitutes a glitch phenomenon.
%Industry analysis indicates that glitch-induced power dissipation may constitute up to 40\% of the overall power in modern digital circuits~\cite{synopsys_what_is_glitch_power}.
%Therefore, accurate analysis of glitch power is also essential in power analysis tools.
Industry analysis indicates that glitches may contribute up to 40\% of total power in modern digital circuits~\cite{synopsys_what_is_glitch_power}, making glitch-power analysis essential.

\subsubsection{Power Waveform}
%Although dynamic, leakage, and glitch power offer discrete metrics, they cannot adequately characterize time-dependent power fluctuations. 
Dynamic, leakage, and glitch power summarize component-level consumption but do not capture time-dependent fluctuations.
%Therefore, the per-cycle power waveform, as a time-domain representation, is crucial for diagnosing cycle-specific power anomalies.
The per-cycle power waveform provides this time-domain view and is crucial for diagnosing cycle-specific anomalies.
% It reveals when power deviates from its nominal behavior, enabling direct correlation between anomalous peaks and the corresponding clock cycles.
%Such cycle-level visibility supports fine-grained root-cause analysis, \eg, attributing abnormal power to signal activities, input-state conditions, or transient events within a narrow time window.
This cycle-level visibility supports fine-grained root-cause analysis, \eg, attributing abnormal power to signal activity, input-state conditions, or transient events within a narrow window.
% The mechanisms by which dynamic and leakage power contribute to the power waveform exhibit fundamental differences.
% The dynamic energy dissipation $E_{\mathrm{dyn}}$ occurs exclusively during event occurs. 
% The leakage power value $p_{\mathrm{leak}}^{(g)}$ remains constant when the gate state is unchanged. 
% This process necessitates iterating through all cycles between the two consecutive events.

\subsection{Power Analysis Modes} \label{sec:power_ana_mode}
%Typically, power analysis can be conducted in two distinct modes: the averaged mode and the time-based mode. 
Power analysis is typically performed in averaged mode or time-based mode.
%The averaged mode provides a simplified assessment by calculating average power consumption over a specified period, which is fast but only provides coarse-grained power information.
The averaged mode is fast but reports only coarse-grained power over a specified interval.
The time-based mode enables a comprehensive evaluation of power consumption by examining its temporal variations, thereby effectively capturing transient behaviors and dynamic characteristics of the circuit.
% Time-based mode processes transition events across gates to capture temporal behavior, including dynamic, leakage, glitch, and waveform information, but this event-by-event computation is costly for large designs.
In this mode, the power analysis tool computes power on an event-by-event basis, enabling detailed characterization of power consumption, including dynamic, leakage, and glitch components, as well as the power waveform.
Although the time-based mode enables power analysis with fine temporal granularity, it incurs substantial computational overhead for large-scale designs due to the necessity of processing every transition event across all gates.
% To this end, this work targets specifically on \emph{the GPU acceleration of the gate-level time-based power analysis} to achieve an accurate and ultra-fast power analysis.

In a standard gate-level time-based power-analysis flow, design/model parsing, activity-file parsing, and static timing analysis (STA) provide the inputs required by event-driven power computation~\cite{ptpx_user_guide}.
Design/model parsing reads the gate-level netlist, parasitics, constraints, and Liberty power models, including the internal power LUTs defined in the Liberty files; activity-file parsing reads per-pin transition events from the activity file; and STA provides the slew information required for LUT interpolation and glitch detection.
These inputs are then used in the subsequent event-driven power-computation stage.

\subsection{Time-Based Power Computation with a Naive GPU Implementation} \label{sec:pre_naive_implementation_of_time_based_power_computation}
%To our knowledge, there is currently no open-source implementation of a time-based power analysis tool. 
%Consequently, we build our own naive GPU implementation.
Because no open-source time-based power-analysis implementation is available to our knowledge, we build a naive GPU implementation.
% In the time-based power computation, each gate in the design is treated independently.  
% For a given gate, power is computed on a per-event basis over all events occurring on its pins.   
% \Cref{fig:per_event_power_computation_overview} provides an overview of the power computation process for a given gate, which consists of two components: dynamic power and leakage power.
%Dynamic power is incurred only at discrete event times, whereas leakage power persists continuously over time, so the two contribute to the power waveform in fundamentally different ways.
Dynamic power occurs at discrete event times, whereas leakage power persists continuously; thus, they contribute to the waveform differently.
%Therefore, we implement two naive GPU kernels, Naive-D and Naive-L, each dedicated to one component, to compute them separately.
Therefore, the naive implementation computes them separately using two GPU kernels: Naive-D for dynamic power and Naive-L for leakage power.
% Naive-D assigns one thread per event to compute dynamic power, while Naive-L assigns one thread per clock cycle to compute leakage power.
% The detailed procedures for computing these two components are presented in the remainder of this subsection by introducing two naive GPU kernels.

\subsubsection{Naive-D: Event-Parallel Dynamic Power Kernel} \label{sec:pre_naive_d_event_parallel_dynamic_power_kernel}
\begin{figure*}[tb!]
    \centering
    \includegraphics[width=1.0\linewidth]{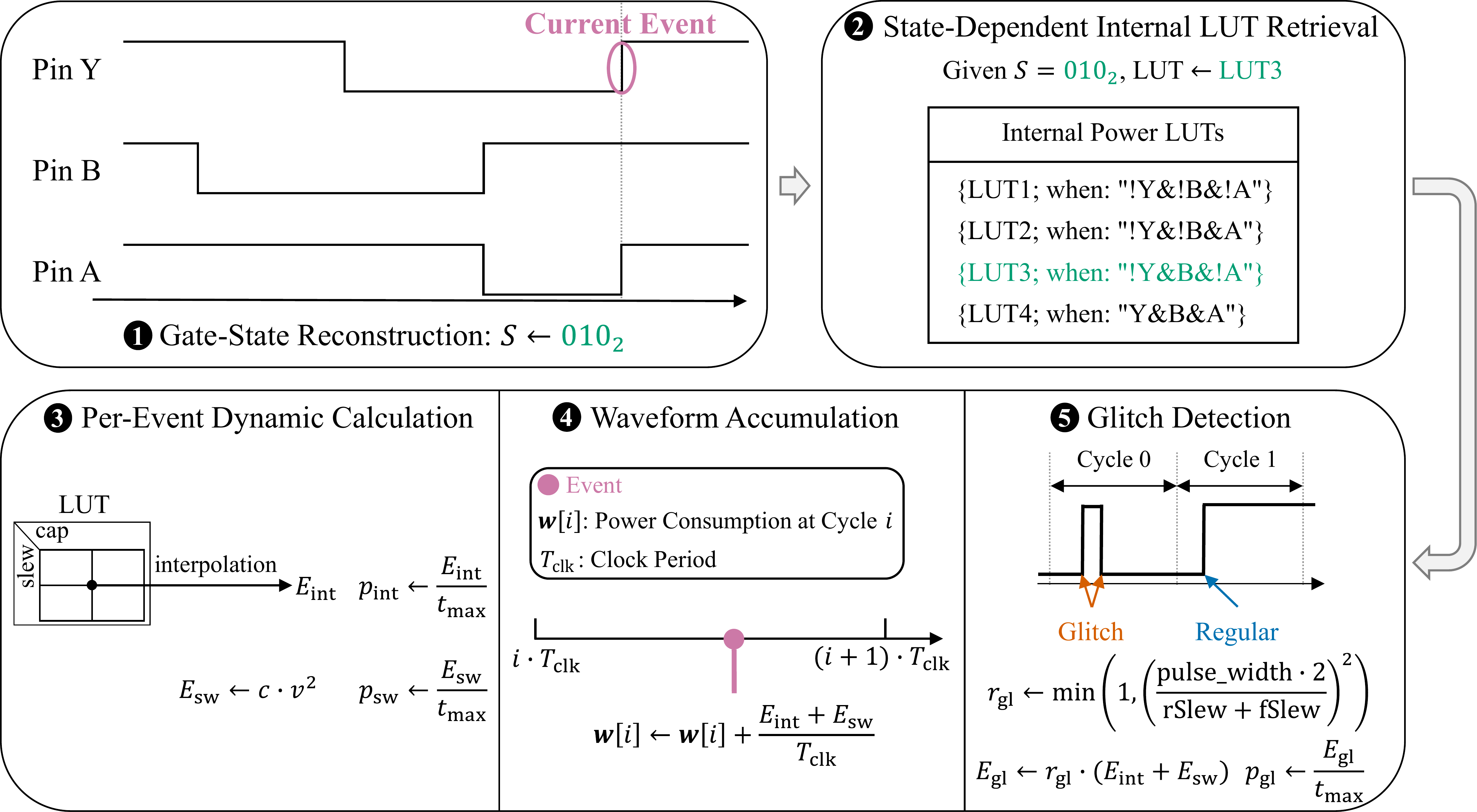}
    \caption{Illustration of the per-event dynamic-power computation flow for Naive-D.}
    \label{fig:overview_of_naive_1t1e_dynamic_kernel} 
\end{figure*}

In this kernel, each GPU thread is mapped to a single event on a specific pin of a gate and computes its dynamic-power contribution.
%This per-event computation includes internal and switching energy calculation, glitch detection, and calculation of the event's contribution to the corresponding power component over the analysis window.
For the assigned event, the computation first reconstructs the preceding gate state~(\Cref{fig:overview_of_naive_1t1e_dynamic_kernel}~\bcircled{\bfseries\scriptsize 1}), scans the state-dependent array to retrieve the matching internal-power LUT~(\Cref{fig:overview_of_naive_1t1e_dynamic_kernel}~\bcircled{\bfseries\scriptsize 2}), and computes the per-event internal and switching energy~(\Cref{fig:overview_of_naive_1t1e_dynamic_kernel}~\bcircled{\bfseries\scriptsize 3}).
% It also accumulates the event's energy contribution into the per-cycle power waveform.
%The detailed procedure is illustrated in \Cref{fig:overview_of_naive_1t1e_dynamic_kernel}.
%It first reconstructs the gate state before the event~(\Cref{fig:overview_of_naive_1t1e_dynamic_kernel}~\bcircled{\bfseries\scriptsize 1}), and scans all items in the state-dependent array to find the internal power LUT corresponding to this gate state for energy calculation~(\Cref{fig:overview_of_naive_1t1e_dynamic_kernel}~\bcircled{\bfseries\scriptsize 2}).
%It then calculates the event's internal energy and switching energy~(\Cref{fig:overview_of_naive_1t1e_dynamic_kernel}~\bcircled{\bfseries\scriptsize 3}).
% It also accumulates the event energy into the per-cycle waveform~\bcircled{\bfseries\scriptsize 4} and performs glitch detection~\bcircled{\bfseries\scriptsize 5}.
%The internal energy $E_{\mathrm{int}}$ is determined by interpolating the retrieved LUT using the input slew and output capacitance as interpolation variables.
The internal energy $E_{\mathrm{int}}$ is obtained by interpolating the retrieved LUT using input slew and output capacitance.
And the switching energy $E_{\mathrm{sw}}$ of a single 0-to-1 event at an output pin is:
\begin{equation}
    \label{eq:switching_energy_calc}
    E_{\mathrm{sw}}=c \cdot v^{2},
\end{equation}
where $c$ is the total output load capacitance, including load pin capacitance and interconnect capacitance, and $v$ is the power supply voltage~\cite{weste2015cmos}.

Next, the event contribution is accumulated into the per-cycle waveform using atomic updates to combine contributions across threads~(\Cref{fig:overview_of_naive_1t1e_dynamic_kernel}~\bcircled{\bfseries\scriptsize 4}).
Finally, glitches are detected with a clock-cycle-based approach~\cite{ptpx_user_guide}~(\Cref{fig:overview_of_naive_1t1e_dynamic_kernel}~\bcircled{\bfseries\scriptsize 5}).
This method, which is widely used in commercial tools, provides an intuitive yet efficient solution.
It detects glitch occurrences by counting the number of events within a single clock cycle. 
If the number of events exceeds one in a clock cycle, the events in that cycle are regarded as glitch events.
% Note that the clock pins should be excluded in glitch detection as the clock signal events twice per cycle.
%After an event is marked as a glitch event, the dynamic energy of it is scaled by a glitch scaling ratio $r_{\mathrm{gl}}$ which is obtained according to the glitch pulse width and the sum of rise and fall slews~\cite{ptpx_user_guide}:
For each glitch event, its dynamic energy is scaled into glitch energy $E_{\mathrm{gl}}$ by a ratio $r_{\mathrm{gl}}$ determined by the glitch pulse width and the rise/fall slews~\cite{ptpx_user_guide}:
\begin{equation}
    \label{eq:glitch_scaling_ratio_calc}
    r_{\mathrm{gl}} = \min \left(1, \left( \frac{\mathrm{pulse\_width} \cdot 2}{\mathrm{rSlew} + \mathrm{fSlew}} \right)^{2} \right),
\end{equation}
where $\mathrm{pulse\_width}$ is the glitch pulse width, $\mathrm{rSlew}$ and $\mathrm{fSlew}$ are rise slew and fall slew, respectively.
%After obtaining the event energy $E$, its contribution $p$ to the corresponding power component over the full analysis window is calculated as:
After obtaining event energy $E$, its contribution $p$ over the full analysis window is:
\begin{equation}
    \label{eq:average_power_calc}
    p=\frac{E}{t_{\max}},
\end{equation}
where $E$ denotes $E_{\mathrm{int}}$, $E_{\mathrm{sw}}$, or $E_{\mathrm{gl}}$, and $t_{\max}$ is the maximum event time.
Accumulating these per-event contributions gives the design-level internal, switching, and glitch power components.

\subsubsection{Naive-L: Cycle-Parallel Leakage Power Kernel} \label{sec:pre_naive_l_cycle_parallel_leakage_power_kernel}

\begin{figure*}[tb!]
    \centering
    \includegraphics[width=1.0\linewidth]{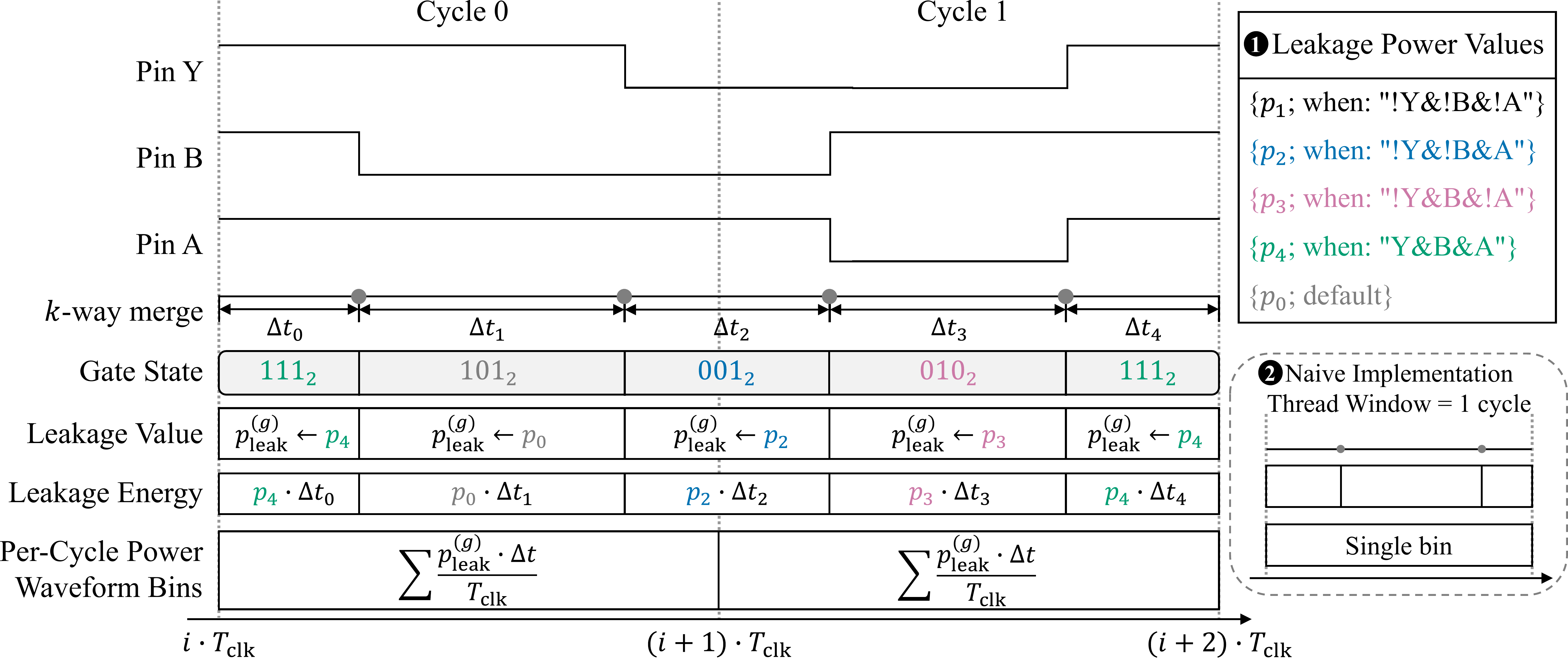}
    \caption{Illustration of the time-based power computation flow for Naive-L.}
    \label{fig:overview_of_naive_leakage_kernel} 
\end{figure*}

% In general, Naive-L partitions the time domain into cycle windows, each consisting of a contiguous range of clock cycles.
% Each thread is assigned one such window and computes the corresponding leakage power.
In general, Naive-L partitions the temporal dimension into cycle windows of $n_{\mathrm{cyc}}^{\mathrm{fix}}$ contiguous cycles, where $n_{\mathrm{cyc}}^{\mathrm{fix}}$ is globally fixed across all gates.
%Each thread is assigned one such window for a given gate and computes the leakage power over the window.
%This computation comprises leakage power calculation and accumulation of each cycle's contribution to the power waveform.
Each thread processes one window for one gate, computing leakage power and accumulating its per-cycle waveform contribution.
% In this kernel, each GPU thread is assigned to a cycle of a gate and computes the corresponding leakage power.
The detailed procedure is depicted in \Cref{fig:overview_of_naive_leakage_kernel}.
The thread first finds each pin's event-index range in its assigned window and reconstructs the gate state at the window start.
It then performs an on-the-fly $k$-way merge over per-pin event streams.
%It then performs an on-the-fly $k$-way merge over the per-pin event streams by repeatedly selecting the minimum next event timestamp across pins.
At each timestamp, it scans the state-dependent array~(\Cref{fig:overview_of_naive_leakage_kernel}~\bcircled{\bfseries\scriptsize 1}) to retrieve the current $p_{\mathrm{leak}}^{(g)}$, which remains constant over the preceding inter-event interval.
The leakage energy of that interval is:
\begin{equation}
    \label{eq:leakage_energy_calc}
    E_{\mathrm{leak}}=p_{\mathrm{leak}}^{(g)} \cdot \Delta t,
\end{equation}
where $\Delta t$ represents the time interval between the two consecutive events.

For power waveform accumulation, each inter-event interval is split at clock-cycle boundaries, and the leakage contribution is added to every overlapped cycle bin in proportion to the overlap duration between the interval and that cycle.

After processing each timestamp, the gate state is updated by applying all events occurring at that timestamp, and the procedure repeats until the end of its assigned window.

The workload in Naive-L is two-fold: (i) the event-side cost, arising from the on-the-fly $k$-way merge over per-pin event streams and the associated gate-state updates; and (ii) the cycle-side cost, arising from updating the per-cycle power waveform bins for the thread's assigned cycles.

\subsubsection{Parallel Granularity of Naive Kernels} \label{sec:pre_parallel_granularity_of_naive_baselines}
In Naive-D, each thread is assigned exactly one event.
% Increasing the number of events handled by a single thread would require maintaining a consistent state evolution across events from all pins of the assigned gate, which in turn necessitates enforcing a well-defined ordering across multiple pin event streams.
Increasing the number of events handled by a single thread would require maintaining a consistent gate-state evolution across multiple pins, and thus a non-trivial kernel redesign.
Consequently, the parallel granularity of Naive-D is inherently fixed to one event per thread.
% , and becomes tunable only after redesigning the task mapping and state management.
In contrast, Naive-L offers a tunable parallel parameter $n_{\mathrm{cyc}}^{\mathrm{fix}}$.
For the naive implementation, we use the one-cycle-per-thread configuration ($n_{\mathrm{cyc}}^{\mathrm{fix}}=1$), which represents the finest-grained setting~(\Cref{fig:overview_of_naive_leakage_kernel}~\bcircled{\bfseries\scriptsize 2}).
Together, these two kernels form the naive implementation for power computation in this work: \emph{Naive-D adopts one-event-per-thread parallelism, whereas Naive-L employs one-cycle-per-thread parallelism.}

\section{Characterization and Motivation} \label{sec:characterization_and_motivation}
% This section outlines the challenges associated with the naive implementation introduced in \Cref{sec:pre_naive_implementation_of_time_based_power_computation}, 
This section outlines three key challenges in leveraging GPUs for time-based power computation, derived from an analysis of the naive implementation introduced in \Cref{sec:pre_naive_implementation_of_time_based_power_computation}.
% including inefficient state-dependent power retrieval, unbalanced event distribution across gates, and redundant work in separate kernels.

\subsection{Challenge \#1: Inefficient State-Dependent Power Retrieval} \label{sec:challenge_1_inefficient_power_retrieval}
As discussed in \Cref{sec:pre_leak_power_and_dyn_power}, the internal power and leakage power in a digital design can be state-dependent. 
One straightforward way to analyze state-dependent power consumption is to exhaustively traverse all state patterns and find the match. 
A significant limitation of this method is its computational inefficiency, as each event necessitates two separate matching procedures for leakage and internal power, respectively. 
% Each procedure incurs a time complexity of $O(n_{\mathrm{pin}}^{(g)} \cdot n_{\mathrm{sd}}^{(g)})$, where $n_{\mathrm{pin}}^{(g)}$ denotes the number of pins of the assigned gate and $n_{\mathrm{sd}}^{(g)}$ denotes the number of state-dependent power items, resulting in substantial runtime overhead.
Given a gate $g$ with pin count $n_{\mathrm{pin}}^{(g)}$ and $n_{\mathrm{sd}}^{(g)}$ state-dependent power entries, each procedure runs in $O\!\left(n_{\mathrm{pin}}^{(g)} \cdot n_{\mathrm{sd}}^{(g)}\right)$ time, incurring nontrivial runtime overhead.
Therefore, an efficient data structure is required to reduce the overhead of retrieving state-dependent power information.

\subsection{Challenge \#2: Imbalanced Event Distribution Across Gates} \label{sec:challenge_2_imbalanced_event_dist}

\begin{figure}[tb!]
    \centering
    \definecolor{myorange}{RGB}{255,166,26}
\definecolor{myred}{RGB}{227,107,98}
\definecolor{myyellow}{RGB}{255,214,75}
\definecolor{mymiddleblue}{RGB}{139,201,246} 
\definecolor{mylight}{RGB}{38,192,159}  
\definecolor{mygreen}{RGB}{205,252,197}   
\definecolor{myblue}{RGB}{29,114,221}    %1D72DD
\definecolor{mybrown}{RGB}{120, 80, 40} 

\begin{tikzpicture}
\pgfplotstableread[col sep=space]{
  name                   p50  p90  p99
  Adder(random)         46318  220141  265077
  Multiplier(random)    12301.2  51392  188101
  Rocket(dhrystone)     11  529574  560760
  SmallBOOM(dhrystone)  253.667  111763  152265
  SmallBOOM(median)     161  167546  192573
  SmallBOOM(mt-vvadd)   392.5  193103  210717
  GigaBOOM(dhrystone)   532.308  55295.1  62687.2
  GigaBOOM(median)      249.25  79230.2  83313.4
  GigaBOOM(mt-vvadd)    237.091  98496.4  101372
}{\dataTableGateEventDistAcrossDatasets}

\begin{axis}[
    width=\linewidth,
    height=0.26\linewidth, % 你想要的高度比例
    ymode=log,
    ymin=1,
    enlarge x limits=0.03,
    symbolic x coords={
        Adder(random),
        Multiplier(random),
        Rocket(dhrystone),
        SmallBOOM(dhrystone),
        SmallBOOM(median),
        SmallBOOM(mt-vvadd),
        GigaBOOM(dhrystone),
        GigaBOOM(median),
        GigaBOOM(mt-vvadd)
    },
    xtick=data,
    xticklabels={
        \shortstack{Adder\\(random)}, 
        \shortstack{Multiplier\\(random)}, 
        \shortstack{Rocket\\(dhrystone)}, 
        \shortstack{SmallBOOM\\(dhrystone)}, 
        \shortstack{SmallBOOM\\(median)}, 
        \shortstack{SmallBOOM\\(mt-vvadd)}, 
        \shortstack{GigaBOOM\\(dhrystone)},
        \shortstack{GigaBOOM\\(median)},
        \shortstack{GigaBOOM\\(mt-vvadd)}
    },
    ylabel={Per-gate event counts},
    ylabel style={font=\footnotesize},
    ymajorgrids,
    x tick label style={font=\footnotesize},
    y tick label style={font=\footnotesize},
    grid style={dashed},
    legend style={at={(0.99,0.01)}, font=\footnotesize, legend columns=-1, anchor=south east}
]
    % interval [p50, p99] as whisker, centered at p90
    \addplot+[only marks, color=mybrown, mark=triangle*, mark options={fill=mybrown}] table[x=name, y=p99] {\dataTableGateEventDistAcrossDatasets};
    \addlegendentry{$p_{99}$}

    \addplot+[
        only marks,
        color=myblue,
        mark=*,
        mark options={fill=myblue},
        error bars/.cd,
        y dir=both,
        y explicit,
    ] table[
        x=name,
        y=p90,
        y error plus expr=\thisrow{p99}-\thisrow{p90},
        y error minus expr=\thisrow{p90}-\thisrow{p50},
    ] {\dataTableGateEventDistAcrossDatasets};
    \addlegendentry{$p_{90}$}
    
    \addplot+[only marks, color=myorange, mark=square*, mark options={fill=myorange}] table[x=name, y=p50] {\dataTableGateEventDistAcrossDatasets};
    \addlegendentry{$p_{50}$}

\end{axis}
\end{tikzpicture}
    \caption{Per-gate event-count percentiles ($p_{50}$, $p_{90}$, $p_{99}$) across nine datasets, shown on a logarithmic scale. The statistics of the evaluated designs are shown in \Cref{tab:design_statistics}, with the benchmark names indicated in parentheses.}
    \label{fig:dist_of_per_gate_event_counts}
    % \vspace{-0.4cm}
\end{figure}
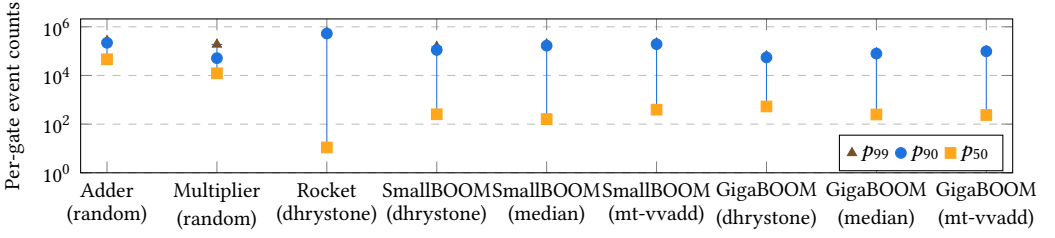

In time-based power analysis, there are two dimensions of parallelism to leverage.
Spatially, gate-level parallelism can be leveraged since the power is calculated separately for each gate.
Temporally, event-level or cycle-level parallelism is also available, as the events or cycles of a gate are also independent in the power computation.
It is straightforward to leverage the spatial parallelism, as the power computations among different gates are independent. 
% However, the number of events per gate can be highly skewed, which complicates the effective use of temporal parallelism.
In contrast, temporal parallelism within each gate is harder to exploit efficiently, because highly skewed per-gate event counts make a global temporal granularity poorly balanced.  
\Cref{fig:dist_of_per_gate_event_counts} reports the $p_{50}$, $p_{90}$, and $p_{99}$ (\ie the $50$th/$90$th/$99$th percentiles) of per-gate event counts across nine datasets.  
We observe that event counts at $p_{90}$ and $p_{99}$ are substantially larger than at $p_{50}$, reaching up to four orders of magnitude higher on \texttt{Rocket}'s \texttt{dhrystone} benchmark.  
This suggests a highly skewed distribution of event counts across gates.
However, Naive-L adopts a uniform partitioning granularity across gates, with a fixed cycles-per-thread setting.
% In the naive implementation, one cycle is assigned to each thread.
As a result, the number of events processed per thread varies widely, leading to substantial variation in power computation workload across threads. 
This implies that a globally fixed cycles-per-thread parameter $n_{\mathrm{cyc}}^{\mathrm{fix}}$ cannot simultaneously accommodate all gates, and therefore fails to distribute the workload effectively. 
If $n_{\mathrm{cyc}}^{\mathrm{fix}}$ is too small, sparse gates with few events may still be allocated many threads.  
Many of these threads perform little useful work and then terminate, incurring substantial scheduling overhead.  
Conversely, if $n_{\mathrm{cyc}}^{\mathrm{fix}}$ is too large, the thread budget for dense gates with many events is reduced, preventing their heavier workloads from being effectively distributed and increasing the thread work depth.
Therefore, we need fine-grained control over the parallel granularity of each gate according to its workload. 

\subsection{Challenge \#3: Redundant Work in Separate Kernels} \label{sec:challenge_3_redundant_computation}
Both Naive-D and Naive-L perform gate-state reconstruction, leading to substantial redundant work when the two kernels are executed separately.
Specifically, each thread must reconstruct the gate state at its assigned start time to enable state-dependent power retrieval.
This is accomplished by performing a binary search on the event stream of each pin to identify the event index corresponding to its start time.
% incurring $O\!\left(\sum_{i=1}^{n_{\mathrm{pin}}} \log_{2}(k_i)\right)$ work per thread.
Consider a gate $g$ with $n_{\mathrm{pin}}^{(g)}$ pins.
Let $k_i$ denote the event count at the $i$-th pin. 
The work required to find an event index corresponding to a specific time at pin $i$ is $\log_{2}k_i$, with each binary search iteration counting as one work unit. 
For a given thread, the work required to search the event indices associated with a specific time for gate $g$ is:
\begin{equation}
W_{\mathrm{search}}^{(g)} = \sum_{i=1}^{n_{\mathrm{pin}}^{(g)}} \log_{2}k_i
\end{equation}
When allocating $N_{\mathrm{thr}}^{(g)}$ threads to each gate $g$, the total work required to reconstruct the gate states of the entire design is:
\begin{equation}
    \label{eq:total_work_of_gate_state_reconstruction}
    \sum_{g \in G} N_{\mathrm{thr}}^{(g)} \cdot W_{\mathrm{search}}^{(g)}
\end{equation}
Since two separate kernels are launched, the total number of threads $N_{\mathrm{thr}}^{(g)}$ increases, thereby incurring non-negligible overhead for gate-state reconstruction.

In addition, Naive-D performs per-event glitch detection.
This requires identifying the neighboring events within the same clock cycle via per-pin binary searches to locate cycle boundaries, incurring $O(\log_{2}k_{i})$ work per event on pin $i$.
% This requires identifying the neighboring events within the same clock cycle, incurring $O(\log_{2}k_{i})$ work per event on pin $i$.
% , followed by constant-time neighbor checks.
However, Naive-L already has access to cycle-level event count, because it explicitly traverses the event streams of all pins over time, thereby exposing the number of events that occur in each clock cycle.

Consequently, running Naive-D and Naive-L independently not only duplicates gate-state reconstruction but also recomputes glitch-related cycle statistics that could be reused, motivating a unified formulation that shares gate-state reconstruction and cycle-level information.

\section{Multi-Level GPU Optimizations} \label{sec:multi_level_gpu_opt_framework}

\begin{figure*}[tb!]
    \centering
    \includegraphics[width=0.90\linewidth]{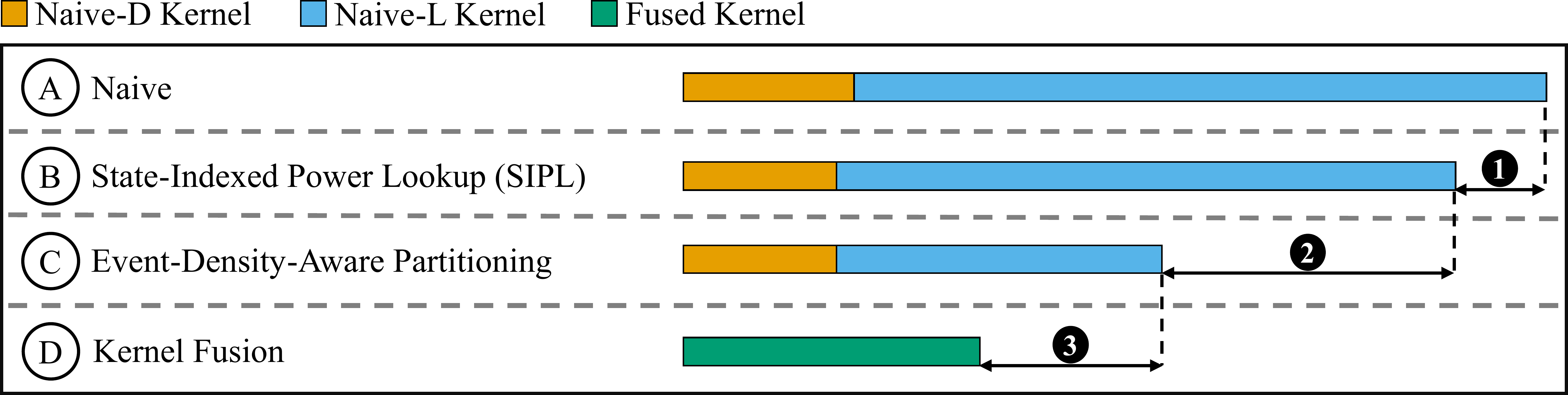}
    \caption{Overview of a series of incremental optimizations applied to the power computation kernels: timelines from the naive implementation to the optimized version. Intended for illustrative purposes only; detailed experimental results are presented in \Cref{sec:incremental_opt_evaluation}.}
    \label{fig:stepwise_optimization_timelines} 
\end{figure*}

In this section, we introduce our GPU optimization techniques for addressing the challenges discussed in \Cref{sec:characterization_and_motivation}.
For clarity in presenting our proposed methods, we have designed \Cref{fig:stepwise_optimization_timelines} as a multi-timeline visualization where each timeline graphically represents a distinct optimization technique.
We begin with the naive implementation proposed in \Cref{sec:pre_naive_implementation_of_time_based_power_computation} (\Cref{fig:stepwise_optimization_timelines} \textcircled{\scriptsize A}), where dynamic power is computed by the Naive-D kernel (one event per thread) and leakage power is computed by the Naive-L kernel (one cycle per thread).

\subsection{Efficient State-Dependent Power Retrieval with State-Indexed Power Lookup} \label{sec:efficient_power_retrieval_with_bsim}

\begin{figure}[tb!]
    \centering
    \includegraphics[width=.9\linewidth]{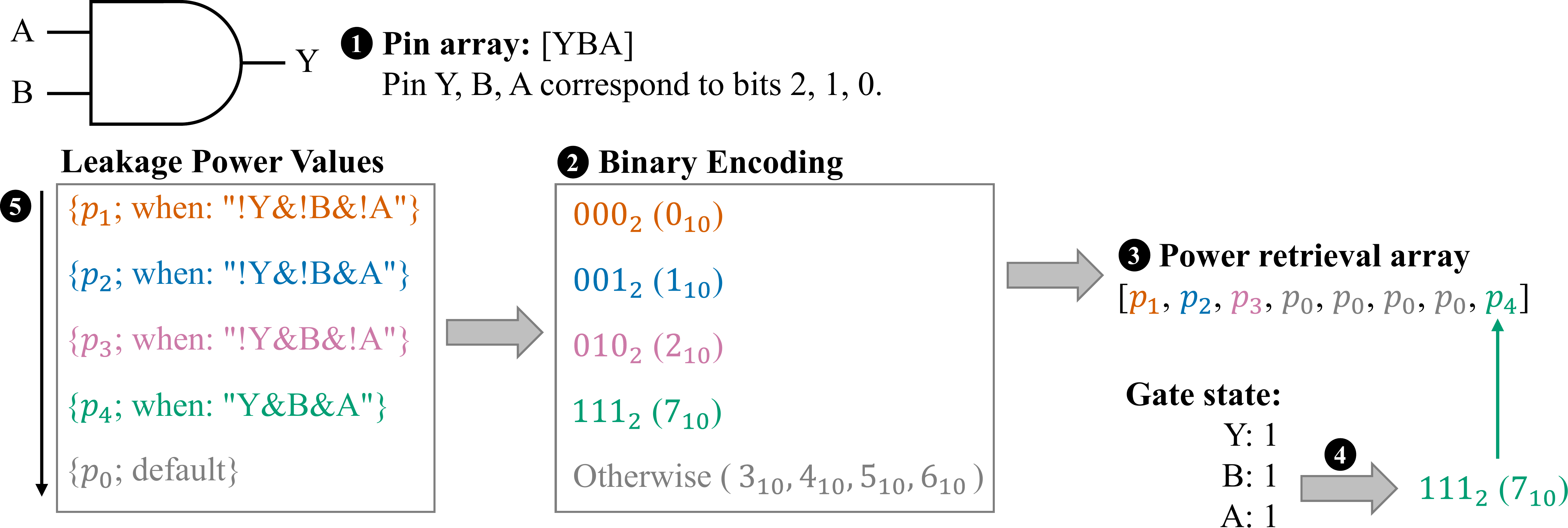}
    \caption{State-Indexed Power Lookup applied to the leakage power of an \texttt{AND} gate.}
    \label{fig:binary_index_mapping}
\end{figure}
As discussed in \Cref{sec:challenge_1_inefficient_power_retrieval}, each event requires two $O\!\left(n_{\mathrm{pin}}^{(g)} \cdot n_{\mathrm{sd}}^{(g)}\right)$ state-dependent power retrieval processes for leakage and internal power, respectively.
To reduce this per-event overhead, we convert the traditional state-pattern traversal process into a \emph{table lookup} using \textit{State-Indexed Power Lookup} (SIPL).
% we propose an approach called \textit{State-Indexed Power Lookup (SIPL)} for efficient power retrieval, as illustrated in \Cref{fig:binary_index_mapping}.
Specifically, during preprocessing, we process each cell in the Liberty file and organize its pins into a fixed-order array on the host side~(\Cref{fig:binary_index_mapping}~\bcircled{\bfseries\scriptsize 1}); this pin order is preserved in all subsequent computations.
Each Liberty state pattern is then binary-encoded over this pin order, with each bit set according to the corresponding pin state~(\Cref{fig:binary_index_mapping}~\bcircled{\bfseries\scriptsize 2}).
The resulting binary-encoded value is used as an integer index into a $2^{n_{\mathrm{pin}}^{(g)}}$-entry power retrieval array, where $n_{\mathrm{pin}}^{(g)}$ denotes the cell pin count, and the corresponding power item is stored at the indexed entry~(\Cref{fig:binary_index_mapping}~\bcircled{\bfseries\scriptsize 3}).

Some Liberty state patterns specify only a subset of pins.
Pins omitted from a pattern are treated as don't-care pins, meaning that the same power item applies regardless of their logic values.
Thus, one state pattern may correspond to multiple SIPL indices.
To populate the retrieval array, we process each Liberty state pattern by enumerating all binary assignments of its don't-care pins while keeping its specified pin values fixed.
Each complete pin assignment is then binary-encoded into an array index, and the indexed entry is filled with the power item of the current state pattern.

During power computation, we encode the current gate state using the same pin order and use the encoded value as the array index, allowing us to retrieve the power item directly (\Cref{fig:binary_index_mapping}~\bcircled{\bfseries\scriptsize 4}).
The same construction is also applied to internal power LUTs, with the difference being that the elements in the power retrieval array are LUTs.

One important consideration of this method is that the number of pins can be too large. 
For example, the number of pins of an SRAM macro can be over one hundred. % need a check
Directly mapping its state patterns into an array will result in the array size exceeding $2^{100}$, which exceeds the memory capacity. 
To avoid excessive retrieval-array sizes, we use SIPL only for gates with at most $16$ pins and fall back to direct state-pattern traversal for gates with more than $16$ pins.
In the fallback path~(\Cref{fig:binary_index_mapping}~\bcircled{\bfseries\scriptsize 5}), the original state patterns are scanned, each pattern is checked against the current gate state, and the power item associated with the matching pattern is retrieved.
The choice of the 16-pin threshold is justified in \Cref{sec:bsim_fallback_threshold_tuning}.
With this threshold, only 0.05\% of gates invoke this fallback strategy in our experiments.
By replacing state-pattern traversal with direct indexed access, SIPL reduces the complexity of state-dependent power retrieval from $O\!\left(n_{\mathrm{pin}}^{(g)} \cdot n_{\mathrm{sd}}^{(g)}\right)$ to $O\!\left(n_{\mathrm{pin}}^{(g)}\right)$.
As shown in \Cref{fig:stepwise_optimization_timelines}, our SIPL data structure (\textcircled{\scriptsize B}) reduces GPU kernel execution time by \bcircled{\bfseries\scriptsize 1} over \textcircled{\scriptsize A}.

\subsection{Adaptive Per-Gate Temporal Partitioning} \label{sec:adaptive_temporal_part}

In this subsection, we present an event-density-aware, per-gate partitioning scheme to accommodate the imbalanced event distribution across gates.
% We first describe the partitioning procedure and then characterize the resulting thread count.
% As discussed in \Cref{sec:challenge_2_imbalanced_event_dist}, the event counts across gates can be highly skewed.
% Under such inter-gate heterogeneity, using a globally fixed cycles-per-thread parameter $n_{\mathrm{cyc}}^{\mathrm{fix}}$ for Naive-L leads to systematic misallocation of parallelism.
% a small $n_{\mathrm{cyc}}^{\mathrm{fix}}$ oversubscribes threads for sparse gates, whereas a large $n_{\mathrm{cyc}}^{\mathrm{fix}}$ undersubscribes threads for dense gates.
% To address this issue, we propose an adaptive per-gate temporal partitioning strategy that allocates parallelism using a gate-specific event-density model, thereby improving parallelism utilization and reducing over-parallelization overhead.

\subsubsection{Event-Density-Aware Per-Gate Cycle Partitioning}
% We adapt the temporal partitioning granularity on a per-gate basis to cope with the wide variation in event count. 
As discussed in \Cref{sec:challenge_2_imbalanced_event_dist}, the event counts across gates can be highly skewed.
Under such inter-gate heterogeneity, using a globally fixed cycles-per-thread setting for Naive-L leads to systematic misallocation of parallelism.
To accommodate the wide variation in per-gate event count, we adapt the temporal partitioning granularity on a \emph{per-gate basis}.
Specifically, we introduce a target event count per thread budget $n_{\mathrm{evt,tgt}}$.
For each gate $g$, we select the cycles-per-thread parameter $n_{\mathrm{cyc}}^{(g)}$ such that each thread is expected to process approximately $n_{\mathrm{evt,tgt}}$ events.
Let $N_{\mathrm{cyc}}$ denote the total number of cycles and let $N_{\mathrm{evt}}^{(g)}$ denote the total number of events associated with gate $g$.
% We partition the $N_{\mathrm{cyc}}$ cycles into contiguous windows of $n_{\mathrm{cyc}}^{(g)}$ cycles and assign each window to a single thread.
We partition the events of gate $g$ along the temporal dimension by dividing the $N_{\mathrm{cyc}}$ cycles into contiguous windows of length $n_{\mathrm{cyc}}^{(g)}$.
Each thread processes the events of gate $g$ whose timestamps fall within its assigned $n_{\mathrm{cyc}}^{(g)}$-cycle window.
To estimate the expected per-thread workload, we define the average event density of gate $g$ to be
\begin{equation}
\rho^{(g)} = \frac{N_{\mathrm{evt}}^{(g)}}{N_{\mathrm{cyc}}}.
\end{equation}
% Under this model, the expected event workload per thread is approximately $\rho^{(g)} \cdot n_{\mathrm{cyc}}^{(g)}$.  
Treating $\rho^{(g)}$ as a coarse average event density over the full time range, a thread assigned $n_{\mathrm{cyc}}^{(g)}$ cycles is expected to process approximately $\rho^{(g)} \cdot n_{\mathrm{cyc}}^{(g)}$ events.
Accordingly, a natural choice is $n_{\mathrm{cyc}}^{(g)} \approx \frac{n_{\mathrm{evt,tgt}}}{\rho^{(g)}}$, obtained by matching the expected per-thread event count to the target budget $n_{\mathrm{evt,tgt}}$.  
% A direct inversion would suggest $n_{\mathrm{cyc}}^{(g)} \approx \frac{n_{\mathrm{evt,tgt}}}{\rho^{(g)}}$.  
However, this choice becomes ill-conditioned when $\rho^{(g)}$ is small and may lead to excessively large windows.
As discussed in \Cref{sec:pre_naive_l_cycle_parallel_leakage_power_kernel}, the workload in Naive-L is inherently two-fold, comprising the event-side cost and the cycle-side cost.
Therefore, for sparse gates with small $\rho^{(g)}$, choosing the parallel granularity solely based on event density can under-provision threads for the cycle-side work, leaving clock-cycle traversal and power waveform updates to only a small number of threads.
To stabilize the granularity in the sparse regime, we introduce a parallelism floor $N_{\mathrm{thr,floor}}$, which induces a window size
\begin{equation}
n_{\mathrm{cyc,floor}}=\left\lceil \frac{N_{\mathrm{cyc}}}{N_{\mathrm{thr,floor}}}\right\rceil.
\end{equation}
We then convert this window size into a density-floor term
\begin{equation}
\rho_{\mathrm{floor}}=\frac{n_{\mathrm{evt,tgt}}}{n_{\mathrm{cyc,floor}}}.
\end{equation}
We combine the measured density and the density-floor term into an effective density
\begin{equation}
\rho_{\mathrm{eff}}^{(g)}=\rho^{(g)}+\rho_{\mathrm{floor}}.
\end{equation}
Intuitively, $\rho_{\mathrm{floor}}$ acts as a density floor that prevents the inferred window size from growing too large when the measured density $\rho^{(g)}$ is small.  

The final cycles-per-thread granularity is obtained by inverting $\rho_{\mathrm{eff}}^{(g)}$ with integer rounding,
\begin{equation}
n_{\mathrm{cyc}}^{(g)}
=
\mathrm{round}\!\left(\frac{n_{\mathrm{evt,tgt}}}{\rho_{\mathrm{eff}}^{(g)}}\right).
\end{equation}
% Here, $\mathrm{round}(\cdot)$ produces an integer granularity that closely approximates the continuous optimum, and we subsequently clamp $n_{\mathrm{cyc}}^{(g)}$ to the valid range $[1,\,N_{\mathrm{cyc}}]$ to ensure feasibility.
Given a thread with local index $\tau^{(g)}$ within gate $g$ and clock period $T_{\mathrm{clk}}$, the time window $[t_{\mathrm{start}}, t_{\mathrm{end}})$ assigned to it is:
\begin{align}
    t_{\mathrm{start}} &= \tau^{(g)} \cdot n_{\mathrm{cyc}}^{(g)} \cdot T_{\mathrm{clk}} \label{eq:thread_start_time} \\
    t_{\mathrm{end}}   &= (\tau^{(g)}+1) \cdot n_{\mathrm{cyc}}^{(g)} \cdot T_{\mathrm{clk}} \label{eq:thread_end_time}
\end{align}

\subsubsection{Thread-Count Characterization}
We now characterize the thread-count behavior induced by the proposed event-density-aware, per-gate cycle partitioning scheme.
% In this paper, $N_{\mathrm{thr,floor}}$ denotes a \emph{target} (desired) level of parallelism rather than the exact realized thread count.
% Given a desired parallelism $N_{\mathrm{thr,floor}}$ and a per-thread event budget $n_{\mathrm{evt,tgt}}$, we first define the target cycles per thread as
% $$
% n_{\mathrm{cyc,floor}}=\left\lceil \frac{N_{\mathrm{cyc}}}{N_{\mathrm{thr,floor}}}\right\rceil.
% $$
% The event density of $g$ and the baseline density are
% $$
% \rho^{(g)}=\frac{N_{\mathrm{evt}}}{N_{\mathrm{cyc}}},\qquad
% \rho_{\mathrm{floor}}=\frac{n_{\mathrm{evt,tgt}}}{n_{\mathrm{cyc,floor}}}.
% $$
% We then form an effective density $\rho_{\mathrm{eff}}=\rho^{(g)}+\rho_{\mathrm{floor}}$ and choose the final cycles per thread as

The number of cycles per thread is
\begin{equation}
n_{\mathrm{cyc}}^{(g)} = \mathrm{round}\!\left(\frac{n_{\mathrm{evt,tgt}}}{\rho_{\mathrm{eff}}^{(g)}}\right) \approx \frac{n_{\mathrm{evt,tgt}}}{\rho^{(g)}+\rho_{\mathrm{floor}}}.
\end{equation}
% where $\rho_{\mathrm{eff}}^{(g)}=\rho^{(g)}+\rho_{\mathrm{floor}}$. To interpret the induced thread count, we ignore rounding and approximate
% $$
% n_{\mathrm{cyc}}^{(g)} \approx \frac{n_{\mathrm{evt,tgt}}}{\rho^{(g)}+\rho_{\mathrm{floor}}}.
% $$
Therefore, the number of threads assigned to $g$ is
\begin{equation}
N_{\mathrm{thr}}^{(g)}
= \left\lceil \frac{N_{\mathrm{cyc}}}{n_{\mathrm{cyc}}^{(g)}}\right\rceil \approx \frac{N_{\mathrm{cyc}} \cdot (\rho^{(g)}+\rho_{\mathrm{floor}})}{n_{\mathrm{evt,tgt}}}
= \frac{N_{\mathrm{cyc}} \cdot \rho^{(g)}}{n_{\mathrm{evt,tgt}}}+\frac{N_{\mathrm{cyc}} \cdot \rho_{\mathrm{floor}}}{n_{\mathrm{evt,tgt}}}.
\end{equation}
% Substituting $\rho^{(g)} = \frac{N_{\mathrm{evt}}^{(g)}}{N_{\mathrm{cyc}}}$ gives
% $$
% \frac{N_{\mathrm{cyc}} \cdot \rho^{(g)}}{n_{\mathrm{evt,tgt}}} = \frac{N_{\mathrm{evt}}^{(g)}}{n_{\mathrm{evt,tgt}}}.
% $$
% Substituting $\rho_{\mathrm{floor}} = \frac{n_{\mathrm{evt,tgt}}}{n_{\mathrm{cyc,floor}}}$ gives
% $$
% \frac{N_{\mathrm{cyc}} \cdot \rho_{\mathrm{floor}}}{n_{\mathrm{evt,tgt}}} = \frac{N_{\mathrm{cyc}}}{n_{\mathrm{cyc,floor}}}.
% $$
Substituting $\rho^{(g)} = \frac{N_{\mathrm{evt}}^{(g)}}{N_{\mathrm{cyc}}}$ and $\rho_{\mathrm{floor}} = \frac{n_{\mathrm{evt,tgt}}}{n_{\mathrm{cyc,floor}}}$ gives
\begin{equation}
N_{\mathrm{thr}}^{(g)} \approx \frac{N_{\mathrm{evt}}^{(g)}}{n_{\mathrm{evt,tgt}}}+\frac{N_{\mathrm{cyc}}}{n_{\mathrm{cyc,floor}}}.
\end{equation}
Moreover, because $n_{\mathrm{cyc,floor}}=\left\lceil N_{\mathrm{cyc}}/N_{\mathrm{thr,floor}}\right\rceil$, we obtain
\begin{equation}
\frac{N_{\mathrm{cyc}}}{n_{\mathrm{cyc,floor}}}\approx N_{\mathrm{thr,floor}},
\end{equation}
% where the approximation error is solely due to the ceiling operation.
Consequently, the number of threads allocated to gate $g$ can be approximated as
\begin{equation}
N_{\mathrm{thr}}^{(g)} \approx \frac{N_{\mathrm{evt}}^{(g)}}{n_{\mathrm{evt,tgt}}}+N_{\mathrm{thr,floor}}.
\end{equation}

This additive form clarifies the behavior in two limiting regimes.
When the workload is sparse ($N_{\mathrm{evt}}^{(g)}$ is small and thus $\rho^{(g)}$ is small), the event-driven term $\frac{N_{\mathrm{evt}}^{(g)}}{n_{\mathrm{evt,tgt}}}$ becomes negligible, and the thread count $N_{\mathrm{thr}}^{(g)}$ is dominated by the parallelism floor $N_{\mathrm{thr,floor}}$, thereby avoiding over-parallelization while still maintaining sufficient parallelism.
Conversely, when the workload is dense ($N_{\mathrm{evt}}^{(g)}$ is large and thus $\rho^{(g)}$ is large), the event-driven term $\frac{N_{\mathrm{evt}}^{(g)}}{n_{\mathrm{evt,tgt}}}$ dominates $N_{\mathrm{thr}}^{(g)}$, yielding a finer cycle granularity and allocating more parallelism (larger $N_{\mathrm{thr}}^{(g)}$) to gates with heavier workloads.

% Therefore, the proposed rule provides a smooth transition between event-budgeted partitioning (dense regime) and parallelism-preserving partitioning (sparse regime).  
In summary, the proposed event-density-aware per-gate cycle partitioning scheme reallocates parallelism according to each gate's event density.
It avoids over-parallelization for sparse gates while increasing parallelism when event processing dominates, thereby improving overall parallelism utilization and reducing the overhead associated with excessive parallelization.
As demonstrated in \Cref{fig:stepwise_optimization_timelines}, this approach (\textcircled{\scriptsize C}) achieves \bcircled{\bfseries\scriptsize 2} reduction in kernel execution time compared to \textcircled{\scriptsize B}.

\subsection{Kernel Fusion for Unified Power Computation} \label{sec:kernel_fusion}

\begin{algorithm}[tbp]
    \caption{Fused Power Computation Kernel}
    \label{algo:fused_power_computation_kernel_pseudocode}
    \renewcommand{\algorithmicrequire}{\textbf{Input:}}
    \renewcommand{\algorithmicensure}{\textbf{Output:}}
    \begin{algorithmic}[1]
    \Require The current gate $g$, the local thread index within gate $g$, the cycles-per-thread granularity of gate $g$, the clock period $T_{\mathrm{clk}}$, the maximum of all event times, the design-level per-cycle power waveform $\vec{w}$.
    \Ensure Thread-local leakage, internal, switching, and glitch power values, along with their contributions to the design-level per-cycle power waveform.
    \State Initialize $\vec{p} \gets \vec{0}$ \label{line:init_p}
    \State Initialize $t_{\mathrm{start}}$ and $t_{\mathrm{end}}$ \label{line:init_t} 
    \State Initialize gate state $S$ at $t_{\mathrm{start}}$ \label{line:init_s}
    \State Initialize per-pin glitch counter $\vec{c} \gets \vec{0}$ \label{line:init_glitch_counter}

    \For{$t_{\mathrm{cur}} \gets$ the next minimum unprocessed event timestamp in $[t_{\mathrm{start}}, t_{\mathrm{end}})$} \label{line:start_of_power_calc_loop}
        % \State Get the minimum unprocessed event time $t_{\mathrm{cur}}$ \label{line:get_minimum_unprocessed_t}
        \State Get leakage power value $p_{\mathrm{leak}}^{(g)}$ by $S$ \label{line:get_leak_p}
        \State Update leakage power $p_{\mathrm{leak}}$ and waveform $\vec{w}$ \label{line:add_leak_to_p_and_w}

        \ForAll {pin has event at $t_{\mathrm{cur}}$ \textbf{of} $g$} \label{line:dynamic_power_start}
            \State Get the internal power LUT by $S$ \label{line:get_int_lut}
            \State Calculate dynamic (internal and switching) energy of the current pin \label{line:calc_dyn_ene}
            \State $r_{\mathrm{gl}} \gets \mathrm{GlitchDetectUpdate}(g,\text{pin},t_{\mathrm{cur}},\mathbf{c},T_{\mathrm{clk}})$ \label{line:glitch_update_and_ratio}
            \If{$r_{\mathrm{gl}} > 0$} \label{line:begin_of_glitch_if}
                \State Scale the dynamic energy by $r_{\mathrm{gl}}$ \label{line:scale_dyn_ene}
                \State Update glitch power $p_{\mathrm{gl}}$ \label{line:add_dyn_to_glitch}
            \EndIf \label{line:end_of_glitch_if}
            \State Update internal power $p_{\mathrm{int}}$ and waveform $\vec{w}$ \label{line:add_int_power_to_p_and_w}
            \State Update switching power $p_{\mathrm{sw}}$ and waveform $\vec{w}$ \label{line:add_sw_power_to_p_and_w}
        \EndFor\label{line:dynamic_power_end}

        \State Update gate state $S$ by events occurring at $t_{\mathrm{cur}}$ \label{line:update_gate_state}
    \EndFor \label{line:end_of_power_calc_loop}

    % \State Atomically add vector $\vec{p}$ to the row $\vec{R}^{(g)}$[\texttt{threadIdx.x}]\label{line:atomic_add_to_buffer}
\end{algorithmic}

\end{algorithm}

\begin{figure*}[tb!]
    \centering
    \includegraphics[width=1.0\linewidth]{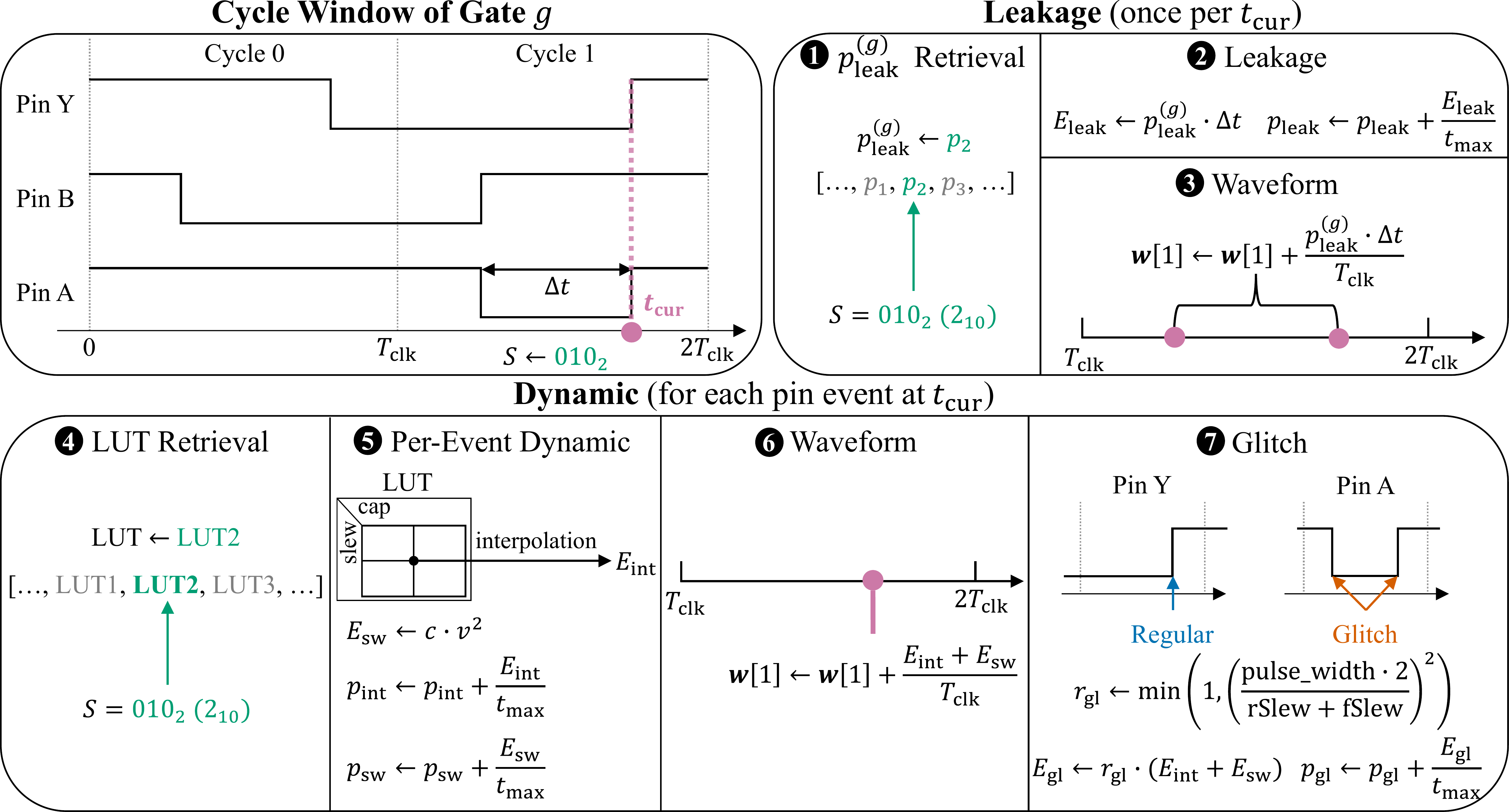}
    \caption{Procedure for computing and accumulating leakage, internal, switching, and glitch power contributions at an event timestamp $t_{\mathrm{cur}}$ in the fused power computation kernel.}
    \label{fig:overview_of_fused_kernel_one_event_power_calculation} 
\end{figure*}

As discussed in \Cref{sec:challenge_3_redundant_computation}, both Naive-D and Naive-L require each thread to reconstruct the gate state, resulting in redundant work when the kernels are executed separately.
Moreover, Naive-L already exposes cycle-level event-count information that can be reused for glitch detection, making the corresponding work in Naive-D redundant as well.
Motivated by these observations, we propose a \emph{fused kernel} that shares gate state and cycle-level information.
This design computes both dynamic and leakage power within a single kernel invocation, thereby reducing redundant work.
The resulting per-thread control flow is shown in \Cref{algo:fused_power_computation_kernel_pseudocode}.
Each thread is assigned a cycle window according to the per-gate partitioning scheme described in \Cref{sec:adaptive_temporal_part}.
% Each power component corresponds to a global array to store the per-cycle waveform, namely, $\vec{w}_{\mathrm{leak}}$, $\vec{w}_{\mathrm{int}}$, $\vec{w}_{\mathrm{sw}}$, and $\vec{w}_{\mathrm{gl}}$ are associated with leakage, internal, switching and glitch power waveforms, respectively.
A global array $\vec{w}$ stores the design-level per-cycle power waveform.
Each thread maintains a power results buffer $\vec{p}=[p_{\mathrm{leak}}, p_{\mathrm{int}}, p_{\mathrm{sw}}, p_{\mathrm{gl}}]$ initialized to zeros, which accumulates the thread-local contributions of leakage, internal, switching, and glitch power before reduction~(\Cref{line:init_p}).
The time window of the current thread $[t_{\mathrm{start}}, t_{\mathrm{end}})$ is derived using \Cref{eq:thread_start_time} and \Cref{eq:thread_end_time}~(\Cref{line:init_t}).
The gate state $S$ at $t_{\mathrm{start}}$ and the per-pin glitch counter $\vec{c}$ are \emph{initialized once and then reused throughout the thread's time window}~(\Cref{line:init_s}--\Cref{line:init_glitch_counter}).
Here, $\vec{c}$ records the number of events on each pin within the current clock cycle.
The outer \texttt{for} loop enumerates all event timestamps $t_{\mathrm{cur}}$ in chronological order via an on-the-fly $k$-way merge of the per-pin event streams.
Each iteration performs the per-event-timestamp computation illustrated in  \Cref{fig:overview_of_fused_kernel_one_event_power_calculation} and terminates when no unprocessed event timestamps remain~(\Cref{line:start_of_power_calc_loop}--\Cref{line:end_of_power_calc_loop}).
% Following the power computation of all assigned events, each thread atomically adds $\vec{p}$ to the reduction buffer of the current gate $\vec{R}^{(g)}$~(\Cref{line: atomic_add_to_buffer}).
The detailed power computation process will be presented in the rest of this subsection.

\subsubsection{Leakage Power~(\Cref{line:get_leak_p}--\Cref{line:add_leak_to_p_and_w})}
For leakage power, the leakage power value $p_{\mathrm{leak}}^{(g)}$ is retrieved based on the gate state $S$ by our SIPL method~(\Cref{fig:overview_of_fused_kernel_one_event_power_calculation}~\bcircled{\bfseries\scriptsize 1}).
Subsequently, the leakage energy $E_{\mathrm{leak}}$ dissipated from the previous event to the current event is calculated using \Cref{eq:leakage_energy_calc}.
Then the leakage power is calculated using \Cref{eq:average_power_calc} and added to $p_{\mathrm{leak}}$~(\Cref{fig:overview_of_fused_kernel_one_event_power_calculation}~\bcircled{\bfseries\scriptsize 2}).
In addition, the leakage energy dissipation across cycles is cumulatively incorporated into $\vec{w}$ via \texttt{atomicAdd}~(\Cref{fig:overview_of_fused_kernel_one_event_power_calculation}~\bcircled{\bfseries\scriptsize 3}).

\subsubsection{Dynamic Power~(\Cref{line:dynamic_power_start}--\Cref{line:dynamic_power_end})}
% \Cref{line: dynamic_power_start}--\Cref{line: dynamic_power_end} is to iterate through the pins toggle at $t_{cur}$ to calculate dynamic power, including internal power and switching power.
Since multiple pins of a gate may have events at the same timestamp, \Cref{line:dynamic_power_start} iterates over all pins with an event at $t_{\mathrm{cur}}$ and computes their dynamic power contributions.
The calculation of dynamic power comprises two components: internal power and switching power.
Internal power calculation requires retrieving the LUT associated with the gate state $S$, which is obtained via the SIPL-based retrieval method proposed in \Cref{sec:efficient_power_retrieval_with_bsim}~(\Cref{fig:overview_of_fused_kernel_one_event_power_calculation}~\bcircled{\bfseries\scriptsize 4}).
After retrieving the internal power LUT, the internal energy $E_{\mathrm{int}}$ is computed by interpolating the LUT using input slew and output capacitance, whereas the switching energy $E_{\mathrm{sw}}$ is calculated according to \Cref{eq:switching_energy_calc}.
Following this, if the current event is not a glitch, the event's contribution to the corresponding dynamic-power component is calculated using \Cref{eq:average_power_calc}~(\Cref{fig:overview_of_fused_kernel_one_event_power_calculation}~\bcircled{\bfseries\scriptsize 5}).
Additionally, the event's dynamic energy is accumulated into the power waveform $\vec{w}$ via \texttt{atomicAdd} in the clock cycle in which the event occurs, as illustrated in \Cref{fig:overview_of_fused_kernel_one_event_power_calculation}~\bcircled{\bfseries\scriptsize 6}.

\subsubsection{Glitch Power~(\Cref{line:glitch_update_and_ratio}--\Cref{line:end_of_glitch_if})}
% We implement the clock-cycle-based approach introduced in \Cref{sec:pre_naive_d_event_parallel_dynamic_power_kernel} for glitch detection. 
% $\mathrm{GlitchDetectUpdate}(\cdot)$ maintains an auxiliary per-pin cycle identifier internally (equivalent to storing the last cycle index of each pin) to detect clock-cycle boundaries and reset the corresponding entry in $\vec{c}$ when the cycle changes, and computes $r_{\mathrm{gl}}$ as follows:
$\mathrm{GlitchDetectUpdate}(\cdot)$~(\Cref{line:glitch_update_and_ratio}) performs clock-cycle-based glitch detection by maintaining the per-pin event counter $\vec{c}$, following the method introduced in \Cref{sec:pre_naive_d_event_parallel_dynamic_power_kernel}~(\Cref{fig:overview_of_fused_kernel_one_event_power_calculation}~\bcircled{\bfseries\scriptsize 7}).
If a pin has more than one event within the current clock cycle, the current event is treated as a glitch and a scaling ratio $r_{\mathrm{gl}}$ calculated using \Cref{eq:glitch_scaling_ratio_calc} is returned; otherwise, $-1$ is returned to mark it as a regular one.
% $$
% r_{\mathrm{gl}} = 
% \begin{cases}
%     \min \left(1, \left( \frac{\mathrm{pulse\_width} \cdot 2}{\mathrm{rSlew} + \mathrm{fSlew}} \right)^{2} \right) & \text{if it is a glitch event}  \\
%     -1 & \text{if it is not a glitch event}
% \end{cases}
% $$
After an event is marked as a glitch ($r_{\mathrm{gl}} > 0$), its dynamic energy is scaled by the glitch scaling ratio $r_{\mathrm{gl}}$.
The glitch energy $E_{\mathrm{gl}}$ is the scaled dynamic energy, and the event's contribution to the glitch-power component is calculated using \Cref{eq:average_power_calc}.
% and the glitch energy is atomically added to $\vec{w}_{\mathrm{gl}}$, similar to dynamic energy in \Cref{fig:overview_of_fused_kernel_one_event_power_calculation}~\bcircled{\bfseries\scriptsize 6}.

As depicted in \Cref{fig:stepwise_optimization_timelines}, by reusing the gate state $S$ and the per-pin event counter $\vec{c}$ within each thread, kernel fusion (\textcircled{\scriptsize D}) achieves \bcircled{\bfseries\scriptsize 3} reduction in kernel execution time compared to \textcircled{\scriptsize C}.

\subsection{Thread Mapping and Two-Phase Power Results Reduction}
\begin{figure*}[tb!]
    \centering
    \includegraphics[width=1.0\linewidth]{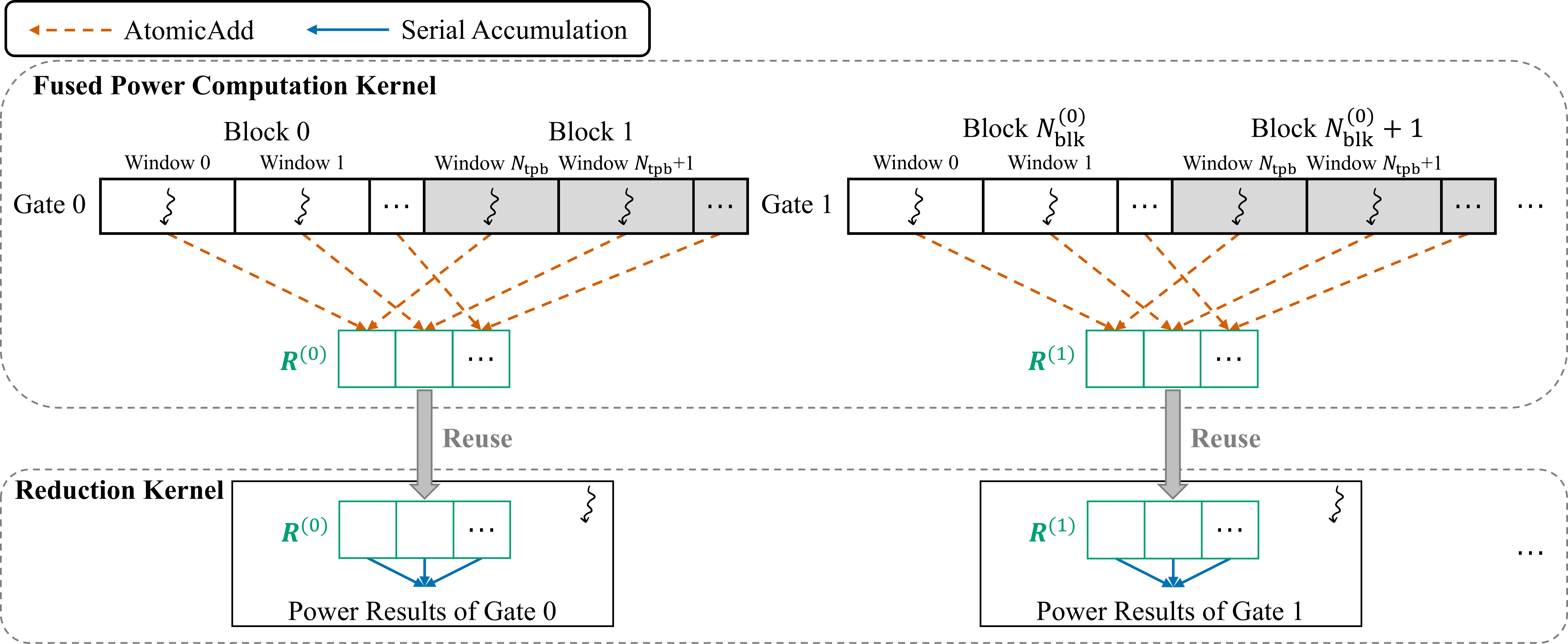}
    \caption{Illustration of the mapping from cycle windows to threads and the two-phase power results reduction.}
    \label{fig:thread_mapping_and_reduction}
\end{figure*}

As shown in \Cref{fig:thread_mapping_and_reduction}, our GPU implementation employs one-dimensional thread organization for power computation, where each thread processes a continuous cycle window of its assigned gate.
The number of threads allocated to each gate, denoted by $N_{\mathrm{thr}}^{(g)}$, is determined by that gate's event density, as introduced in \Cref{sec:adaptive_temporal_part}.
Given a fixed number of threads per block $N_{\mathrm{tpb}}$, the number of blocks assigned to each gate, $N_{\mathrm{blk}}^{(g)}$, is derived from its thread allocation (\ie $N_{\mathrm{blk}}^{(g)} = \lceil N_{\mathrm{thr}}^{(g)} / N_{\mathrm{tpb}} \rceil$).
% Each gate is allocated a uniform number of blocks $m$, where $m$ is derived from the division of the total cycle count by $n_{cycle}$.
% There are two advantages of this approach.
% First, gates may have different power computation patterns, restricting threads within each block to process only one gate helps to reduce thread divergence and mitigate workload imbalance among warps.
% For instance, suppose we assign an inverter gate with only two pins and an SRAM macro with more than 100 pins to two threads of the same warp. 
% For each toggle event, the inverter thread will complete its power computation before the SRAM thread and remain idle until the SRAM thread completes its computation.
% Second, since power consumption must be reported for each individual gate, this thread mapping strategy streamlines the reduction of power results across GPU threads for each individual gate.

The GPU execution involves two distinct kernels: a fused power computation kernel and a reduction kernel.
In the fused power computation kernel, each thread computes partial power results for its assigned gate, storing the results into a thread-local buffer $\vec{p}=[p_{\mathrm{leak}}, p_{\mathrm{int}}, p_{\mathrm{sw}}, p_{\mathrm{gl}}]$.
These partial results need to be reduced to obtain the final per-gate power results.
Allocating global memory privately for the intermediate results of each thread would impose excessive GPU memory overhead. 
% Allocating per-thread private global memory for result recording of each thread would impose excessive GPU memory overhead.
To tackle this issue, we propose a two-phase reduction approach as illustrated in \Cref{fig:thread_mapping_and_reduction}.
In the fused power computation kernel, threads initially accumulate their outputs in a per-gate reduction buffer $\vec{R}^{(g)} \in \mathbb{R}^{N_{\mathrm{tpb}} \times 4}$, where $N_{\mathrm{tpb}}=\texttt{blockDim.x}$. 
This matrix, stored in global memory, is dimensioned to align with the thread block size.
The four columns of $\vec{R}^{(g)}$ correspond to leakage, internal, switching, and glitch power components, respectively.
% Each thread deterministically maps to a unique buffer slot via its block-local thread index (threadIdx.x) and atomically accumulates its result into the designated slot .
Each thread locates its assigned position in $\vec{R}^{(g)}$ using block-local thread index (\texttt{threadIdx.x}), then performs thread-safe accumulation via \texttt{atomicAdd}.
Following the completion of the fused power computation kernel, buffer $\vec{R}^{(g)}$ will be reused. 
A dedicated reduction kernel processes all gate buffers in parallel, with serial accumulation performed internally within each individual buffer, thereby yielding the final per-gate results.
This design ensures conflict-free updates while achieving a balance between computational efficiency and GPU memory usage.

\section{Framework Overview} \label{sec:end_to_end_framework}

\begin{figure}[t]
    \centering
    \includegraphics[width=1.0\linewidth]{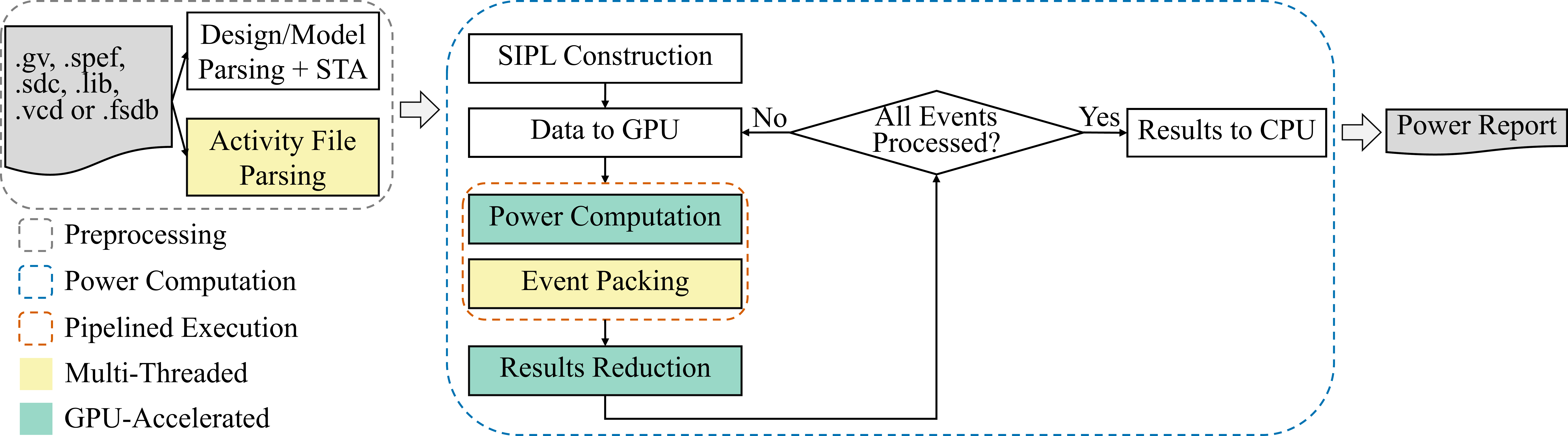}
    \caption{Overview of the proposed GPU-accelerated gate-level time-based power analysis framework.}
    \label{fig:overview_of_framework} 
\end{figure}

After presenting our GPU-specific optimizations, we now introduce our proposed framework, as shown in \Cref{fig:overview_of_framework}.
The framework begins with a two-step preprocessing phase on the CPU side.
In the first step, the design/model parser reads the netlist (.gv), parasitics (.spef), constraints (.sdc), and Liberty (.lib) files to prepare gate-level physical parameters and Liberty power models, while the STA engine derives timing information.
In the second step, the activity file (.fsdb or .vcd) is parsed using a multi-threaded parser that exploits signal-level parallelism.
% , which is implemented using the library provided by Verdi~\cite{verdi}. 
% Note that the multi-threaded VCD parser can be implemented by the approach proposed in~\cite{iccad_2021_fdu_gate_lve_resim} and seamlessly integrated into our framework.
After parsing the activity file, we divide events into uniformly sized chunks to fit within limited GPU memory.

Following the preprocessing phase, the preprocessed design metadata (timing information, Liberty power models, gate-level physical parameters) and the chunked events are passed to the power-computation phase.
This phase begins with two CPU-side setup steps:
(1) Leakage power values and internal power LUTs are structured using the proposed SIPL data structure to facilitate efficient state-dependent power retrieval.
(2) The first event chunk and the design metadata (timing/power/physical) are packed into contiguous buffers and transferred to GPU memory.
% (2) The preprocessed design metadata (timing/power/physical) and the first event chunk are then transferred to the GPU memory.
The system then proceeds to perform power computation, operating iteratively to process the event chunks one by one.
%Upon completion of event handling in the current chunk, the system automatically proceeds to process the next chunk.
% At each iteration, the event packing on the CPU and the power computation on the GPU operate in a CPU-GPU pipelined execution mode, enabling concurrent execution of these two steps.
% After the power computation kernel completes, a reduction kernel performs power results reduction for each gate, and if unprocessed events remain, the system transfers the next event chunk to the GPU.
There are two main stages in each iteration.
The first stage uses a pipelined CPU--GPU execution model that overlaps GPU-side power computation for the current chunk with CPU-side event packing for the next chunk.
During the packing step, multiple CPU threads pack events from all signals in the next chunk into a contiguous buffer for efficient bulk transfer.
The subsequent stage executes a GPU reduction kernel to reduce the power results for each gate.
This process terminates upon completion of all event processing, with power computation results being transferred to the CPU.
The system subsequently generates a comprehensive power analysis report, which includes both design-level and gate-level power consumption, along with the power waveform of the entire design.
At the design level, leakage, internal, switching, and glitch power are obtained by summing the corresponding power contributions across all gates.
% The per-cycle power waveform $\vec{w}$ is obtained by summing, at each cycle, the corresponding entries of $\vec{w}_{\mathrm{leak}}$, $\vec{w}_{\mathrm{int}}$, and $\vec{w}_{\mathrm{sw}}$, yielding the total power consumption per clock cycle.
The peak power is the maximum power value observed across all cycles in the power waveform $\vec{w}$.

\section{Experimental Results}\label{sec:exp}
\subsection{Settings}\label{sec:exp_settings}
\textbf{System Configurations.}
We implement the framework with C++ and CUDA.
The C++ code is compiled by GCC 11.4.0, and the CUDA code is compiled by NVCC 12.4, both with the \texttt{-O3} optimization enabled.
Unless otherwise stated, experiments are conducted on a CentOS 7.9 machine equipped with dual-socket AMD EPYC 7543 CPUs (2.8 GHz, 32 cores per socket, 64 cores total), 512 GB RAM, and an NVIDIA RTX 3090 GPU.

\textbf{Power Analysis Configurations.}
For all designs, we use 128 threads per block, set the SIPL fallback pin threshold to 16, set the target event count per thread $n_{\mathrm{evt,tgt}}$ to 8, and set the parallelism floor $N_{\mathrm{thr,floor}}$ to 512; the rationale for these parameters is discussed in \Cref{sec:parameter_sensitivity_analysis}.
% We use a chunk size of $8 \times 10^{8}$ and a target event count per thread of $n_{\mathrm{evt,tgt}}=8$.
% The chunk size is determined by ensuring that the GPU power computation time can adequately accommodate the CPU event packing time, and the determination of $n_{\mathrm{evt,tgt}}$ will be discussed in \Cref{sec:parameter_sensitivity_analysis}.
The activity file is in .fsdb format, which is parsed using the library provided by Verdi~\cite{verdi}.

\textbf{Baseline.}
We adopt PrimeTime PX Version T-2022.03-SP1~\cite{synopsys_primetime} as our baseline, comparing our end-to-end runtime against it while treating its power analysis results as a golden reference.
The measured end-to-end runtime uses the same accounting boundary for our framework and PrimeTime PX runs: design/model parsing, activity-file parsing, STA, and time-based power computation.
The design/model parsing portion includes parsing the gate-level netlist, parasitics, constraints, and Liberty power models.
For our GPU-accelerated framework, this boundary covers all stages shown in \Cref{fig:overview_of_framework}; in particular, it also accounts for GPU-specific implementation costs such as SIPL construction, CPU-side event packing, and CPU--GPU data transfer.
To ensure a fair comparison, the glitch detection method is set to clock-cycle-based, and the interval of the power waveform is set to the clock cycle, which ensures consistency between the baseline and our implementation.
% We develop a multi-threaded binary activity file parser based on the library provided by the commercial tool.
PrimeTime PX is configured with its maximum supported thread count (16 CPU threads), and our framework is similarly configured with 16 CPU threads and 1 GPU.
It should be noted that the baseline also performs temporal partitioning of the workload, distributing the activity file into discrete time windows containing an identical number of cycles.
A comparison with existing open-source or academic tools is not provided, as, to our knowledge, none currently support full time-based power analysis reporting.

\begin{table}[tb]
    \caption{The design statistics.}
    \label{tab:design_statistics}
    \centering 
    \resizebox{0.41\columnwidth}{!}{%
        \begin{tabular}{crrr}
    \toprule
    \textbf{Design} & \textbf{\#Gates} & \textbf{\#Pins} & \textbf{\#Nets} \\
    \midrule
    Adder              & 1,815     & 5,220      & 2,457     \\
    Multiplier         & 13,267    & 42,523     & 15,439    \\
    Rocket             & 116,617   & 448,063    & 121,379   \\
    SmallBOOM          & 288,789   & 1,116,613  & 300,649   \\
    GigaBOOM           & 1,053,918 & 4,435,524  & 1,183,615 \\
    %OpenSTA Sample     & 252       & xxx  & \\
    \bottomrule
\end{tabular}

    }
\end{table}

\textbf{Experimental Designs.}
% The designs we use in the experiment are listed in \Cref{tab:design_statistics}, including two arithmetic circuits and three RISC-V SoCs~\cite{micro_2020_chipyard}. 
The designs used in our experiment, including two arithmetic circuits and three RISC-V system-on-chip (SoC) designs~\cite{micro_2020_chipyard}, are presented in~\Cref{tab:design_statistics}, covering various design sizes.
All the designs are synthesized using the TSMC N28 process.
The largest design \texttt{GigaBOOM} contains more than one million gates and nets. 
%We validate the performance and accuracy of our framework on five designs which are synthesized under the TSMC 28m process.
%Three of them are RISC-V SoC designs with more than $100,000$ gates, Rocket[XXref], BOOM[XXref], and GigaBOOM[XXref], and each of them includes two testbenches.
%Another two benchmarks are one 64-bit adder with over 1,000 gates and one 64-bit multiplier with over 10,000 gates.

%We use mean absolute error (MAE) and mean absolute percentage error (MAPE) to evaluate the accuracy of the internal, switching, leakage, and glitch of the top-level design.
%For the power of each gate, we use Kendall’s tau (denoted as K-$\tau$) to measure the similarity in the ranking of the gate total power (i.e., $P_{inter}+P_{switch}+P_{leak}$) and glitch power (i.e., $P_{glitch\_inter}+P_{glitch\_switch}$). 
%A higher Kendall’s tau value indicates that the gate power ranking we obtained is more similar to that of the baseline.
%For the power waveform, we use MAE, and MAPE to evaluate the similarity of power ($P_{inter}+P_{switch}+P_{leak}$) across each cycle.

\subsection{Power Analysis Results Compared with PrimeTime PX}
%We evaluate the power analysis results mainly from overall accuracy, runtime speedup.
\subsubsection{Precision} \label{sec:precision_comparison_with_ptpx}
We use PrimeTime PX power results as the reference $P_{\mathrm{ref}}$ and compute relative error as $\frac{|P-P_{\mathrm{ref}}|}{P_{\mathrm{ref}}}$.
%The error is calculated by $\frac{|P-P_{\mathrm{ref}}|}{P_{\mathrm{ref}}}$, in which $P$ is the power result obtained by our framework. 
\Cref{tab:runtime_and_precision_comparison_with_ptpx} lists the accuracy and performance comparison results between the baseline and our framework, together with the activity factor of each workload.
It can be seen that the power computation error is very small.
Specifically, the average errors of leakage, internal, and switching power are 0.12\%, 0.06\%, and 0.05\%, respectively, demonstrating precise power computation.
The average errors of glitch and peak power are 0.11\% and 0.71\%, respectively, which remain low.
The activity-factor column also shows that higher-activity workloads generally produce larger dynamic-power components for the same design.
  % Notably, one source of error in our power computation stems from the inherent accuracy limitations of OpenSTA when compared to commercial STA tools~\cite{date_2024_heter_sta}.
% Nevertheless, our framework consistently achieves highly precise results, demonstrating its ability to deliver accurate power computation.

\begin{table*}[t]
    \centering
    \caption{Results comparison on precision and runtime. (Power Unit: $mW$)}
    \label{tab:runtime_and_precision_comparison_with_ptpx}
    \resizebox{\linewidth}{!}{
        \begin{threeparttable}
\setlength{\tabcolsep}{3.5pt}
\begin{tabular}{c|c|c|c|cc|cc|cc|cc|cc|ccc}
     \toprule
     Design & Benchmark & \#Cycles & \shortstack{AF\tnote{1}} & $P_{\mathrm{leak}}$ & Err. (\%) & $P_{\mathrm{int}}$ & Err. (\%) & $P_{\mathrm{sw}}$ & Err. (\%) & $P_{\mathrm{gl}}$ & Err. (\%) & $P_{\mathrm{peak}}$ & Err. (\%) & RT-ref. ($s$)\tnote{2} & RT ($s$)\tnote{3} & Speedup \\ \midrule
     Adder      & random    & 100,000 & 0.145 & 0.06  & 0.00 & 0.77   & 0.04 & 0.04 & 0.00 & 0.01 & 0.00 & 1.26   & 1.54  & 37.94    & 7.43   & 5.10\\ \midrule
     Multiplier & random    & 100,000 & 0.300 & 0.43  & 0.00 & 3.37   & 0.21 & 1.43 & 0.00 & 2.89 & 0.17 & 8.49   & 0.18  & 147.71   & 75.97  & 1.94\\ \midrule
     Rocket     & dhrystone & 529,602 & 0.047 & 5.13  & 0.06 & 27.29  & 0.04 & 0.37 & 0.00 & 0.65 & 0.05 & 64.49  & 1.05  & 1678.85  & 63.07  & 26.62\\ \midrule
     SmallBOOM  & dhrystone & 502,947 & 0.064 & 15.68 & 0.13 & 106.70 & 0.00 & 2.74 & 0.04 & 4.71 & 0.04 & 187.71 & 0.37  & 6872.54  & 332.35 & 20.68\\ \midrule
     SmallBOOM  & median    & 167,541 & 0.058 & 15.75 & 0.32 & 102.90 & 0.00 & 1.75 & 0.06 & 3.68 & 0.14 & 187.46 & 0.55  & 2188.23  & 93.79  & 23.33\\ \midrule
     SmallBOOM  & mt-vvadd  & 579,293 & 0.056 & 15.74 & 0.26 & 102.20 & 0.00 & 1.52 & 0.00 & 3.54 & 0.06 & 187.92 & 0.15  & 7767.76  & 206.44 & 37.63\\ \midrule
     % SmallBOOM  & qsort     & 517,052 & 15.72 & 0.12 & 106.90 & 0.84 & 2.75 & 1.95 & 4.80 & 3.00 & 170.03 & 3.93  & 8086.74  & 431.79 & 18.73\\ \midrule
     GigaBOOM   & dhrystone & 359,397 & 0.041 & 44.55 & 0.11 & 265.50 & 0.08 & 6.60 & 0.08 & 9.47 & 0.16 & 467.22 & 0.76  & 20550.59 & 620.19 & 33.14\\ \midrule
     GigaBOOM   & median    & 158,453 & 0.037 & 43.96 & 0.09 & 246.30 & 0.12 & 4.53 & 0.15 & 7.24 & 0.19 & 451.42 & 1.07  & 6041.12  & 311.58 & 19.39\\ \midrule
     GigaBOOM   & mt-vvadd  & 541,712 & 0.036 & 43.95 & 0.11 & 243.20 & 0.08 & 3.98 & 0.15 & 6.79 & 0.21 & 459.32 & 0.75  & 20675.98 & 608.24 & 33.99\\ \bottomrule
\end{tabular}
\begin{tablenotes}
    \footnotesize
    \item[] $P_{\mathrm{leak}}$, $P_{\mathrm{int}}$, $P_{\mathrm{sw}}$, $P_{\mathrm{gl}}$, and $P_{\mathrm{peak}}$ correspond to the leakage, internal, switching, glitch, and peak power of the whole design, respectively.
    \item[1] Activity factor denotes the average number of 0-to-1 events per pin per simulated clock cycle.
    \item[2] End-to-end runtime of PrimeTime PX using its maximum supported 16 threads.
    \item[3] End-to-end runtime of our framework using 16 threads + 1 GPU.
\end{tablenotes}
\end{threeparttable}

    }
\end{table*}

\subsubsection{Fine-Grained Spatial and Temporal Comparison}\label{sec:fine_grained_spatial_temporal_cmp}
\begin{figure}[tb!]
    \centering
    %\definecolor{myred}{RGB}{255, 128, 128}
%\definecolor{mygreen}{RGB}{128, 255, 128}     %FFABA4q
%\definecolor{myblue}{RGB}{128, 80, 40}    %1D72DD
\definecolor{myorange}{RGB}{255,166,26}
\definecolor{myred}{RGB}{227,107,98}
\definecolor{myyellow}{RGB}{255,214,75}
\definecolor{mymiddleblue}{RGB}{139,201,246} 
\definecolor{mylight}{RGB}{38,192,159}  
\definecolor{mygreen}{RGB}{205,252,197}   
%FFABA4q
\definecolor{myblue}{RGB}{29,114,221}    %1D72DD
\definecolor{mybrown}{RGB}{120, 80, 40}

\pgfplotstableread[col sep=space]{
	dataset                 total_power  glitch_power  
	Adder(random)           0.998        1.000 
	Multiplier(random)      0.999        1.000   
	Rocket(dhrystone)       0.996        0.998   
	SmallBOOM(dhrystone)    0.994        0.998  
  	SmallBOOM(median)       0.994        0.998 
    SmallBOOM(mt-vvadd)     0.995        0.998  
    GigaBOOM(dhrystone)     0.998        0.999  
    GigaBOOM(median)        0.997        0.998  
    GigaBOOM(mt-vvadd)      0.997        0.998  
}{\dataTableKendallTau}

\pgfplotsset{
    width=0.86\linewidth,
    height=0.36\linewidth
}

\begin{tikzpicture}
\begin{axis}[
    width=\linewidth,
    height=0.24\linewidth, % 你想要的高度比例
    ybar,
    enlarge x limits=0.06,
    symbolic x coords={
        Adder(random),
        Multiplier(random),
        Rocket(dhrystone),
        SmallBOOM(dhrystone),
        SmallBOOM(median),
        SmallBOOM(mt-vvadd),
        GigaBOOM(dhrystone),
        GigaBOOM(median),
        GigaBOOM(mt-vvadd)
    },
    xtick=data,
    xticklabels={
        \shortstack{Adder\\(random)}, 
        \shortstack{Multiplier\\(random)}, 
        \shortstack{Rocket\\(dhrystone)}, 
        \shortstack{SmallBOOM\\(dhrystone)}, 
        \shortstack{SmallBOOM\\(median)}, 
        \shortstack{SmallBOOM\\(mt-vvadd)}, 
        \shortstack{GigaBOOM\\(dhrystone)},
        \shortstack{GigaBOOM\\(median)},
        \shortstack{GigaBOOM\\(mt-vvadd)}
    },
    xlabel near ticks,
    xtick align=inside,
    ylabel={Kendall's $\tau$},
    ylabel near ticks,
    % ytick align=outside,
    ytick={0, 0.5, 1.0},
    bar width = 12pt,
    ymin=0,
    ymax=1.2,
    legend style={at={(0.5, 1.0)}, draw=none,anchor=south,legend columns=5, font=\footnotesize},
    x tick label style={font=\scriptsize},
    y tick label style={font=\footnotesize},
    label style={font=\footnotesize}, % 调整轴标签的字体大小
    point meta=y,
    nodes near coords={\pgfmathprintnumber[fixed,precision=3]{\pgfplotspointmeta}},
    every node near coord/.append style={font=\fontsize{5pt}{5.5pt}\selectfont, color=black}
]
    % use TeX as calculator:
    \addplot +[ybar, fill=myred, draw=black, area legend] table [x={dataset},  y={total_power}] {\dataTableKendallTau};
    \addplot +[      fill=mymiddleblue, draw=black, area legend] table [x={dataset},  y={glitch_power}] {\dataTableKendallTau};
    % \legend{,cluster-1\%, random-1\%, cluster-0.1\%, random-0.1\%}
    \legend{Total Power, Glitch Power}
\end{axis}
\end{tikzpicture}
    \caption{Kendall's Tau of per-gate total power and glitch power ranking correlations between ours and baseline.}
    \label{fig: kendal_tau_bar_char}
    % \vspace{-0.4cm}
\end{figure}
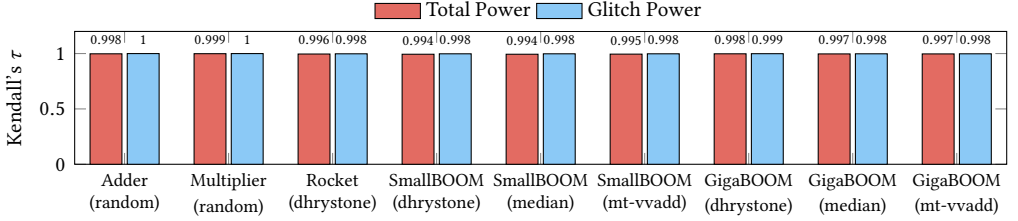

\begin{figure}[t]
  \centering
  \definecolor{myorange}{RGB}{255,166,26}
\definecolor{myred}{RGB}{227,107,98}
\definecolor{myyellow}{RGB}{255,214,75}
\definecolor{mymiddleblue}{RGB}{139,201,246} 
\definecolor{mylight}{RGB}{38,192,159}  
\definecolor{mygreen}{RGB}{205,252,197}   
\definecolor{myblue}{RGB}{29,114,221}
\definecolor{mybrown}{RGB}{120, 80, 40} 
\definecolor{mygray}{RGB}{210,210,210}

\pgfplotstableread[col sep=space]{
dataset            GTleone GTonefive GTfiveten GTgtten GTna GGleone GGonefive GGfiveten GGgtten GGna STleone STonefive STfiveten STgtten STna SGleone SGonefive SGfiveten SGgtten SGna
Adder              99.89   0.11      0.00      0.00    0.00 99.83   0.06      0.06      0.06    0.00 100.00  0.00      0.00      0.00    0.00 100.00  0.00      0.00      0.00    0.00
Multiplier         99.63   0.37      0.00      0.00    0.00 95.96   3.14      0.90      0.00    0.00 100.00  0.00      0.00      0.00    0.00 98.18   1.82      0.00      0.00    0.00
Rocket             98.76   1.01      0.19      0.04    0.00 96.89   1.56      0.64      0.92    0.00 99.26   0.74      0.00      0.00    0.00 93.04   5.86      0.73      0.37    0.00
SmallBOOMDhrystone 97.84   1.83      0.23      0.10    0.00 95.01   2.72      0.94      1.32    0.00 96.39   3.61      0.00      0.00    0.00 80.13   15.31     3.26      1.31    0.00
SmallBOOMMedian    98.66   1.14      0.14      0.05    0.00 94.93   2.72      1.04      1.31    0.00 96.72   3.28      0.00      0.00    0.00 80.46   13.36     4.23      1.96    0.00
SmallBOOMMtVvadd   98.74   1.07      0.14      0.05    0.00 95.02   2.81      0.97      1.21    0.00 96.39   3.61      0.00      0.00    0.00 81.43   12.38     3.58      2.61    0.00
GigaBOOMDhrystone  98.50   1.28      0.15      0.06    0.00 95.14   2.79      0.72      1.34    0.00 97.98   1.73      0.29      0.00    0.00 90.23   8.62      0.86      0.29    0.00
GigaBOOMMedian     99.10   0.76      0.09      0.05    0.00 95.28   2.70      0.67      1.35    0.00 97.98   2.02      0.00      0.00    0.00 87.64   10.63     0.86      0.86    0.00
GigaBOOMMtVvadd    99.24   0.65      0.06      0.05    0.00 95.25   2.75      0.66      1.34    0.00 98.55   1.45      0.00      0.00    0.00 88.22   9.20      1.72      0.86    0.00
}{\dataTablePerGateStdCellErrDist}

\providecommand{\perGateStdCellErrLegendItem}[2]{%
    \tikz[baseline=-0.5ex]\draw[fill=#1, draw=black] (0,-0.05) rectangle (0.45,0.08);\,#2%
}

\pgfplotsset{
    perGateStdCellErrAxis/.style={
        width=0.48\linewidth,
        height=0.20\linewidth,
        ybar stacked,
        ymajorgrids=true,
        grid style={dashed},
        ymin=0,
        ymax=100,
        bar width=4.0pt,
        symbolic x coords={
            Adder,
            Multiplier,
            Rocket,
            SmallBOOMDhrystone,
            SmallBOOMMedian,
            SmallBOOMMtVvadd,
            GigaBOOMDhrystone,
            GigaBOOMMedian,
            GigaBOOMMtVvadd
        },
        xtick=data,
        xticklabels={
            \shortstack{Add\\rand},
            \shortstack{Mul\\rand},
            \shortstack{Rkt\\dhr},
            \shortstack{SB\\dhr},
            \shortstack{SB\\med},
            \shortstack{SB\\vv},
            \shortstack{GB\\dhr},
            \shortstack{GB\\med},
            \shortstack{GB\\vv}
        },
        x tick label style={font=\tiny},
        y tick label style={font=\footnotesize},
        title style={at={(0.5,1)}, anchor=north, yshift=8pt, font=\footnotesize},
        ytick={0,50,100},
        yticklabel={\pgfmathprintnumber{\tick}\%},
        ylabel={\footnotesize Share (\%)},
        ylabel near ticks,
        ylabel shift=-6pt,
        enlarge x limits=0.05,
    },
}

\newcommand{\perGateStdCellErrAxis}[6]{%
\begin{tikzpicture}
\begin{axis}[perGateStdCellErrAxis, title={#1}]
    \addplot +[ybar, fill=mygreen, draw=black, area legend] table [x=dataset, y=#2] {\dataTablePerGateStdCellErrDist};
    \addplot +[fill=mymiddleblue, draw=black, area legend] table [x=dataset, y=#3] {\dataTablePerGateStdCellErrDist};
    \addplot +[fill=myyellow, draw=black, area legend] table [x=dataset, y=#4] {\dataTablePerGateStdCellErrDist};
    \addplot +[fill=myred, draw=black, area legend] table [x=dataset, y=#5] {\dataTablePerGateStdCellErrDist};
    \addplot +[fill=mygray, draw=black, area legend] table [x=dataset, y=#6] {\dataTablePerGateStdCellErrDist};
\end{axis}
\end{tikzpicture}%
}

\centering
{\scriptsize
\perGateStdCellErrLegendItem{mygreen}{$\le 1\%$}\quad
\perGateStdCellErrLegendItem{mymiddleblue}{$1$--$5\%$}\quad
\perGateStdCellErrLegendItem{myyellow}{$5$--$10\%$}\quad
\perGateStdCellErrLegendItem{myred}{$>10\%$}
}

\vspace{0.1em}
\begin{tabular}{@{}c@{\hspace{0.01\linewidth}}c@{}}
\perGateStdCellErrAxis{Gate total}{GTleone}{GTonefive}{GTfiveten}{GTgtten}{GTna} &
\perGateStdCellErrAxis{Gate glitch}{GGleone}{GGonefive}{GGfiveten}{GGgtten}{GGna} \\[-0.4em]
\perGateStdCellErrAxis{Standard-cell total}{STleone}{STonefive}{STfiveten}{STgtten}{STna} &
\perGateStdCellErrAxis{Standard-cell glitch}{SGleone}{SGonefive}{SGfiveten}{SGgtten}{SGna}
\end{tabular}
  \caption{Relative-error distributions of per-gate and standard-cell power values for total and glitch power. Each stacked bar shows the percentage of gates or standard-cell types in each error range.}
  \label{fig:dist_of_gate_and_std_cell_power_error}
\end{figure}
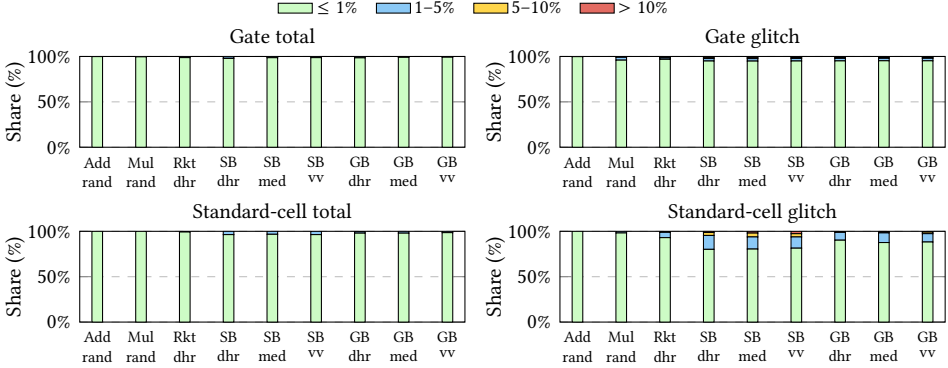

We further validate the power analysis results with a fine-grained spatial and temporal comparison. 
On the one hand, despite the very small error of the power analysis results in \Cref{tab:runtime_and_precision_comparison_with_ptpx}, spatial accuracy is important for analyzing and optimizing power hotspots.
To this end, we use Kendall's Tau~\cite{1966kendall} to evaluate the correlation in gate power dissipation rankings. 
As shown in \Cref{fig: kendal_tau_bar_char}, our framework achieves high ranking correlation, with Kendall's Tau coefficients no lower than 0.994 for total power and 0.998 for glitch power across all designs.
Beyond ranking correlation, \Cref{fig:dist_of_gate_and_std_cell_power_error} reports the relative-error distributions for total and glitch power at the per-gate and standard-cell levels.
Most gates and standard-cell types fall into the low-error ranges, showing that our framework closely matches the baseline in both ranking and absolute power values.
In addition, the hotspot-overlap columns in \Cref{tab:waveform_and_hotspot_accuracy} show that the hotspot overlap@1\% remains high for both per-gate and standard-cell levels, confirming that our framework accurately identifies the highest-power gates and standard-cell types.
% the Kendall's Tau values of per-gate total power ranking on all designs reach 0.98, and the Kendall's Tau values of glitch power also reach 0.95 on all of the designs.
% \begin{figure}[tb!]
%     \centering
%     \caption{Kendall's Tau of gate ranking correlation between ours and baseline.}
%     \label{fig: kendal_tau_bar_char}
% \end{figure}

\begin{table*}[t]
    \centering
    \caption{Hotspot overlap@1\% and cycle-level power waveform accuracy compared with PrimeTime PX.}
    \label{tab:waveform_and_hotspot_accuracy}
    \resizebox{\linewidth}{!}{
        \begin{threeparttable}
\setlength{\tabcolsep}{4pt}
\begin{tabular}{ccccccccc}
    \toprule
    \multirow{2}{*}{\textbf{Design}} &
    \multirow{2}{*}{\textbf{Benchmark}} &
    \multicolumn{4}{c}{\textbf{Hotspot Overlap@1\% (\%)}} &
    \multicolumn{3}{c}{\textbf{Waveform Accuracy}} \\
    \cmidrule(lr){3-6}\cmidrule(lr){7-9}
    & & \textbf{Gate Total} & \textbf{Gate Glitch} & \textbf{Std. Total} & \textbf{Std. Glitch} &
    \textbf{RMSE ($W$)} & \textbf{NRMSE (\%)\tnote{1}} & \textbf{Pearson Corr.} \\
    \midrule
    Adder      & random    & 100.00 & 100.00 & 100.00 & 100.00 & $3.05{\times}10^{-6}$ & 0.614 & 1.000 \\
    Multiplier & random    & 100.00 & 100.00 & 100.00 & 100.00 & $7.98{\times}10^{-6}$ & 0.102 & 1.000 \\
    Rocket     & dhrystone & 99.74  & 99.29  & 100.00 & 100.00 & $1.58{\times}10^{-4}$ & 0.401 & 1.000 \\
    SmallBOOM  & dhrystone & 99.58  & 99.82  & 100.00 & 100.00 & $7.64{\times}10^{-4}$ & 0.704 & 0.999 \\
    SmallBOOM  & median    & 99.03  & 99.37  & 100.00 & 100.00 & $4.26{\times}10^{-4}$ & 0.391 & 1.000 \\
    SmallBOOM  & mt-vvadd  & 99.41  & 98.78  & 100.00 & 100.00 & $1.00{\times}10^{-3}$ & 0.914 & 0.997 \\
    GigaBOOM   & dhrystone & 99.49  & 99.50  & 100.00 & 100.00 & $1.47{\times}10^{-3}$ & 0.538 & 1.000 \\
    GigaBOOM   & median    & 97.57  & 98.77  & 100.00 & 100.00 & $1.01{\times}10^{-3}$ & 0.391 & 0.999 \\
    GigaBOOM   & mt-vvadd  & 99.01  & 98.89  & 100.00 & 100.00 & $2.18{\times}10^{-3}$ & 0.842 & 0.998 \\
    \bottomrule
\end{tabular}
\begin{tablenotes}[flushleft]
    \footnotesize
    \item[1] The NRMSE is computed by normalizing the RMSE by the dynamic range of the corresponding reference waveform.
\end{tablenotes}
\end{threeparttable}

    }
\end{table*}

\begin{figure}[tb!]
    \input{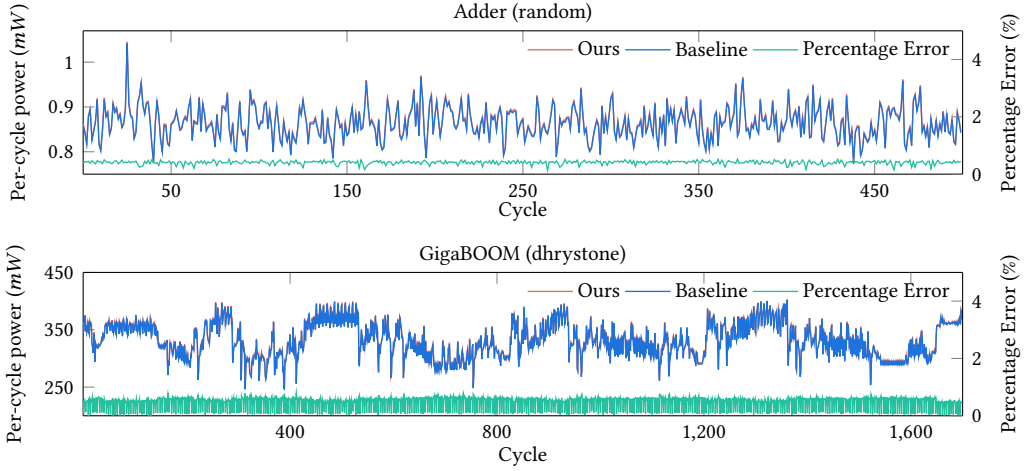}
    \caption{Power traces comparison between ours and PrimeTime PX.}
    \label{fig: power_waveforms_comparison}
\end{figure}

On the other hand, temporal power consumption information is important for diagnosing cycle-specific power issues.
Therefore, we compare the power traces between PrimeTime PX and ours.
For brevity, we select the smallest design \texttt{Adder} and the largest design \texttt{GigaBOOM} for power trace comparison, focusing on a segment with significant fluctuations for detailed analysis.
As shown in \Cref{fig: power_waveforms_comparison}, although there are some small deviations in values, the trends of our power traces are consistent with the baseline.
The waveform-accuracy columns in \Cref{tab:waveform_and_hotspot_accuracy} further confirm that the cycle-level power traces closely match PrimeTime PX, with low root mean square error (RMSE) and normalized RMSE (NRMSE), as well as high Pearson correlation across all benchmarks.

% \begin{figure}[tb!]
%     \centering
%     \includegraphics[width=0.9\linewidth]{figs/power_waveforms.pdf}
%     \caption{Power traces comparison between ours and baseline.}
%     \label{fig: power_waveforms_comparison}
% \end{figure}

The strong consistency in both spatial and temporal dimensions validates the effectiveness of our framework in per-gate power optimization and cycle-level power analysis.

\subsubsection{Speedup}
In terms of efficiency, we compare the end-to-end runtime of the baseline and ours. 
It can be seen from \Cref{tab:runtime_and_precision_comparison_with_ptpx} that we achieve a 1.94$\times$ to 37.63$\times$ end-to-end speedup compared to the multi-threaded PrimeTime PX baseline across various design sizes and benchmarks.
Specifically, our framework achieves greater speedup on larger designs and longer benchmarks, as power computation typically requires more time for such cases.
The runtime results demonstrate that our framework can perform ultra-fast and accurate time-based power analysis on large-scale designs.
%the commercial tool takes nearly 50 hours on the largest design, while the proposed GPU-accelerated flow only takes less than 3 hours, achieving 16X speed up 

%We achieve 102-1178$\times$ power computation kernel speedup and 5-61$\times$ overall speedup compared to the single-threaded baseline across various design sizes.
%Notably, we achieve greater speedup (16-61$\times$ overall, 601-1178$\times$ for kernel) on larger designs containing 100,000 gates, as more time is spent on the power computation kernel for these designs.

\subsection{Performance Analysis} \label{sec:performance_analysis}
% To demonstrate that our proposed GPU-accelerated method can accelerate the process of time-based power analysis,
% we conducted profiling of the runtime on an equivalent single-threaded CPU version of our framework.
% We implemented an equivalent single-threaded CPU version of our framework, and conducted a runtime breakdown of the overall runtime on it using 3 large benchmarks: Rocket (full), SmallBOOM (short) and GigaBOOM (short) (\Cref{fig: cpu_runtime_breakdown}).
% The breakdown results show that only a small portion of the time is spent on preprocessing (read VCD, read .lib file, STA, etc.), while the remainder is dominated by power computation kernel.
% This suggests that the primary runtime bottleneck of time-based power analysis lies in the power computation kernel.
% To assess the effectiveness of our GPU kernel design and the corresponding optimizations, we conduct a comparative analysis between our GPU-based framework and its equivalent multi-threaded CPU implementation (64 threads).
We analyze framework performance through runtime breakdowns, CPU/GPU comparison, GPU memory footprint, and GPU bottleneck analysis.
The CPU baseline in this section is a multi-threaded implementation that exploits gate-level parallelism using a publicly available thread pool~\cite{progschj_threadpool}.
Gates are shuffled for load balance and evenly assigned to worker threads by gate count; each worker computes event-driven power with thread-local accumulators before a final reduction.
% which adheres to an identical workflow.
% we simplified the complex workflow of the commercial tool by implementing a single-threaded CPU version that follows the same workflow as our GPU version. 
%This allows for a fair comparison. 

\subsubsection{Runtime Breakdown}\label{sec:runtime_breakdown}
\begin{figure}[!t]
    \subfloat[]{\hspace{-5.8em} \usetikzlibrary{arrows.meta}
%\usepackage[most]{tcolorbox}

% Adjusts the size of the wheel:
\def\innerradius{0\textwidth}
\def\outerradius{0.08\textwidth}

\definecolor{myblue}{RGB}{29,114,221}    %1D72DD
\definecolor{myyellow}{RGB}{255,255,191} %FFFFBF
\definecolor{myorange}{RGB}{244,106,18}  %F47012
\definecolor{mygray}{RGB}{102,102,102}   %666666
\definecolor{mypink}{RGB}{252,228,215}   %FCE4D7
\definecolor{myred}{RGB}{227,107,98}
\definecolor{myyellow}{RGB}{255,214,75}
\definecolor{mymiddleblue}{RGB}{139,201,246} 
\definecolor{mylight}{RGB}{38,192,159} 

\pgfplotsset{
    width=0.6\linewidth,
    height=0.4\linewidth
}

% The main macro
\newcommand{\wheelchartwithlegend}[1]{
  % Calculate total
  \pgfmathsetmacro{\totalnum}{0}
  \foreach \value/\colour/\name in {#1} {
      \pgfmathparse{\value+\totalnum}
      \global\let\totalnum=\pgfmathresult
  }

    \begin{tikzpicture}

    % Calculate the thickness and the middle line of the wheel
    \pgfmathsetmacro{\wheelwidth}{\outerradius-\innerradius}
    \pgfmathsetmacro{\midradius}{(\outerradius+\innerradius)/2}

    % Rotate so we start from the top
    \begin{scope}[rotate=90]

    % add coordinate to define the upper left starting point of the legend entries
    \coordinate (L-0) at (\outerradius-3mm,-\outerradius-1cm);

    % Loop through each value set. \cumnum keeps track of where we are in the wheel
    \pgfmathsetmacro{\cumnum}{0}
    \foreach [count=\i,remember=\i as \j (initially 0)] \value/\colour/\name in {#1} {
          \pgfmathsetmacro{\newcumnum}{\cumnum + \value/\totalnum*360}

          % Calculate the percent value
          \pgfmathsetmacro{\percentage}{\value/\totalnum*100}
                    %\pgfmathsetmacro{\percentage}{\value}
          % Calculate the mid angle of the colour segments to place the labels
          \pgfmathsetmacro{\midangle}{-(\cumnum+\newcumnum)/2}

          % This is necessary for the labels to align nicely
          \pgfmathparse{
             (-\midangle<180?"west":"east")
          } \edef\textanchor{\pgfmathresult}
          \pgfmathsetmacro\labelshiftdir{1-2*(-\midangle>180)}

          % Draw the color segments. Somehow, the \midrow units got lost, so we add 'pt' at the end. Not nice...
          \fill[\colour] (-\cumnum:\outerradius) arc (-\cumnum:-(\newcumnum):\outerradius) --
          (-\newcumnum:\innerradius) arc (-\newcumnum:-(\cumnum):\innerradius) -- cycle;

          % Draw the data labels
          \draw  [Circle-,thin] node [append after command={(\midangle:\midradius pt) -- (\midangle:\outerradius + 1ex) -- (\tikzlastnode)}] at (\midangle:\outerradius + 1ex) [xshift=\labelshiftdir*0.5cm,inner sep=0pt, outer sep=0pt, ,anchor=\textanchor,font=\footnotesize]{\pgfmathprintnumber{\percentage}\thinspace\%};

          % add legend node
          \node [anchor=north west,text width=3cm,font=\footnotesize] (L-\i) at (L-\j.south west) {\name};
          % draw legend image
          \fill [fill=\colour] ([xshift=-3pt,yshift=1mm]L-\i.north west) rectangle ++(-2mm,5mm);

          % Set the old cumulated angle to the new value
          \global\let\cumnum=\newcumnum
      }
    \end{scope}
  \end{tikzpicture}
 % Closing \fbox
} % Closing \newenvironment

\wheelchartwithlegend{
  % 447.95/mypink/{Data Transfer},
  52.84179/mygray/{Design/Model Parsing},
  359.654/myyellow/{Activity File Parsing},
  1505.34/myorange/{Power Computation},
  55.0535/mylight/{STA}
  % 362.35/mymiddleblue/{Time Slicing}
  %3.49/mygray/{Print result}
} \label{fig:cpu_runtime_breakdown_GigaBOOM}}
    \subfloat[]{\hspace{-7.0em} %\usepackage{blindtext}
%\usepackage[most]{tcolorbox}

% Adjusts the size of the wheel:
\def\innerradius{0\textwidth}
\def\outerradius{0.08\textwidth}

\definecolor{myblue}{RGB}{29,114,221}    %1D72DD
\definecolor{myyellow}{RGB}{255,255,191} %FFFFBF
\definecolor{myorange}{RGB}{244,106,18}  %F47012
\definecolor{mygray}{RGB}{102,102,102}   %666666
\definecolor{mypink}{RGB}{252,228,215}   %FCE4D7
\definecolor{myred}{RGB}{227,107,98}
\definecolor{myyellow}{RGB}{255,214,75}
\definecolor{mymiddleblue}{RGB}{139,201,246} 
\definecolor{mylight}{RGB}{38,192,159}

% The main macro
\newcommand{\wheelchartwithlegend}[1]{
  % Calculate total
  \pgfmathsetmacro{\totalnum}{0}
  \foreach \value/\colour/\name in {#1} {
      \pgfmathparse{\value+\totalnum}
      \global\let\totalnum=\pgfmathresult
  }

    \begin{tikzpicture}

    % Calculate the thickness and the middle line of the wheel
    \pgfmathsetmacro{\wheelwidth}{\outerradius-\innerradius}
    \pgfmathsetmacro{\midradius}{(\outerradius+\innerradius)/2}

    % Rotate so we start from the top
    \begin{scope}[rotate=90]

    % add coordinate to define the upper left starting point of the legend entries
    \coordinate (L-0) at (\outerradius-2mm,-\outerradius-1cm);

    % Loop through each value set. \cumnum keeps track of where we are in the wheel
    \pgfmathsetmacro{\cumnum}{0}
    \foreach [count=\i,remember=\i as \j (initially 0)] \value/\colour/\name in {#1} {
          \pgfmathsetmacro{\newcumnum}{\cumnum + \value/\totalnum*360}

          % Calculate the percent value
          \pgfmathsetmacro{\percentage}{\value/\totalnum*100}
                    %\pgfmathsetmacro{\percentage}{\value}
          % Calculate the mid angle of the colour segments to place the labels
          \pgfmathsetmacro{\midangle}{-(\cumnum+\newcumnum)/2}

          % This is necessary for the labels to align nicely
          \pgfmathparse{
             (-\midangle<180?"west":"east")
          } \edef\textanchor{\pgfmathresult}
          \pgfmathsetmacro\labelshiftdir{1-2*(-\midangle>180)}

          % Draw the color segments. Somehow, the \midrow units got lost, so we add 'pt' at the end. Not nice...
          \fill[\colour] (-\cumnum:\outerradius) arc (-\cumnum:-(\newcumnum):\outerradius) --
          (-\newcumnum:\innerradius) arc (-\newcumnum:-(\cumnum):\innerradius) -- cycle;

          % Draw the data labels
          \draw  [Circle-,thin] node [append after command={(\midangle:\midradius pt) -- (\midangle:0.09\textwidth + 1ex) -- (\tikzlastnode)}] at (\midangle:0.09\textwidth + 1ex) [xshift=\labelshiftdir*0.5cm,inner sep=0pt, outer sep=0pt, ,anchor=\textanchor,font=\footnotesize]{\pgfmathprintnumber{\percentage}\thinspace\%};

          % add legend node
          \node [anchor=north west,text width=3cm,font=\footnotesize] (L-\i) at (L-\j.south west) {\name};
          % draw legend image
          \fill [fill=\colour] ([xshift=-3pt,yshift=1mm]L-\i.north west) rectangle ++(-2mm,5mm);

          % Set the old cumulated angle to the new value
          \global\let\cumnum=\newcumnum
      }
    \end{scope}
  \end{tikzpicture}
 % Closing \fbox
} % Closing \newenvironment

%round2-full/GigaBoom-mt-vvadd_n\=8_20250419_141124.log
\wheelchartwithlegend{
  % 155.54/mymiddleblue/{Data Transfer},
  %227.93/mylight/{Others},
  51.878820/mygray/{Design/Model Parsing},
  350.408/myyellow/{Activity File Parsing},
  % 10.9814/mygray/{First-Chunk Packing},
  42.6241/myorange/{Power Comp. Kernel},
  18.7230851/mypink/{Data Transfer},
  % 5.691994/mygray/{Results Recording},
  51.647/mylight/{STA}
  % 41.8357/myorange/{Power Computation Kernel}
  % 362.35/mymiddleblue/{Event-Equalized Chunking}
  % 1.124/mygray/{Final-Reduction Kernel}
} \label{fig:gpu_runtime_breakdown_GigaBOOM}}
    \caption{The end-to-end runtime breakdown of the \texttt{mt-vvadd} benchmark on \texttt{GigaBOOM}. (a) CPU version (64 threads). (b) GPU version (64 host CPU threads and 1 GPU).}
    \label{fig:runtime_breakdown_GigaBOOM}
\end{figure}
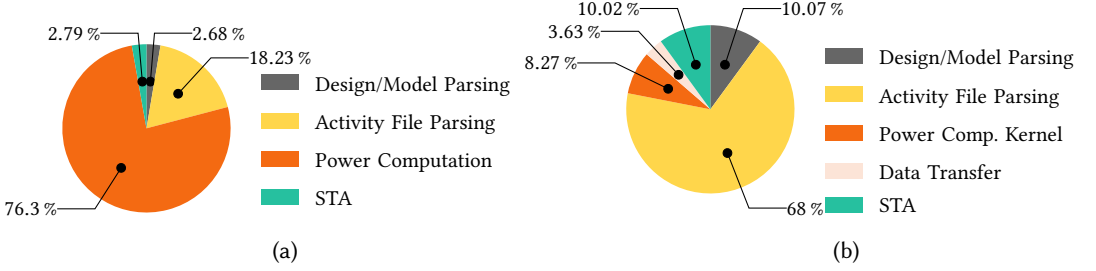
\Cref{fig:runtime_breakdown_GigaBOOM} presents the end-to-end runtime breakdown for both multi-threaded CPU (64 threads) and GPU (64 host CPU threads and 1 GPU) implementations on \texttt{GigaBOOM} with the \texttt{mt-vvadd} benchmark.
For completeness, the GPU-accelerated flow also includes SIPL construction, CPU-side event packing, and a final-reduction kernel.
Because event packing is overlapped with GPU execution and SIPL construction/final reduction account for less than 0.01\%/0.22\% of the end-to-end runtime, these components are omitted from the breakdown for readability.
As demonstrated in \Cref{fig:cpu_runtime_breakdown_GigaBOOM}, the primary performance bottleneck in our multi-threaded CPU version is power computation, which accounts for 76.3\% of the CPU end-to-end runtime.
In contrast, \Cref{fig:gpu_runtime_breakdown_GigaBOOM} reveals that our GPU scheme effectively addresses this bottleneck: the GPU power-computation kernel and CPU--GPU data transfer together account for only 11.9\%.
With the power-computation bottleneck largely removed, the GPU end-to-end runtime is instead dominated by activity-file parsing, which accounts for 68.0\%, shifting the remaining performance limit to the inherent file-parsing overhead in the end-to-end flow.
For comparison, we also report the preprocessing and power-computation portions of PrimeTime PX on the same case.
Its preprocessing consists of design/model parsing and STA, accounting for 0.6\% of the end-to-end runtime.
Because PrimeTime PX reads activity data during time-based power computation~\cite{ptpx_user_guide}, this interleaved activity parsing is included in the power-computation category, which accounts for 99.4\%.
% The comparative analysis demonstrates that the proposed GPU-accelerated framework effectively eliminates the computational bottleneck
% Of note, the final-reduction kernel accounts for merely 0.15\% of the end-to-end runtime, and thus is not included in the GPU runtime breakdown.
% Notably, the reduction kernel accounts for only 0.15\% of the end-to-end runtime and is thus excluded from the runtime breakdown. 
% Meanwhile, data transfer contributes to 3.6\% of the total runtime, as substantial information, including events, LUTs, slews, and other data needed for power computation, must be transferred to the GPU.
% shifting the performance bottleneck to activity file reading, \weihao{which is an unavoidable overhead}.
% that the computational bottleneck has been effectively mitigated through the proposed GPU-accelerated framework, while I/O becomes a more critical step, \weihao{which remains an unavoidable overhead}.

\subsubsection{CPU/GPU Comparison}\label{sec:cpu_gpu_comparison}
\begin{figure}[t]
  \centering
  \definecolor{myorange}{RGB}{255,166,26}
\definecolor{myred}{RGB}{227,107,98}
\definecolor{mymiddleblue}{RGB}{139,201,246}
\definecolor{mylight}{RGB}{38,192,159}

\pgfplotstableread[col sep=space]{
dataset                 cpu_64          gpu_64          cpu_32          gpu_32          cpu_16          gpu_16
Adder(random)           1.470000        0.713000        2.097000        0.714000        3.922000        0.707000
Multiplier(random)      27.847000       7.272000        44.855000       6.563000        78.776000       7.220000
Rocket(dhrystone)       147.222000      12.581000       259.538000      13.604000       502.725000      14.674000
SmallBOOM(dhrystone)    760.055000      47.444000       1289.584000     48.501000       2224.745000     48.500000
SmallBOOM(median)       228.695000      18.604000       357.499000      14.246000       644.156000      19.397000
SmallBOOM(mt-vvadd)     836.306000      34.200000       1276.942000     39.521000       2223.537000     34.978000
GigaBOOM(dhrystone)     1191.259000     86.479000       2012.248000     89.244000       3953.054000     94.285000
GigaBOOM(median)        451.724000      45.472000       721.664000      46.847000       1434.394000     49.064000
GigaBOOM(mt-vvadd)      1505.341000     83.726000       2538.935000     86.715000       4612.746000     89.937000
}{\dataTableCPUScalingPowerAnalysisTime}

\pgfplotsset{
    width=0.86\linewidth,
    height=0.36\linewidth
}

\begin{tikzpicture}
  \begin{axis}[
    width=\linewidth,
    height=0.25\linewidth,
    ymajorgrids=true,
    ybar,
    symbolic x coords={
        Adder(random),
        Multiplier(random),
        Rocket(dhrystone),
        SmallBOOM(dhrystone),
        SmallBOOM(median),
        SmallBOOM(mt-vvadd),
        GigaBOOM(dhrystone),
        GigaBOOM(median),
        GigaBOOM(mt-vvadd)
    },
    ymin=0,
    ymax=2.1,
    ytick={0,0.5,1.0,1.5,2.0},
    bar width=4pt,
    enlarge x limits=0.06,
    ylabel={Runtime},
    ylabel near ticks,
    ylabel shift=-5pt,
    ylabel style={align=center},
    xtick align=inside,
    xtick=data,
    xticklabels={
        \shortstack{Adder\\(random)},
        \shortstack{Multiplier\\(random)},
        \shortstack{Rocket\\(dhrystone)},
        \shortstack{SmallBOOM\\(dhrystone)},
        \shortstack{SmallBOOM\\(median)},
        \shortstack{SmallBOOM\\(mt-vvadd)},
        \shortstack{GigaBOOM\\(dhrystone)},
        \shortstack{GigaBOOM\\(median)},
        \shortstack{GigaBOOM\\(mt-vvadd)}
    },
    legend style={at={(0.5, 1.0)}, draw=none, anchor=south, legend columns=-1, font=\footnotesize},
    x tick label style={font=\scriptsize},
    y tick label style={font=\footnotesize},
    label style={font=\footnotesize},
    every node near coord/.append style={font=\tiny, color=black}
  ]
    \addplot+[
      draw=black,
      fill=myred,
      area legend
    ] table[x=dataset, y expr=\thisrow{cpu_16}/\thisrow{cpu_32}] {\dataTableCPUScalingPowerAnalysisTime};

    \addplot+[
      draw=black,
      fill=myorange,
      area legend
    ] table[x=dataset, y expr=\thisrow{cpu_32}/\thisrow{cpu_32}] {\dataTableCPUScalingPowerAnalysisTime};

    \addplot+[
      draw=black,
      fill=mymiddleblue,
      area legend
    ] table[x=dataset, y expr=\thisrow{cpu_64}/\thisrow{cpu_32}] {\dataTableCPUScalingPowerAnalysisTime};

    \addplot+[
      draw=black,
      fill=mylight,
      area legend
    ] table[x=dataset, y expr=\thisrow{gpu_64}/\thisrow{cpu_32}] {\dataTableCPUScalingPowerAnalysisTime};

    \legend{CPU (16 threads), CPU (32 threads), CPU (64 threads), GPU (64 host CPU threads)}
  \end{axis}
\end{tikzpicture}
  \caption{Power-computation runtime of CPU implementations with 16, 32, and 64 threads and GPU with 64 host CPU threads, normalized to 32-thread CPU.}
  \label{fig:cpu_gpu_thread_scaling}
\end{figure}
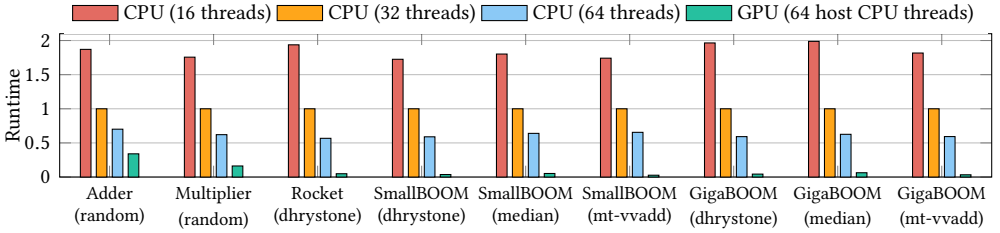

\Cref{fig:cpu_gpu_thread_scaling} compares the power-computation runtime of our multi-threaded CPU implementation under 16, 32, and 64 CPU threads, together with the GPU implementation using 64 host-side CPU threads.
In the GPU implementation, these host-side CPU threads are used for CPU-side work such as event packing.
The measured runtime incorporates all steps within the power computation phase depicted in \Cref{fig:overview_of_framework}, excluding preprocessing.
The results show that increasing the CPU thread count improves performance, but the gains remain sublinear: increasing the CPU thread count from 16 to 32 and from 32 to 64 achieves geometric-mean speedups of 1.84$\times$ and 1.61$\times$, respectively.
This sublinear scaling is mainly due to the highly skewed per-gate event-count distribution, as shown in \Cref{fig:dist_of_per_gate_event_counts}.
In comparison, the GPU implementation with 64 host-side CPU threads achieves a 10.10$\times$ geometric-mean speedup over the 64-thread CPU implementation, effectively removing the primary power-computation bottleneck in time-based power analysis.
% Specifically, \Cref{tab:runtime_comparison_with_my_mt_cpu} compares the runtime of power computation (the orange part in \Cref{fig: cpu_runtime_breakdown_GigaBOOM}) between our CPU and GPU implementations. 
% Our GPU version achieves an average speedup of 89.72$\times$ compared to CPU, which means the primary runtime bottleneck of time-based power analysis is effectively accelerated by our GPU acceleration method. 

\subsubsection{GPU Memory Footprint}\label{sec:gpu_memory_footprint}
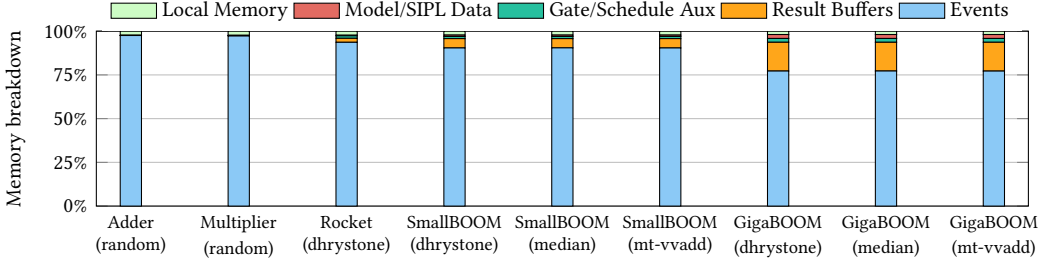
\begin{figure}[t]
    \centering
    % PGFPlots datatable: max CUDA memory breakdown across datasets (GB)
\pgfplotstableread[col sep=space]{
dataset events_device gates_device gate_aux_device delay_device power_result_device schedule_device tile_result_device lut_managed lut_table_device fallback_lut_index_device bsim_lut_index_device bsim_leakage_index_device kernel1all_local_memory_max_resident_gb
  64bAdder  11.9209  0.000473298  0.000170153  5.07385e-05  0.00189645  0.00126156  0.0043273  2.92659e-05  2.20612e-05  2.18302e-06  4.23193e-06  5.21541e-07  0.294037
  64bMultiplier  11.9209  0.00345964  0.00138609  0.000480607  0.00210976  0.00387068  0.031631  0.000101  0.000101484  7.33882e-06  1.77696e-05  1.93715e-06  0.294037
  RocketTile-dhrystone  11.9209  0.0304103  0.0146052  0.00556622  0.0120362  0.167095  0.278037  0.00432578  0.00171673  0.000265092  0.00041198  2.11e-05  0.294037
  BoomTile-dhrystone  11.9209  0.0753076  0.0363974  0.015376  0.0147466  0.125639  0.688527  0.00961204  0.00289884  0.000665456  0.000594899  2.33948e-05  0.294037
  BoomTile-median  11.9209  0.0753076  0.0363974  0.015376  0.00849981  0.125639  0.688527  0.00961204  0.00289884  0.000665456  0.000594899  2.33948e-05  0.294037
  BoomTile-mt-vvadd  11.9209  0.0753076  0.0363974  0.015376  0.0161693  0.125639  0.688527  0.00961204  0.00289884  0.000665456  0.000594899  2.33948e-05  0.294037
  GigaBoom-dhrystone  11.9209  0.274831  0.144582  0.0593645  0.026325  0.192299  2.51274  0.00941947  0.00301013  0.000631295  0.00064797  2.78354e-05  0.294037
  GigaBoom-median  11.9209  0.274831  0.144582  0.0593645  0.0225822  0.192607  2.51274  0.00941947  0.00301013  0.000631295  0.00064797  2.78354e-05  0.294037
  GigaBoom-mt-vvadd  11.9209  0.274831  0.144582  0.0593645  0.0297209  0.192299  2.51274  0.00941947  0.00301013  0.000631295  0.00064797  2.78354e-05  0.294037
}{\dataTableAdditionalMemoryBreakdown}

\pgfplotstablecreatecol[
    create col/expr={
        \thisrow{events_device}
        + \thisrow{gates_device}
        + \thisrow{gate_aux_device}
        + \thisrow{delay_device}
        + \thisrow{power_result_device}
        + \thisrow{schedule_device}
        + \thisrow{tile_result_device}
        + \thisrow{lut_managed}
        + \thisrow{lut_table_device}
        + \thisrow{fallback_lut_index_device}
        + \thisrow{bsim_lut_index_device}
        + \thisrow{bsim_leakage_index_device}
        + \thisrow{kernel1all_local_memory_max_resident_gb}
    }
]{total_gpu_memory_gb}{\dataTableAdditionalMemoryBreakdown}

\definecolor{myorange}{RGB}{255,166,26}
\definecolor{myred}{RGB}{227,107,98}
\definecolor{myyellow}{RGB}{255,214,75}
\definecolor{mymiddleblue}{RGB}{139,201,246}
\definecolor{mylight}{RGB}{38,192,159}
\definecolor{mygreen}{RGB}{205,252,197}
\definecolor{myblue}{RGB}{29,114,221}
\definecolor{mybrown}{RGB}{120, 80, 40}

\begin{tikzpicture}
\begin{axis}[
    width=\linewidth,
    height=0.28\linewidth,
    ybar stacked,
    ymin=0,
    ymax=100,
    ytick={0,25,50,75,100},
    yticklabel={\pgfmathprintnumber{\tick}\%},
    ylabel={Memory breakdown},
    ylabel near ticks,
    symbolic x coords={
        64bAdder,
        64bMultiplier,
        RocketTile-dhrystone,
        BoomTile-dhrystone,
        BoomTile-median,
        BoomTile-mt-vvadd,
        GigaBoom-dhrystone,
        GigaBoom-median,
        GigaBoom-mt-vvadd
    },
    xtick=data,
    xticklabels={
        \shortstack{Adder\\(random)}, 
        \shortstack{Multiplier\\(random)}, 
        \shortstack{Rocket\\(dhrystone)}, 
        \shortstack{SmallBOOM\\(dhrystone)}, 
        \shortstack{SmallBOOM\\(median)}, 
        \shortstack{SmallBOOM\\(mt-vvadd)}, 
        \shortstack{GigaBOOM\\(dhrystone)},
        \shortstack{GigaBOOM\\(median)},
        \shortstack{GigaBOOM\\(mt-vvadd)}
    },
    x tick label style={font=\scriptsize},
    y tick label style={font=\footnotesize},
    label style={font=\footnotesize},
    ymajorgrids=true,
    bar width=8pt,
    enlarge x limits=0.04,
    reverse legend=true,
    legend style={
        at={(0.5,1.0)},
        draw=none,
        anchor=south,
        legend columns=-1,
        font=\footnotesize,
        cells={anchor=west}
    },
]
    \addplot+[ybar, fill=mymiddleblue, draw=black, area legend]
        table[
            x=dataset,
            y expr={100 * \thisrow{events_device} / \thisrow{total_gpu_memory_gb}}
        ] {\dataTableAdditionalMemoryBreakdown};
    \addlegendentry{Events}

    \addplot+[ybar, fill=myorange, draw=black, area legend]
        table[
            x=dataset,
            y expr={
                100 * (
                \thisrow{tile_result_device}
                + \thisrow{power_result_device}
                ) / \thisrow{total_gpu_memory_gb}
            }
        ] {\dataTableAdditionalMemoryBreakdown};
    \addlegendentry{Result Buffers}

    \addplot+[ybar, fill=mylight, draw=black, area legend]
        table[
            x=dataset,
            y expr={
                100 * (
                \thisrow{gate_aux_device}
                + \thisrow{schedule_device}
                ) / \thisrow{total_gpu_memory_gb}
            }
        ] {\dataTableAdditionalMemoryBreakdown};
    \addlegendentry{Gate/Schedule Aux}

    \addplot+[ybar, fill=myred, draw=black, area legend]
        table[
            x=dataset,
            y expr={
                100 * (
                \thisrow{gates_device}
                + \thisrow{delay_device}
                + \thisrow{lut_managed}
                + \thisrow{lut_table_device}
                + \thisrow{fallback_lut_index_device}
                + \thisrow{bsim_lut_index_device}
                + \thisrow{bsim_leakage_index_device}
                ) / \thisrow{total_gpu_memory_gb}
            }
        ] {\dataTableAdditionalMemoryBreakdown};
    \addlegendentry{Model/SIPL Data}

    \addplot+[ybar, fill=mygreen, draw=black, area legend]
        table[
            x=dataset,
            y expr={100 * \thisrow{kernel1all_local_memory_max_resident_gb} / \thisrow{total_gpu_memory_gb}}
        ] {\dataTableAdditionalMemoryBreakdown};
    \addlegendentry{Local Memory}
\end{axis}
\end{tikzpicture}
    \vspace{-1em}
    \caption{
        Breakdown of peak GPU memory across benchmark designs.
        % Memory usage is normalized to the total GPU memory for each benchmark, which includes tracked CUDA device allocations and a local-memory component derived from the power computation kernel's per-thread local-memory usage and maximum resident-thread count; SIPL-related LUT and index data are included in the model/SIPL data category.
    }
    \label{fig:gpu_memory_breakdown}
\end{figure}

We also analyze the GPU memory footprint to quantify the memory overhead of the framework.
The measured peak GPU memory ranges from 12.22~GB to 15.44~GB across the evaluated benchmarks.
As shown in \Cref{fig:gpu_memory_breakdown}, event storage is the dominant component, accounting for 77.2\%--97.5\% of peak memory, while result buffers become the second-largest component on larger designs and account for up to 16.5\%.
In contrast, model/SIPL data accounts for at most 2.3\% of peak memory, indicating that SIPL does not change memory scalability of the flow.

\subsubsection{GPU Kernel Bottleneck Characterization} \label{sec:gpu_kernel_bottleneck_characterization}
To characterize the fused kernel, we collect Nsight Compute profiling metrics across the evaluated benchmarks.
\Cref{tab:gpu_profiling} shows that the fused kernel has low memory-throughput utilization (25.65\%--45.01\%) and low compute-throughput utilization (31.38\%--38.10\%), indicating that it is not primarily limited by memory bandwidth or arithmetic throughput.
Instead, eligible warps per scheduler remain low, ranging from 0.40 to 0.50, and Long Scoreboard accounts for 49.42\%--59.60\% of all issue-stall reasons.
These results indicate that the kernel is mainly limited by low eligible-warp availability, primarily due to long-latency dependency stalls associated with global-memory loads of event data and power-model metadata.

\begin{table*}[t]
    \centering
    \caption{Nsight Compute profiling metrics of the fused power computation kernel.}
    \label{tab:gpu_profiling}
    \resizebox{\linewidth}{!}{
        \begin{threeparttable}
    \setlength{\tabcolsep}{4pt}
    \begin{tabular}{cccccccc}
        \toprule
        {Design}
            & {Benchmark}
            & {\shortstack{Memory/Compute\\Throughput}}
            & {\shortstack{DRAM Memory\\Throughput}}
            & {\shortstack{Achieved\\Occupancy\tnote{1}}}
            & {\shortstack{Eligible/Issued\\Warps}}
            & {\shortstack{Stall Share\\Long Scoreboard/Wait}}
            & {\shortstack{L1/L2\\Hit Rate}} \\
        {}
            & {}
            & {(\%)}
            & {(Gbyte/second)}
            & {(\%)}
            & {warps/sched.}
            & {(\%)}
            & {(\%)} \\
        \midrule
        Adder      & random    & {30.06/38.10} & {273.89} & {37.41} & {0.50/0.38} & {49.42/23.03} & {71.23/70.80} \\
        Multiplier & random    & {25.65/36.75} & {233.73} & {36.46} & {0.48/0.37} & {49.44/23.07} & {72.89/62.36} \\
        Rocket     & dhrystone & {43.26/31.38} & {385.42} & {37.23} & {0.40/0.31} & {59.60/19.22} & {53.98/65.73} \\
        SmallBOOM  & dhrystone & {30.26/35.62} & {244.95} & {37.40} & {0.47/0.35} & {53.99/21.59} & {61.75/75.14} \\
        SmallBOOM  & median    & {37.83/34.81} & {289.82} & {37.96} & {0.47/0.34} & {55.39/20.65} & {57.59/76.85} \\
        SmallBOOM  & mt-vvadd  & {40.94/34.15} & {313.46} & {39.25} & {0.46/0.34} & {56.88/19.65} & {56.27/76.89} \\
        GigaBOOM   & dhrystone & {35.11/34.55} & {208.21} & {37.67} & {0.47/0.34} & {54.22/20.78} & {60.20/81.84} \\
        GigaBOOM   & median    & {40.99/33.35} & {215.30} & {37.67} & {0.45/0.34} & {55.47/19.99} & {55.22/85.56} \\
        GigaBOOM   & mt-vvadd  & {45.01/32.46} & {223.23} & {38.90} & {0.44/0.32} & {57.34/18.58} & {52.30/86.35} \\
        \midrule
        Median     &           & {37.83/34.55} & {244.95} & {37.67} & {0.47/0.34} & {55.39/20.65} & {57.59/76.85} \\
        \bottomrule
    \end{tabular}
    \begin{tablenotes}[flushleft]
        \footnotesize
        \item[] For benchmarks with multiple kernel invocations, profiling metrics are aggregated using kernel-duration-weighted averages.
        \item[1] The theoretical maximum occupancy is 41.67\%.
    \end{tablenotes}
\end{threeparttable}

% Metric definitions:
% - ``Memory/Compute Throughput'' denotes
%   \texttt{gpu__compute_memory_throughput.avg.pct_of_peak_sustained_elapsed}.
%   /
%   \texttt{sm__throughput.avg.pct_of_peak_sustained_elapsed}.
% - ``DRAM Memory Throughput'' denotes
%   \texttt{dram__bytes.sum.per_second}.
% - ``Achieved Occupancy'' denotes
%   \texttt{sm__warps_active.avg.pct_of_peak_sustained_active}.
% - ``Eligible/Issued Warps'' denotes
%   \texttt{smsp__warps_eligible.avg.per_cycle_active} and
%   \texttt{smsp__issue_active.avg.per_cycle_active}, respectively.
% - ``Stall Share Long Scoreboard/Wait'' denotes the percentage of all
%   issue-stall reasons attributed to Long Scoreboard and Wait, respectively.
%   The two numerators are
%   \texttt{smsp__average_warps_issue_stalled_long_scoreboard_per_issue_active.ratio}
%   and
%   \texttt{smsp__average_warps_issue_stalled_wait_per_issue_active.ratio};
%   the denominator is the sum of the corresponding
%   \texttt{smsp__average_warps_issue_stalled_*_per_issue_active.ratio}
%   metrics over all issue-stall reasons.
% - ``L1/L2 Hit Rate'' denotes
%   \texttt{l1tex__t_sector_hit_rate.pct} /
%   \texttt{lts__t_sector_hit_rate.pct}.
%
% Interpretation:
% Lower memory and compute throughput, together with low eligible warps and dominant Long Scoreboard stalls, indicates that performance is not primarily limited by raw DRAM bandwidth or compute throughput.
% Instead, the main loss occurs at the warp-readiness and scheduler-issue level: many resident warps are not eligible for issue because they are waiting on long-latency memory dependencies.

    }
\end{table*}

Because waveform-bin atomic updates may become a bottleneck under high switching activity, we run a stress test on \texttt{GigaBOOM} using 100,000-cycle synthetic high-activity waveforms with 1, 2, 4, and 8 toggles per cycle.
Compared with a no-atomic variant that replaces \texttt{atomicAdd} with regular global-memory writes, waveform-bin atomic updates introduce only 2.03\%--4.19\% relative overhead.
This limited overhead is primarily due to the low eligible-warp availability observed in profiling, which limits the number of warps issuing atomic updates concurrently.
Moreover, practical activity traces often span hundreds of thousands of cycles, allowing concurrent updates to be spread across many waveform bins and further limiting simultaneous conflicts on the same bin.
These results indicate that waveform-bin atomic updates have limited impact on fused-kernel runtime.

% \begin{figure}[t]
%     \centering
%     \input{pgfplot/kernel_time_with_without_waveform_bin_atomic_update}
%     \caption{Fused kernel runtime with and without waveform-bin atomic updates on \texttt{GigaBOOM} under synthetic high-activity waveforms. Each point is normalized to the no-atomic variant at the same toggle count.}
%     \label{fig:kernel_time_with_without_waveform_bin_atomic_update}
% \end{figure}

\subsubsection{Cross-GPU Platform Evaluation} \label{sec:cross_gpu_platform_evaluation}
\begin{figure*}[t]
    \centering
    \definecolor{myorange}{RGB}{255,166,26}
\definecolor{myred}{RGB}{227,107,98}
\definecolor{mymiddleblue}{RGB}{139,201,246}

\pgfplotstableread[col sep=space]{
dataset                rtx3090       a30           h100
Adder(random)          73.017900     104.831000    39.134100
Multiplier(random)     1707.240000   3062.220000   1235.660000
Rocket(dhrystone)      7596.660000   11029.300000  3689.870000
SmallBOOM(dhrystone)   25236.300000  41624.600000  14050.200000
SmallBOOM(median)      5559.690000   8790.590000   2875.920000
SmallBOOM(mt-vvadd)    17063.500000  26137.700000  8501.440000
GigaBOOM(dhrystone)    44820.300000  75383.800000  26083.600000
GigaBOOM(median)       13902.600000  23264.500000  8046.880000
GigaBOOM(mt-vvadd)     41999.100000  67677.700000  23043.000000
}{\dataTableCrossGPUKernelRuntime}

\pgfplotsset{
    width=0.86\linewidth,
    height=0.36\linewidth
}

\begin{tikzpicture}
  \begin{axis}[
    width=\linewidth,
    height=0.25\linewidth,
    ymajorgrids=true,
    ybar,
    symbolic x coords={
        Adder(random),
        Multiplier(random),
        Rocket(dhrystone),
        SmallBOOM(dhrystone),
        SmallBOOM(median),
        SmallBOOM(mt-vvadd),
        GigaBOOM(dhrystone),
        GigaBOOM(median),
        GigaBOOM(mt-vvadd)
    },
    ymin=0,
    ymax=2.0,
    ytick={0,0.5,1.0,1.5,2.0},
    bar width=4pt,
    enlarge x limits=0.06,
    ylabel={Runtime},
    ylabel near ticks,
    ylabel shift=-5pt,
    ylabel style={align=center},
    xtick align=inside,
    xtick=data,
    xticklabels={
        \shortstack{Adder\\(random)},
        \shortstack{Multiplier\\(random)},
        \shortstack{Rocket\\(dhrystone)},
        \shortstack{SmallBOOM\\(dhrystone)},
        \shortstack{SmallBOOM\\(median)},
        \shortstack{SmallBOOM\\(mt-vvadd)},
        \shortstack{GigaBOOM\\(dhrystone)},
        \shortstack{GigaBOOM\\(median)},
        \shortstack{GigaBOOM\\(mt-vvadd)}
    },
    legend style={at={(0.5, 1.0)}, draw=none, anchor=south, legend columns=-1, font=\footnotesize},
    x tick label style={font=\scriptsize},
    y tick label style={font=\footnotesize},
    label style={font=\footnotesize},
    every node near coord/.append style={font=\tiny, color=black}
  ]
    \addplot+[
      draw=black,
      fill=myred,
      area legend
    ] table[x=dataset, y expr=\thisrow{rtx3090}/\thisrow{rtx3090}] {\dataTableCrossGPUKernelRuntime};

    \addplot+[
      draw=black,
      fill=myorange,
      area legend
    ] table[x=dataset, y expr=\thisrow{a30}/\thisrow{rtx3090}] {\dataTableCrossGPUKernelRuntime};

    \addplot+[
      draw=black,
      fill=mymiddleblue,
      area legend
    ] table[x=dataset, y expr=\thisrow{h100}/\thisrow{rtx3090}] {\dataTableCrossGPUKernelRuntime};

    \legend{RTX 3090, A30, H100}
  \end{axis}
\end{tikzpicture}
    \caption{Fused kernel runtime on RTX 3090, A30, and H100, normalized to RTX 3090.}
    \label{fig:cross_gpu_plat_kernel_time_vs_rtx3090}
\end{figure*}
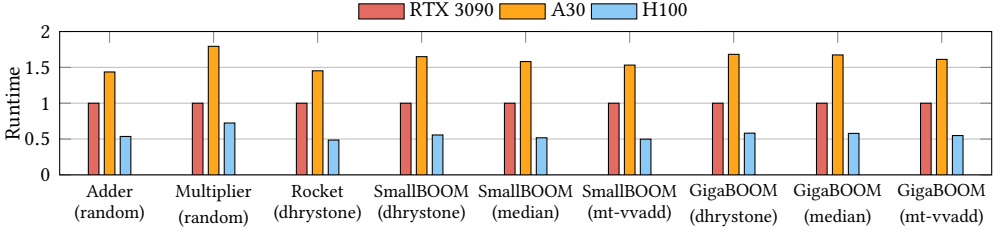

To evaluate the portability of the proposed framework, we additionally measure the execution time of the fused power computation kernel on two other GPU platforms: an NVIDIA A30 GPU with 24~GB memory hosted by an Intel Xeon 6348 CPU, and an NVIDIA H100 GPU with 80~GB memory hosted by an Intel Xeon 8468 CPU.
% We compare these results with the RTX 3090 platform used in the main experiments.
As shown in \Cref{fig:cross_gpu_plat_kernel_time_vs_rtx3090}, A30 is slower than RTX 3090, with normalized runtime ranging from 1.44$\times$ to 1.79$\times$ (geometric mean: 1.60$\times$), while H100 is faster, with normalized runtime ranging from 0.49$\times$ to 0.72$\times$ (geometric mean: 0.56$\times$).
These results confirm that the proposed framework is not limited to a single GPU architecture.
The performance trend is consistent with the relative capability of the platforms, but the measured runtime does not scale simply with peak compute throughput or peak memory bandwidth.
Together with the profiling results in \Cref{tab:gpu_profiling}, this non-linear scaling further suggests that the fused kernel is mainly affected by long-latency dependencies and warp readiness.

\subsection{Optimization Strategies Evaluation}

\subsubsection{Incremental and Ablation Evaluation}\label{sec:incremental_opt_evaluation}
\begin{figure}[tb!]
    \centering
    %\definecolor{myred}{RGB}{255, 128, 128}
%\definecolor{mygreen}{RGB}{128, 255, 128}     %FFABA4q
%\definecolor{myblue}{RGB}{128, 80, 40}    %1D72DD
\definecolor{myorange}{RGB}{255,166,26}
\definecolor{myred}{RGB}{227,107,98}
\definecolor{myyellow}{RGB}{255,214,75}
\definecolor{mymiddleblue}{RGB}{139,201,246} 
\definecolor{mylight}{RGB}{38,192,159}  
\definecolor{mygreen}{RGB}{205,252,197}   
%FFABA4q
\definecolor{myblue}{RGB}{29,114,221}    %1D72DD
\definecolor{mybrown}{RGB}{120, 80, 40} 
\definecolor{mypink}{RGB}{252,228,215}   %FCE4D7

% ====== PGFPlots filecontents (auto-generated) ======
\pgfplotstableread[col sep=space]{
dataset               naive           bsim            auto_selection  fusion
Adder(random)         1.000000        0.851311        0.696660        0.657974
Multiplier(random)    1.000000        0.897622        0.870697        0.835547
Rocket(dhrystone)     1.000000        0.888685        0.582506        0.367371
SmallBOOM(dhrystone)  1.000000        0.879283        0.606937        0.387968
SmallBOOM(median)     1.000000        0.922842        0.597485        0.302776
SmallBOOM(mt-vvadd)   1.000000        0.922655        0.592278        0.280866
GigaBOOM(dhrystone)   1.000000        0.883891        0.438566        0.314457
GigaBOOM(median)      1.000000        0.942670        0.415378        0.256944
GigaBOOM(mt-vvadd)    1.000000        0.965613        0.412770        0.234319
}{\dataTableIncrementalEvaluation}
% ===================================================

\pgfplotsset{
    width=0.86\linewidth,
    height=0.36\linewidth
}

\begin{tikzpicture}
\begin{axis}[
    width=\linewidth,
    height=0.25\linewidth, % 你想要的高度比例
    ymajorgrids=true,
    ybar,
    symbolic x coords={
        Adder(random),
        Multiplier(random),
        Rocket(dhrystone),
        SmallBOOM(dhrystone),
        SmallBOOM(median),
        SmallBOOM(mt-vvadd),
        GigaBOOM(dhrystone),
        GigaBOOM(median),
        GigaBOOM(mt-vvadd)
    },
    xlabel near ticks,
    xtick align=inside,
    xtick=data,
    xticklabels={
        \shortstack{Adder\\(random)}, 
        \shortstack{Multiplier\\(random)}, 
        \shortstack{Rocket\\(dhrystone)}, 
        \shortstack{SmallBOOM\\(dhrystone)}, 
        \shortstack{SmallBOOM\\(median)}, 
        \shortstack{SmallBOOM\\(mt-vvadd)}, 
        \shortstack{GigaBOOM\\(dhrystone)},
        \shortstack{GigaBOOM\\(median)},
        \shortstack{GigaBOOM\\(mt-vvadd)}
    },
    enlarge x limits=0.06,
    ylabel={Runtime},
    ylabel near ticks,
    ylabel shift=-5pt,
    ylabel style={align=center},  % 启用多行对齐
    % ytick align=outside,
    ytick={0, 0.25, 0.5, 0.75, 1.0},
    bar width = 4pt,
    ymin=0,
    ymax=1.1,
    legend style={at={(0.5, 1.0)}, draw=none,anchor=south,legend columns=-1, font=\footnotesize},
    x tick label style={font=\scriptsize},
    y tick label style={font=\footnotesize},
    label style={font=\footnotesize}, % 调整轴标签的字体大小
    every node near coord/.append style={font=\tiny, color=black}
]
    % use TeX as calculator:
    \addplot +[ybar, fill=myred, draw=black, area legend] table [x=dataset,  y={naive}] {\dataTableIncrementalEvaluation};
    \addplot +[      fill=myorange, draw=black, area legend] table [x=dataset,  y={bsim}] {\dataTableIncrementalEvaluation};
    \addplot +[      fill=mymiddleblue, draw=black, area legend] table [x=dataset,  y={auto_selection}] {\dataTableIncrementalEvaluation};
    \addplot +[      fill=mylight, draw=black, area legend] table [x=dataset,  y={fusion}] {\dataTableIncrementalEvaluation};

    \legend{Naive Implementation, SIPL, Event-Density-Aware, Kernel Fusion}
\end{axis}
\end{tikzpicture}
    \caption{Power computation kernel runtime under incremental optimizations. Normalized to "naive".}
    \label{fig:incremental_evaluation}
\end{figure}
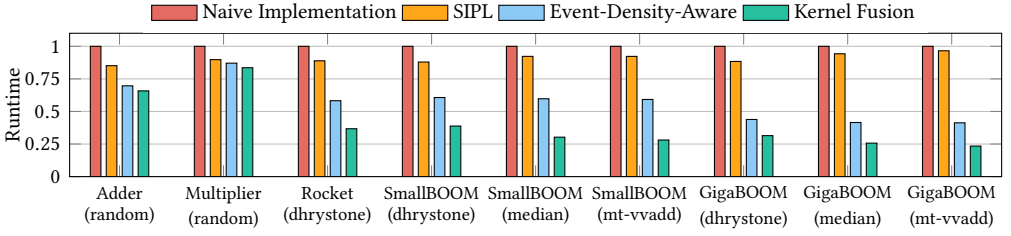
\begin{figure}[tb!]
    \centering
    %\definecolor{myred}{RGB}{255, 128, 128}
%\definecolor{mygreen}{RGB}{128, 255, 128}     %FFABA4q
%\definecolor{myblue}{RGB}{128, 80, 40}    %1D72DD
\definecolor{myorange}{RGB}{255,166,26}
\definecolor{myred}{RGB}{227,107,98}
\definecolor{myyellow}{RGB}{255,214,75}
\definecolor{mymiddleblue}{RGB}{139,201,246} 
\definecolor{mylight}{RGB}{38,192,159}  
\definecolor{mygreen}{RGB}{205,252,197}   
%FFABA4q
\definecolor{myblue}{RGB}{29,114,221}    %1D72DD
\definecolor{mybrown}{RGB}{120, 80, 40} 
\definecolor{mypink}{RGB}{252,228,215}   %FCE4D7

% ====== PGFPlots filecontents (auto-generated) ======
\pgfplotstableread[col sep=space]{
dataset                full            no_bsim         no_auto_selection no_fusion
Adder(random)          1.000000        1.247854        1.856539        1.058796
Multiplier(random)     1.000000        1.331428        1.198468        1.042068
Rocket(dhrystone)      1.000000        1.231413        2.504840        1.582859
SmallBOOM(dhrystone)   1.000000        1.249622        2.167786        1.566284
SmallBOOM(median)      1.000000        1.175060        2.670989        1.965970
SmallBOOM(mt-vvadd)    1.000000        1.163034        2.806702        2.113027
GigaBOOM(dhrystone)    1.000000        1.304226        2.876608        1.398830
GigaBOOM(median)       1.000000        1.228172        3.548151        1.625328
GigaBOOM(mt-vvadd)     1.000000        1.186748        3.758949        1.763060
}{\dataTableAblationStudy}
% ===================================================

\pgfplotsset{
    width=0.86\linewidth,
    height=0.36\linewidth
}

\begin{tikzpicture}
\begin{axis}[
    width=\linewidth,
    height=0.25\linewidth, % 你想要的高度比例
    ymajorgrids=true,
    ybar,
    symbolic x coords={
        Adder(random),
        Multiplier(random),
        Rocket(dhrystone),
        SmallBOOM(dhrystone),
        SmallBOOM(median),
        SmallBOOM(mt-vvadd),
        GigaBOOM(dhrystone),
        GigaBOOM(median),
        GigaBOOM(mt-vvadd)
    },
    xlabel near ticks,
    xtick align=inside,
    xtick=data,
    xticklabels={
        \shortstack{Adder\\(random)}, 
        \shortstack{Multiplier\\(random)}, 
        \shortstack{Rocket\\(dhrystone)}, 
        \shortstack{SmallBOOM\\(dhrystone)}, 
        \shortstack{SmallBOOM\\(median)}, 
        \shortstack{SmallBOOM\\(mt-vvadd)}, 
        \shortstack{GigaBOOM\\(dhrystone)},
        \shortstack{GigaBOOM\\(median)},
        \shortstack{GigaBOOM\\(mt-vvadd)}
    },
    enlarge x limits=0.06,
    ylabel={Runtime},
    ylabel near ticks,
    ylabel shift=-5pt,
    ylabel style={align=center},  % 启用多行对齐
    % ytick align=outside,
    ytick={0, 1.0, 2.0, 3.0, 4.0},
    bar width = 4pt,
    ymin=0,
    ymax=4.0,
    legend style={at={(0.5, 1.0)}, draw=none,anchor=south,legend columns=-1, font=\footnotesize},
    x tick label style={font=\scriptsize},
    y tick label style={font=\footnotesize},
    label style={font=\footnotesize}, % 调整轴标签的字体大小
    every node near coord/.append style={font=\tiny, color=black}
]
    % use TeX as calculator:
    \addplot +[ybar, fill=myred, draw=black, area legend] table [x=dataset,  y={full}] {\dataTableAblationStudy};
    \addplot +[      fill=myorange, draw=black, area legend] table [x=dataset,  y={no_bsim}] {\dataTableAblationStudy};
    \addplot +[      fill=mymiddleblue, draw=black, area legend] table [x=dataset,  y={no_auto_selection}] {\dataTableAblationStudy};
    \addplot +[      fill=mylight, draw=black, area legend] table [x=dataset,  y={no_fusion}] {\dataTableAblationStudy};

    \legend{Complete, w/o SIPL, w/o Event-Density-Aware, w/o Kernel Fusion}
\end{axis}
\end{tikzpicture}
    \caption{Power computation kernel runtime under leave-one-out ablations. Normalized to the complete method.}
    \label{fig:ablation_study}
\end{figure}
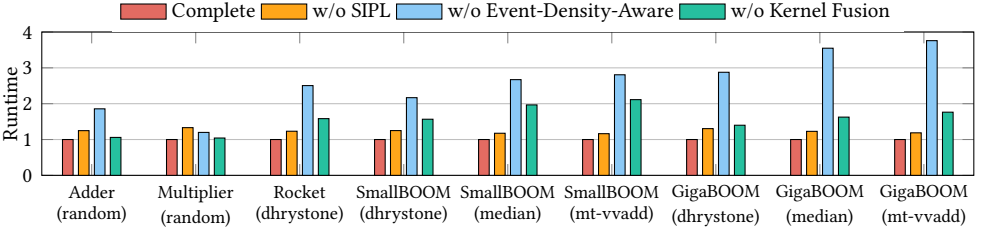
To evaluate the effectiveness of our proposed optimizations in reducing the execution time of the power computation kernel, we build three incremental versions of the naive implementation, adding one optimization at a time to isolate and quantify each contribution.
In the naive implementation, the Naive-L kernel sets the global cycle count per thread $n_{\mathrm{cyc}}^{\mathrm{fix}}$ to 1.
The evaluation results presented in \Cref{fig:incremental_evaluation} reveal two key findings:
First, the SIPL data structure, event-density-aware per-gate cycle partitioning, and kernel fusion versions bring 9.45\%, 43.67\%, and 63.12\% geometric mean runtime reductions compared to the naive implementation.
Second, the event-density-aware per-gate cycle partitioning achieves higher speedups on the other designs than on \texttt{Adder} and \texttt{Multiplier}, because, as shown in \Cref{fig:dist_of_per_gate_event_counts}, their per-gate event count distributions are more skewed than those of \texttt{Adder} and \texttt{Multiplier}.

To clarify each optimization's role in the complete implementation, we conduct a leave-one-out ablation study, as shown in \Cref{fig:ablation_study}.
Removing event-density-aware partitioning causes the largest slowdown, increasing kernel runtime by 147.4\% in geometric mean and up to 275.9\%.
Removing kernel fusion and SIPL increases geometric-mean runtime by 52.9\% and 23.4\%, respectively.
Although SIPL alone gives a modest 9.45\% geometric-mean reduction in the incremental study, removing it from the complete method still causes a 23.4\% slowdown, showing that state-dependent power retrieval remains important after other overheads are reduced.

We further quantify the trade-off between kernel fusion, register usage, occupancy, and runtime in \Cref{tab:fusion_and_occupancy_sensitivity}.
The fused kernel uses 89 registers per thread, which limits the theoretical occupancy to 41.67\% and results in 37.67\% achieved occupancy.
Limiting register usage increases achieved occupancy to 66.53\% and 89.74\%, but it also increases normalized kernel time to 1.60$\times$ and 2.09$\times$, respectively.
The no-fusion variant reduces register pressure and increases occupancy for the separated kernels, but its normalized runtime remains 1.52$\times$ due to the redundant gate-state reconstruction and glitch-detection bookkeeping discussed in \Cref{sec:challenge_3_redundant_computation}.

\begin{table*}[t]
    \centering
    \caption{Register usage, occupancy, and runtime sensitivity of the power computation kernel under fusion and occupancy-control variants.}
    \label{tab:fusion_and_occupancy_sensitivity}
    \resizebox{\linewidth}{!}{
        \begin{threeparttable}
    \setlength{\tabcolsep}{4pt}
    \begin{tabular}{lcccc}
        \toprule
        {Configuration}
            & {Registers per Thread}
            & {Theoretical Occupancy (\%)}
            & {Achieved Occupancy (\%)\tnote{1}}
            & {Normalized Kernel Time\tnote{2}} \\
        \midrule
        Fusion              & {89}    & {41.67}        & {37.67}       & {1.00} \\
        Fusion + Occ. 75\%  & {56}    & {75.00}        & {66.53}       & {1.60} \\
        Fusion + Occ. 100\% & {40}    & {100.00}       & {89.74}       & {2.09} \\
        No Fusion           & {68/40} & {58.33/100.00} & {56.50/88.49} & {1.52} \\
        \bottomrule
    \end{tabular}
    \begin{tablenotes}[flushleft]
        \footnotesize
        \item[] For the no-fusion variant, paired values correspond to the dynamic and leakage power computation kernels, respectively.
        \item[1] Achieved occupancy is first aggregated per benchmark using kernel-duration-weighted averaging. The table reports the median across the nine benchmarks.
        \item[2] Normalized kernel time is first computed for each benchmark relative to the fusion baseline and then aggregated using the geometric mean across the nine benchmarks. For the no-fusion variant, kernel time is the sum of the dynamic and leakage power computation kernel times.
    \end{tablenotes}
\end{threeparttable}

% Metric definitions:
% - ``Registers per Thread'' denotes
%   \texttt{launch__registers_per_thread}.
% - ``Theoretical Occupancy'' denotes
%   \texttt{sm__maximum_warps_per_active_cycle_pct}.
% - ``Achieved Occupancy'' denotes
%   \texttt{sm__warps_active.avg.pct_of_peak_sustained_active}.

    }
\end{table*}

\subsubsection{Event-Density-Aware Per-Gate Partitioning vs. Globally Fixed Partitioning}\label{sec:density_based_cyc_part_vs_global_cycle}
\begin{figure}[tb!]
    \centering
    %\definecolor{myred}{RGB}{255, 128, 128}
%\definecolor{mygreen}{RGB}{128, 255, 128}     %FFABA4q
%\definecolor{myblue}{RGB}{128, 80, 40}    %1D72DD
\definecolor{myorange}{RGB}{255,166,26}
\definecolor{myred}{RGB}{227,107,98}
\definecolor{myyellow}{RGB}{255,214,75}
\definecolor{mymiddleblue}{RGB}{139,201,246} 
\definecolor{mylight}{RGB}{38,192,159}  
\definecolor{mygreen}{RGB}{205,252,197}   
%FFABA4q
\definecolor{myblue}{RGB}{29,114,221}    %1D72DD
\definecolor{mybrown}{RGB}{120, 80, 40} 
\definecolor{mypink}{RGB}{252,228,215}   %FCE4D7

% ====== PGFPlots filecontents (auto-generated) ======
\pgfplotstableread[col sep=space]{
dataset              n=auto          n=2             n=4             n=8             n=16            n=32
Adder(random)        1.000000        1.323041        1.163794        1.111756        1.181663        1.333917
Multiplier(random)   1.000000        1.127915        1.145482        1.233436        1.397260        1.622281
Rocket(dhrystone)    1.000000        1.678062        1.358586        1.302705        1.291266        1.410034
SmallBOOM(dhrystone) 1.000000        1.746132        1.587227        1.668205        1.966926        2.517755
SmallBOOM(median)    1.000000        1.866296        1.595101        1.677559        1.952659        2.599398
SmallBOOM(mt-vvadd)  1.000000        1.898299        1.526277        1.536102        1.690006        2.136467
GigaBOOM(dhrystone)  1.000000        1.926674        1.503057        1.347190        1.433659        1.727851
GigaBOOM(median)     1.000000        2.186227        1.595121        1.502264        1.632676        1.993177
GigaBOOM(mt-vvadd)   1.000000        2.273127        1.583383        1.453206        1.553042        1.827069
}{\dataTableNCycleTuning}
% ===================================================

\pgfplotsset{
    width=0.86\linewidth,
    height=0.36\linewidth
}

\begin{tikzpicture}
\begin{axis}[
    width=\linewidth,
    height=0.25\linewidth, % 你想要的高度比例
    ymajorgrids=true,
    ybar,
    symbolic x coords={
        Adder(random),
        Multiplier(random),
        Rocket(dhrystone),
        SmallBOOM(dhrystone),
        SmallBOOM(median),
        SmallBOOM(mt-vvadd),
        GigaBOOM(dhrystone),
        GigaBOOM(median),
        GigaBOOM(mt-vvadd)
    },
    xlabel near ticks,
    xtick align=inside,
    xtick=data,
    xticklabels={
        \shortstack{Adder\\(random)}, 
        \shortstack{Multiplier\\(random)}, 
        \shortstack{Rocket\\(dhrystone)}, 
        \shortstack{SmallBOOM\\(dhrystone)}, 
        \shortstack{SmallBOOM\\(median)}, 
        \shortstack{SmallBOOM\\(mt-vvadd)}, 
        \shortstack{GigaBOOM\\(dhrystone)},
        \shortstack{GigaBOOM\\(median)},
        \shortstack{GigaBOOM\\(mt-vvadd)}
    },
    enlarge x limits=0.06,
    ylabel={Runtime},
    ylabel near ticks,
    ylabel shift=-5pt,
    ylabel style={align=center},  % 启用多行对齐
    % ytick align=outside,
    ytick={0.5, 1.0, 1.5, 2.0, 2.5},
    bar width = 3pt,
    ymin=0.0,
    ymax=2.7,
    legend style={at={(0.5, 1.0)}, draw=none,anchor=south,legend columns=-1, font=\footnotesize},
    x tick label style={font=\scriptsize},
    y tick label style={font=\footnotesize},
    label style={font=\footnotesize}, % 调整轴标签的字体大小
    every node near coord/.append style={font=\tiny, color=black}
]
    % use TeX as calculator:
    \addplot +[ybar, fill=myred, draw=black, area legend] table [x=dataset,  y={n=auto}] {\dataTableNCycleTuning};
    \addplot +[      fill=myorange, draw=black, area legend] table [x=dataset,  y={n=2}] {\dataTableNCycleTuning};
    \addplot +[      fill=mymiddleblue, draw=black, area legend] table [x=dataset,  y={n=4}] {\dataTableNCycleTuning};
    \addplot +[      fill=mylight, draw=black, area legend] table [x=dataset,  y={n=8}] {\dataTableNCycleTuning};
    \addplot +[      fill=mypink, draw=black, area legend] table [x=dataset,  y={n=16}] {\dataTableNCycleTuning};
    \addplot +[      fill=mybrown, draw=black, area legend] table [x=dataset,  y={n=32}] {\dataTableNCycleTuning};

    \legend{Event-Density-Aware, $n_{\mathrm{cyc}}^{\mathrm{fix}}=2$, $n_{\mathrm{cyc}}^{\mathrm{fix}}=4$, $n_{\mathrm{cyc}}^{\mathrm{fix}}=8$, $n_{\mathrm{cyc}}^{\mathrm{fix}}=16$, $n_{\mathrm{cyc}}^{\mathrm{fix}}=32$}
\end{axis}
\end{tikzpicture}
    \caption{Fused power computation kernel execution time with different global cycle count per thread ($n_{\mathrm{cyc}}^{\mathrm{fix}}$). Normalized against event-density-aware per-gate partitioning with $n_{\mathrm{evt,tgt}}=8$.}
    \label{fig:n_cycle_tuning}
    % \vspace{-0.4cm}
\end{figure}
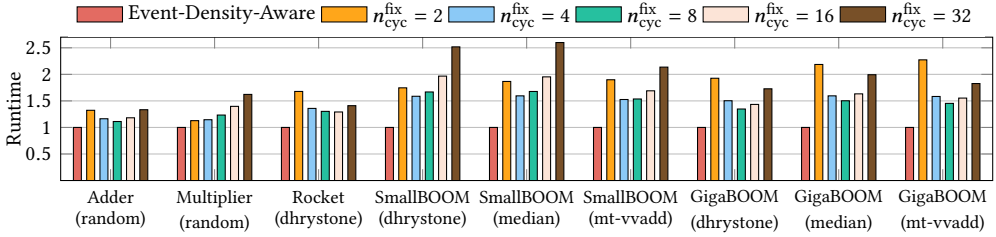
We compare the execution time of the fused power computation kernel under two cycle-partitioning schemes: the event-density-aware per-gate partitioning and the globally fixed partitioning.
The event-density-aware per-gate partitioning sets the target event count per thread as 8 ($n_{\mathrm{evt,tgt}}=8$), while the globally fixed partitioning adopts an exponentially stepped parameter sweep (global cycle count per thread $n_{\mathrm{cyc}}^{\mathrm{fix}}$ ranges from 2 to 32).
As shown in \Cref{fig:n_cycle_tuning}, the event-density-aware per-gate partitioning consistently outperforms the globally fixed partitioning, demonstrating the benefits of fine-grained, density-aware parallelism allocation.

\begin{figure}[!t]
    \captionsetup[subfigure]{skip=0pt}
    \subfloat[]{ 
        \hspace{-.4in}
        \definecolor{myorange}{RGB}{255,166,26}
\definecolor{myred}{RGB}{227,107,98}
\definecolor{myyellow}{RGB}{255,214,75}
\definecolor{mymiddleblue}{RGB}{139,201,246} 
\definecolor{mylight}{RGB}{38,192,159}  
\definecolor{mygreen}{RGB}{205,252,197}   
\definecolor{myblue}{RGB}{29,114,221}    %1D72DD
\definecolor{mybrown}{RGB}{120, 80, 40}

\begin{tikzpicture}

\pgfplotstableread[col sep=space]{
  set  sparse_threads_base  sparse_threads_auto  active_threads_base  active_threads_auto  sparse_H_base  sparse_H_auto  TperG_p90_base  TperG_p90_auto  TperG_p99_base  TperG_p99_auto  EperT_p90_base  EperT_p90_auto  EperT_p99_base  EperT_p99_auto
  Adder(random)         0  0  2.26893e+07  1.90767e+07  0  0  12501  25001  12501  33334  17.6099  8.79518  21.2045  8.80998
  Multiplier(random)    0  0  1.65864e+08  3.35959e+08  0  0  2083.67  8333.5  2083.67  16666.8  24.6657  8.49271  90.2671  11.2851
  Rocket(dhrystone)     2.6208e+09  4.37912e+07  7.7197e+09  2.83992e+09  3.16223e+10  5.36263e+08  33098.5  66196.5  33098.5  67970.5  16.0001  8.00006  18.3373  8.24193
  SmallBOOM(dhrystone)  9.26078e+09  6.09055e+08  1.81556e+10  9.75646e+09  7.66366e+10  4.92608e+09  6985.33  14026.7  6985.33  20158.1  15.9957  7.33384  24.7858  7.80599
  SmallBOOM(median)     1.94933e+09  9.24169e+07  6.0484e+09  2.85431e+09  2.22e+10  1.05537e+09  10472  20943.5  10472  22859  15.9992  7.99984  20.3511  8.22987
  SmallBOOM(mt-vvadd)   1.31059e+10  4.95075e+08  2.09124e+10  9.53697e+09  1.11023e+11  4.25685e+09  12069  24137.8  12069  26566.2  15.9999  7.99994  19.0712  8.16432
  GigaBOOM(dhrystone)   2.8173e+10  2.93046e+09  4.73525e+10  2.22358e+10  2.48466e+11  2.52398e+10  3456.15  7327.08  3456.15  8402.31  15.9616  6.61703  21.1514  7.36738
  GigaBOOM(median)      1.01089e+10  8.32314e+08  2.08771e+10  8.32805e+09  9.32457e+10  7.76506e+09  4952.25  10117  4952.25  10552  15.9977  7.50009  18.1356  7.78374
  GigaBOOM(mt-vvadd)    4.77504e+10  3.2994e+09  7.13682e+10  2.64434e+10  4.26923e+11  2.94375e+10  6156.09  12489.1  6156.09  12764.3  15.9995  7.63684  17.669  7.86864
}{\dataTableThreadAllocStatsCmpAcrossDatasets}

\begin{axis}[
    width=1.0\linewidth,
    height=0.24\linewidth, % 你想要的高度比例
    symbolic x coords={
        Adder(random),
        Multiplier(random),
        Rocket(dhrystone),
        SmallBOOM(dhrystone),
        SmallBOOM(median),
        SmallBOOM(mt-vvadd),
        GigaBOOM(dhrystone),
        GigaBOOM(median),
        GigaBOOM(mt-vvadd)
    },
    xtick={
        Adder(random),
        Multiplier(random),
        Rocket(dhrystone),
        SmallBOOM(dhrystone),
        SmallBOOM(median),
        SmallBOOM(mt-vvadd),
        GigaBOOM(dhrystone),
        GigaBOOM(median),
        GigaBOOM(mt-vvadd)
    },
    xticklabels={
        \shortstack{Adder\\(random)}, 
        \shortstack{Multiplier\\(random)}, 
        \shortstack{Rocket\\(dhrystone)}, 
        \shortstack{SmallBOOM\\(dhrystone)}, 
        \shortstack{SmallBOOM\\(median)}, 
        \shortstack{SmallBOOM\\(mt-vvadd)}, 
        \shortstack{GigaBOOM\\(dhrystone)},
        \shortstack{GigaBOOM\\(median)},
        \shortstack{GigaBOOM\\(mt-vvadd)}
    },
    x tick label style={font=\scriptsize},
    y tick label style={font=\footnotesize},
    xlabel style={font=\small},
    ylabel style={font=\footnotesize},
    title style={font=\small},
    ylabel={Event per thread ratio},
    ymode=log,
    ymin=0.05, ymax=1.2,
    ymajorgrids=true,
    enlarge x limits=0.03,
    legend style={at={(0.99,0.01)}, font=\footnotesize, legend columns=-1, anchor=south east}
]

    % ----------------------------- (B) E/thread tail compressed: ratio lines (log) -----------------------------
    % reference line ratio=1
    \addplot[thick, dashed, color=mybrown, mark=none] coordinates {
        ({Adder(random)},1)
        ({GigaBOOM(mt-vvadd)},1)
    };
    \addlegendentry{ratio=1}

    % p90
    \addplot+[
        thick,
        color=mylight,
        mark=*,
        mark options={solid, fill=mylight},
    ] table[
        x=set,
        y expr=\thisrow{EperT_p90_auto}/\thisrow{EperT_p90_base}
    ] {\dataTableThreadAllocStatsCmpAcrossDatasets};
    \addlegendentry{$p_{90}$}

    % p99
    \addplot+[
        thick,
        color=myred,
        mark=square*,
        mark options={solid, fill=myred},
    ] table[
        x=set,
        y expr=\thisrow{EperT_p99_auto}/\thisrow{EperT_p99_base}
    ] {\dataTableThreadAllocStatsCmpAcrossDatasets};
    \addlegendentry{$p_{99}$}

\end{axis}
\end{tikzpicture} 
        \label{fig:thread_allocation_stat_heavy_gates}
    }
    \\[0pt]
    \subfloat[]{
        \hspace{-.4in}
        \definecolor{myorange}{RGB}{255,166,26}
\definecolor{myred}{RGB}{227,107,98}
\definecolor{myyellow}{RGB}{255,214,75}
\definecolor{mymiddleblue}{RGB}{139,201,246} 
\definecolor{mylight}{RGB}{38,192,159}  
\definecolor{mygreen}{RGB}{205,252,197}   
\definecolor{myblue}{RGB}{29,114,221}    %1D72DD
\definecolor{mybrown}{RGB}{120, 80, 40}

\begin{tikzpicture}

\pgfplotstableread[col sep=space]{
  set  sparse_threads_base  sparse_threads_auto  active_threads_base  active_threads_auto  sparse_H_base  sparse_H_auto  TperG_p90_base  TperG_p90_auto  TperG_p99_base  TperG_p99_auto  EperT_p90_base  EperT_p90_auto  EperT_p99_base  EperT_p99_auto
  Adder(random)         0  0  2.26893e+07  1.90767e+07  0  0  12501  25001  12501  33334  17.6099  8.79518  21.2045  8.80998
  Multiplier(random)    0  0  1.65864e+08  3.35959e+08  0  0  2083.67  8333.5  2083.67  16666.8  24.6657  8.49271  90.2671  11.2851
  Rocket(dhrystone)     2.6208e+09  4.37912e+07  7.7197e+09  2.83992e+09  3.16223e+10  5.36263e+08  33098.5  66196.5  33098.5  67970.5  16.0001  8.00006  18.3373  8.24193
  SmallBOOM(dhrystone)  9.26078e+09  6.09055e+08  1.81556e+10  9.75646e+09  7.66366e+10  4.92608e+09  6985.33  14026.7  6985.33  20158.1  15.9957  7.33384  24.7858  7.80599
  SmallBOOM(median)     1.94933e+09  9.24169e+07  6.0484e+09  2.85431e+09  2.22e+10  1.05537e+09  10472  20943.5  10472  22859  15.9992  7.99984  20.3511  8.22987
  SmallBOOM(mt-vvadd)   1.31059e+10  4.95075e+08  2.09124e+10  9.53697e+09  1.11023e+11  4.25685e+09  12069  24137.8  12069  26566.2  15.9999  7.99994  19.0712  8.16432
  GigaBOOM(dhrystone)   2.8173e+10  2.93046e+09  4.73525e+10  2.22358e+10  2.48466e+11  2.52398e+10  3456.15  7327.08  3456.15  8402.31  15.9616  6.61703  21.1514  7.36738
  GigaBOOM(median)      1.01089e+10  8.32314e+08  2.08771e+10  8.32805e+09  9.32457e+10  7.76506e+09  4952.25  10117  4952.25  10552  15.9977  7.50009  18.1356  7.78374
  GigaBOOM(mt-vvadd)    4.77504e+10  3.2994e+09  7.13682e+10  2.64434e+10  4.26923e+11  2.94375e+10  6156.09  12489.1  6156.09  12764.3  15.9995  7.63684  17.669  7.86864
}{\dataTableThreadAllocStatsCmpAcrossDatasets}

\begin{axis}[
    width=1.0\linewidth,
    height=0.24\linewidth, % 你想要的高度比例
    symbolic x coords={
        Adder(random),
        Multiplier(random),
        Rocket(dhrystone),
        SmallBOOM(dhrystone),
        SmallBOOM(median),
        SmallBOOM(mt-vvadd),
        GigaBOOM(dhrystone),
        GigaBOOM(median),
        GigaBOOM(mt-vvadd)
    },
    xtick={
        Adder(random),
        Multiplier(random),
        Rocket(dhrystone),
        SmallBOOM(dhrystone),
        SmallBOOM(median),
        SmallBOOM(mt-vvadd),
        GigaBOOM(dhrystone),
        GigaBOOM(median),
        GigaBOOM(mt-vvadd)
    },
    xticklabels={
        \shortstack{Adder\\(random)}, 
        \shortstack{Multiplier\\(random)}, 
        \shortstack{Rocket\\(dhrystone)}, 
        \shortstack{SmallBOOM\\(dhrystone)}, 
        \shortstack{SmallBOOM\\(median)}, 
        \shortstack{SmallBOOM\\(mt-vvadd)}, 
        \shortstack{GigaBOOM\\(dhrystone)},
        \shortstack{GigaBOOM\\(median)},
        \shortstack{GigaBOOM\\(mt-vvadd)}
    },
    x tick label style={font=\scriptsize},
    y tick label style={font=\footnotesize},
    xlabel style={font=\small},
    ylabel style={font=\footnotesize},
    title style={font=\small},
    ylabel={Thread count ratio},
    ymode=log,
    ymin=0.05, ymax=10,
    ymajorgrids=true,
    enlarge x limits=0.03,
    legend style={at={(0.99,0.01)}, font=\footnotesize, legend columns=-1, anchor=south east}
]

    % ----------------------------- (A) sparse active threads + total active threads: ratio -----------------------------
    % reference line ratio=1
    \addplot[thick, dashed, color=mybrown, mark=none] coordinates {
        ({Adder(random)},1)
        ({GigaBOOM(mt-vvadd)},1)
    };
    \addlegendentry{ratio=1}

    % total active threads ratio  (requires columns: active_threads_base/active_threads_auto)
    \addplot+[
        thick,
        color=myorange,
        mark=triangle*,
        mark options={solid, fill=myorange},
    ] table[
        x=set,
        y expr={ifthenelse(\thisrow{active_threads_base}==0, 1,
    \thisrow{active_threads_auto}/\thisrow{active_threads_base})}
    ] {\dataTableThreadAllocStatsCmpAcrossDatasets};
    \addlegendentry{total launched threads}

    % sparse active threads ratio
    % \addplot+[
    %     thick,
    %     color=myblue,
    %     mark=*,
    %     mark options={solid, fill=myblue},
    % ] table[
    %     x=set,
    %     y expr={ifthenelse(\thisrow{sparse_threads_base}==0, 1,
    % \thisrow{sparse_threads_auto}/\thisrow{sparse_threads_base})}
    % ] {\dataTableThreadAllocStatsCmpAcrossDatasets};
    % \addlegendentry{sparse launched threads}

\end{axis}
\end{tikzpicture} 
        \label{fig:thread_allocation_stat_threads_ratio}
    }
    
    \caption{Ratios (event-density-aware / global-fixed) of (a) events per thread at the tail ($p_{90}$ and $p_{99}$) and (b) launched thread counts.}
    \label{fig:thread_allocation_stat}
\end{figure}
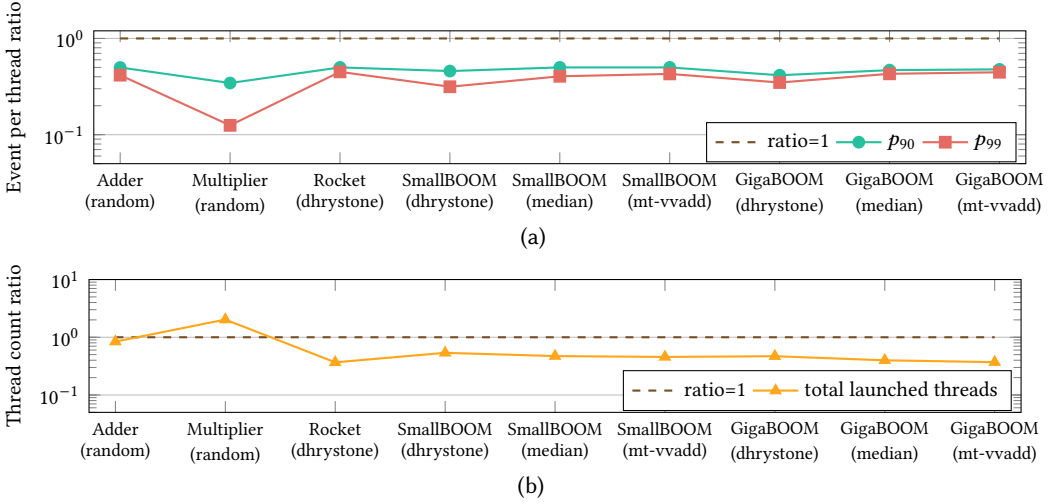

To investigate the underlying reasons, we compare the thread statistics of event-density-aware partitioning ($n_{\mathrm{evt,tgt}}=8$) and the best globally fixed setting ($n_{\mathrm{cyc}}^{\mathrm{fix}}=8$).
\Cref{fig:thread_allocation_stat} reports, for each dataset, the ratios (event-density-aware / global-fixed) of tail events per thread and thread counts.
% This comparison reveals the primary reason for the observed improvement.
% (1) \textbf{Less Total Work.}
% \Cref{fig:thread_allocation_stat_threads_ratio} shows that, for most designs, the event-density-aware scheme launches fewer threads than the globally fixed scheme.
% As given in \Cref{eq:total_work_of_gate_state_reconstruction}, a larger thread count increases the overhead of per-gate preparation, especially gate-state reconstruction.
% Because the power-computation work is fixed for a given workload, the difference in total work mainly comes from such thread-proportional preparation overhead.
% Therefore, reducing the number of launched threads directly reduces the total work.
% \textbf{Lower Thread Depth.}
As shown in \Cref{fig:thread_allocation_stat_heavy_gates}, the event-density-aware scheme lowers events per thread at the tail ($p_{90}$ and $p_{99}$) compared with the globally fixed scheme.
This indicates that the longest per-thread processing chains are shortened, i.e., the effective thread depth is reduced.
Combined with the thread count reduction in \Cref{fig:thread_allocation_stat_threads_ratio}, these results suggest that the event-density-aware scheme redistributes parallelism toward dense gates and compresses tail depth without increasing the total thread count.

\subsection{Parameter Sensitivity Analysis} \label{sec:parameter_sensitivity_analysis}
\subsubsection{SIPL Fallback Threshold}\label{sec:bsim_fallback_threshold_tuning}
\begin{figure}[tb!]
    \centering
    % PGFPlots datatable: BSIM threshold filtering, wide format
\pgfplotstableread[col sep=space]{
dataset total_gate total_event T4_filtered_gate T4_filtered_gate_ratio T4_filtered_event T4_filtered_event_ratio T4_dense_event_coverage T8_filtered_gate T8_filtered_gate_ratio T8_filtered_event T8_filtered_event_ratio T8_dense_event_coverage T16_filtered_gate T16_filtered_gate_ratio T16_filtered_event T16_filtered_event_ratio T16_dense_event_coverage T32_filtered_gate T32_filtered_gate_ratio T32_filtered_event T32_filtered_event_ratio T32_dense_event_coverage
  64bAdder  1815  151422120  0  0  0  0  1  0  0  0  0  1  0  0  0  0  1  0  0  0  0  1
  64bMultiplier  13267  2547321745  1466  0.1105  894746781  0.35125  0.64875  0  0  0  0  1  0  0  0  0  1  0  0  0  0  1
  RocketTile-dhrystone  116617  21942494322  41128  0.352676  2638651015  0.120253  0.879747  52  0.000445904  137125800  0.00624933  0.993751  52  0.000445904  137125800  0.00624933  0.993751  20  0.000171502  54413260  0.00247981  0.99752
  BoomTile-dhrystone  288789  71012308963  101009  0.349767  11444409604  0.161161  0.838839  242  0.000837982  2470467776  0.0347893  0.965211  242  0.000837982  2470467776  0.0347893  0.965211  114  0.000394752  1642488998  0.0231296  0.97687
  BoomTile-median  288789  21368171101  101009  0.349767  2492464841  0.116644  0.883356  242  0.000837982  742263548  0.0347369  0.965263  242  0.000837982  742263548  0.0347369  0.965263  114  0.000394752  494498602  0.0231418  0.976858
  BoomTile-mt-vvadd  288789  72181069422  101009  0.349767  7577949638  0.104985  0.895015  242  0.000837982  2508751712  0.0347564  0.965244  242  0.000837982  2508751712  0.0347564  0.965244  114  0.000394752  1680918722  0.0232875  0.976712
  GigaBoom-dhrystone  1053918  128877104875  545709  0.517791  19046331846  0.147787  0.852213  553  0.000524709  3434064646  0.026646  0.973354  553  0.000524709  3434064646  0.026646  0.973354  297  0.000281806  2211806801  0.0171621  0.982838
  GigaBoom-median  1053918  51513733782  545709  0.517791  5317864578  0.103232  0.896768  553  0.000524709  1382314451  0.0268339  0.973166  553  0.000524709  1382314451  0.0268339  0.973166  297  0.000281806  878155164  0.017047  0.982953
  GigaBoom-mt-vvadd  1053918  171258780369  545709  0.517791  15313121985  0.0894151  0.910585  553  0.000524709  4628943588  0.0270289  0.972971  553  0.000524709  4628943588  0.0270289  0.972971  297  0.000281806  2941672187  0.0171768  0.982823
}{\dataTableAdditionalBsimThresholdFiltering}

\definecolor{myorange}{RGB}{255,166,26}
\definecolor{myred}{RGB}{227,107,98}
\definecolor{myyellow}{RGB}{255,214,75}
\definecolor{mymiddleblue}{RGB}{139,201,246}
\definecolor{mylight}{RGB}{38,192,159}
\definecolor{mygreen}{RGB}{205,252,197}
\definecolor{myblue}{RGB}{29,114,221}
\definecolor{mybrown}{RGB}{120, 80, 40}

\def\bsimGateRatioHeatmapColor{myblue}

\def\drawBsimGateRatioHeatmapCell#1#2#3#4{%
    \pgfplotstablegetelem{#2}{#4}\of{\dataTableAdditionalBsimThresholdFiltering}
    \pgfmathsetmacro{\cellValue}{100 * \pgfplotsretval}
    \pgfmathsetmacro{\cellLeft}{#1}
    \pgfmathsetmacro{\cellRight}{#1 + 1}
    \pgfmathsetmacro{\cellTop}{-#3}
    \pgfmathsetmacro{\cellBottom}{-#3 - 1}
    \pgfmathsetmacro{\cellCenterX}{#1 + 0.5}
    \pgfmathsetmacro{\cellCenterY}{-#3 - 0.5}
    \pgfmathtruncatemacro{\cellShade}{\cellValue == 0 ? 0 : min(75,6 + 69 * sqrt(\cellValue / 52))}
    \path[
        draw=black,
        line width=0.2pt,
        fill=\bsimGateRatioHeatmapColor!\cellShade!white
    ] (\cellLeft,\cellTop) rectangle (\cellRight,\cellBottom);
    \node[font=\tiny] at (\cellCenterX,\cellCenterY)
        {\pgfmathprintnumber[fixed,precision=3]{\cellValue}\%};
}

\resizebox{\linewidth}{!}{%
\begin{tikzpicture}[x=0.1\linewidth,y=0.42cm]
    \node[anchor=east,font=\footnotesize] at (-0.08,-0.5) {4};
    \node[anchor=east,font=\footnotesize] at (-0.08,-1.5) {8};
    \node[anchor=east,font=\footnotesize] at (-0.08,-2.5) {16};
    \node[anchor=east,font=\footnotesize] at (-0.08,-3.5) {32};
    \node[rotate=90,font=\footnotesize] at (-0.55,-2) {Pin threshold};

    \foreach \datasetX/\rowIndex in {0/0,1/1,2/2,3/3,4/4,5/5,6/6,7/7,8/8}{
        \drawBsimGateRatioHeatmapCell{\datasetX}{\rowIndex}{0}{T4_filtered_gate_ratio}
        \drawBsimGateRatioHeatmapCell{\datasetX}{\rowIndex}{1}{T8_filtered_gate_ratio}
        \drawBsimGateRatioHeatmapCell{\datasetX}{\rowIndex}{2}{T16_filtered_gate_ratio}
        \drawBsimGateRatioHeatmapCell{\datasetX}{\rowIndex}{3}{T32_filtered_gate_ratio}
    }

    \node[font=\scriptsize,align=center] at (0.5,-4.92) {\shortstack{Adder\\(random)}};
    \node[font=\scriptsize,align=center] at (1.5,-4.92) {\shortstack{Multiplier\\(random)}};
    \node[font=\scriptsize,align=center] at (2.5,-4.92) {\shortstack{Rocket\\(dhrystone)}};
    \node[font=\scriptsize,align=center] at (3.5,-4.92) {\shortstack{SmallBOOM\\(dhrystone)}};
    \node[font=\scriptsize,align=center] at (4.5,-4.92) {\shortstack{SmallBOOM\\(median)}};
    \node[font=\scriptsize,align=center] at (5.5,-4.92) {\shortstack{SmallBOOM\\(mt-vvadd)}};
    \node[font=\scriptsize,align=center] at (6.5,-4.92) {\shortstack{GigaBOOM\\(dhrystone)}};
    \node[font=\scriptsize,align=center] at (7.5,-4.92) {\shortstack{GigaBOOM\\(median)}};
    \node[font=\scriptsize,align=center] at (8.5,-4.92) {\shortstack{GigaBOOM\\(mt-vvadd)}};
\end{tikzpicture}
}
    \vspace{-1em}
    \caption{Heatmap of fallback-gate ratios for different SIPL pin-count thresholds across benchmarks.}
    \label{fig:bsim_pin_threshold_filtering_gate_ratio}
\end{figure}

\begin{figure}[tb!]
    \centering
    %\definecolor{myred}{RGB}{255, 128, 128}
%\definecolor{mygreen}{RGB}{128, 255, 128}     %FFABA4q
%\definecolor{myblue}{RGB}{128, 80, 40}    %1D72DD
\definecolor{myorange}{RGB}{255,166,26}
\definecolor{myred}{RGB}{227,107,98}
\definecolor{myyellow}{RGB}{255,214,75}
\definecolor{mymiddleblue}{RGB}{139,201,246} 
\definecolor{mylight}{RGB}{38,192,159}  
\definecolor{mygreen}{RGB}{205,252,197}   
%FFABA4q
\definecolor{myblue}{RGB}{29,114,221}    %1D72DD
\definecolor{mybrown}{RGB}{120, 80, 40} 
\definecolor{mypink}{RGB}{252,228,215}   %FCE4D7

% ====== PGFPlots filecontents (auto-generated) ======
\pgfplotstableread[col sep=space]{
dataset                 bsim_thr=4      bsim_thr=8      bsim_thr=16     bsim_thr=32
Adder(random)           1.000000        1.003197        1.002341        1.002313
Multiplier(random)      1.000000        0.934295        0.935764        0.934655
Rocket(dhrystone)       1.000000        0.970238        0.970349        nan
SmallBOOM(dhrystone)    1.000000        0.955685        0.956564        0.956131
SmallBOOM(median)       1.000000        0.983436        0.978445        0.981616
SmallBOOM(mt-vvadd)     1.000000        0.983634        0.983437        0.984891
GigaBOOM(dhrystone)     1.000000        0.951835        0.951477        0.950432
GigaBOOM(median)        1.000000        0.972459        0.968839        0.971571
GigaBOOM(mt-vvadd)      1.000000        0.986820        0.984794        0.982705
}{\dataTableBSIMThrTuning}
% ===================================================

\pgfplotsset{
    width=0.86\linewidth,
    height=0.36\linewidth
}

\begin{tikzpicture}
\begin{axis}[
    width=\linewidth,
    height=0.25\linewidth, % 你想要的高度比例
    ymajorgrids=true,
    ybar,
    symbolic x coords={
        Adder(random),
        Multiplier(random),
        Rocket(dhrystone),
        SmallBOOM(dhrystone),
        SmallBOOM(median),
        SmallBOOM(mt-vvadd),
        GigaBOOM(dhrystone),
        GigaBOOM(median),
        GigaBOOM(mt-vvadd)
    },
    xlabel near ticks,
    xtick align=inside,
    xtick=data,
    xticklabels={
        \shortstack{Adder\\(random)}, 
        \shortstack{Multiplier\\(random)}, 
        \shortstack{Rocket\\(dhrystone)}, 
        \shortstack{SmallBOOM\\(dhrystone)}, 
        \shortstack{SmallBOOM\\(median)}, 
        \shortstack{SmallBOOM\\(mt-vvadd)}, 
        \shortstack{GigaBOOM\\(dhrystone)},
        \shortstack{GigaBOOM\\(median)},
        \shortstack{GigaBOOM\\(mt-vvadd)}
    },
    enlarge x limits=0.06,
    ylabel={Runtime},
    ylabel near ticks,
    ylabel shift=-5pt,
    ylabel style={align=center},  % 启用多行对齐
    % ytick align=outside,
    ytick={0.5, 0.75, 1.0, 1.25},
    bar width = 4pt,
    ymin=0.5,
    ymax=1.1,
    legend style={at={(0.5, 1.0)}, draw=none, anchor=south, legend columns=-1, font=\footnotesize},
    x tick label style={font=\scriptsize},
    y tick label style={font=\footnotesize},
    label style={font=\footnotesize}, % 调整轴标签的字体大小
    every node near coord/.append style={font=\tiny, color=black}
]
    % use TeX as calculator:
    \addplot +[ybar, fill=myred, draw=black, area legend] table [x=dataset,  y={bsim_thr=4}] {\dataTableBSIMThrTuning};
    \addplot +[      fill=myorange, draw=black, area legend] table [x=dataset,  y={bsim_thr=8}] {\dataTableBSIMThrTuning};
    \addplot +[      fill=mymiddleblue, draw=black, area legend] table [x=dataset,  y={bsim_thr=16}] {\dataTableBSIMThrTuning};
    \addplot +[      fill=mylight, draw=black, area legend] table [x=dataset,  y={bsim_thr=32}] {\dataTableBSIMThrTuning};

    \legend{Pin threshold=4, Pin threshold=8, Pin threshold=16, Pin threshold=32}
    \node[font={\scriptsize\bfseries}, text=red!70!black, rotate=90, anchor=west, xshift=-3pt, yshift=-9pt] at (axis cs:{Rocket(dhrystone)},0.5) {OOM};
\end{axis}
\end{tikzpicture}
    \caption{Fused kernel runtime with different SIPL pin thresholds. Normalized to pin threshold=4. OOM denotes out of memory.}
    \label{fig:bsim_thr_tuning}
\end{figure}
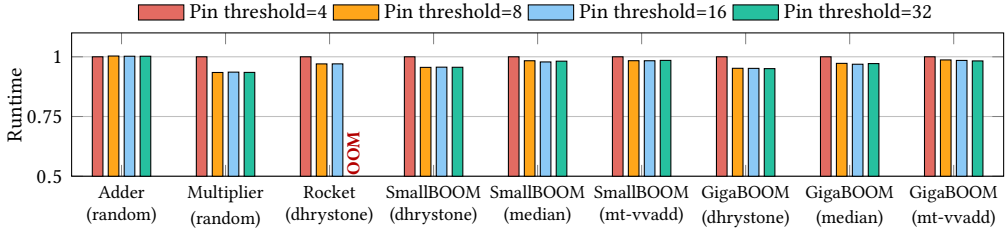

We tune the SIPL fallback threshold, which controls the maximum pin count handled by direct SIPL indexing.
As shown in \Cref{fig:bsim_pin_threshold_filtering_gate_ratio}, threshold 4 yields a high fallback-gate ratio (34.1\% on average), while thresholds 8 and 16 reduce it to 0.05\% on average.
\Cref{fig:bsim_thr_tuning} shows that thresholds 8 and 16 reduce geometric-mean runtime by 2.9\% over threshold 4, whereas threshold 32 gives no meaningful gain and causes out-of-memory on one benchmark.
We therefore choose threshold 16 to keep fallback traversal rare while avoiding the memory risk.

\subsubsection{Event-Density-Aware Partitioning Parameters}\label{sec:partitioning_parameter_tuning}
\begin{figure}[tb!]
    \centering
    %\definecolor{myred}{RGB}{255, 128, 128}
%\definecolor{mygreen}{RGB}{128, 255, 128}     %FFABA4q
%\definecolor{myblue}{RGB}{128, 80, 40}    %1D72DD
\definecolor{myorange}{RGB}{255,166,26}
\definecolor{myred}{RGB}{227,107,98}
\definecolor{myyellow}{RGB}{255,214,75}
\definecolor{mymiddleblue}{RGB}{139,201,246} 
\definecolor{mylight}{RGB}{38,192,159}  
\definecolor{mygreen}{RGB}{205,252,197}   
%FFABA4q
\definecolor{myblue}{RGB}{29,114,221}    %1D72DD
\definecolor{mybrown}{RGB}{120, 80, 40} 
\definecolor{mypink}{RGB}{252,228,215}   %FCE4D7

% ====== PGFPlots filecontents (auto-generated) ======
\pgfplotstableread[col sep=space]{
dataset                 e_target=2     e_target=4     e_target=8     e_target=16    e_target=32
Adder(random)           1.000000        0.869582        0.915300        1.092224        1.245061
Multiplier(random)      1.000000        0.969286        0.929283        0.965591        1.049271
Rocket(dhrystone)       1.000000        0.866139        0.907302        1.063512        1.149713
SmallBOOM(dhrystone)    1.000000        0.870051        0.848242        0.886608        0.938220
SmallBOOM(median)       1.000000        0.855954        0.833675        0.878868        0.926507
SmallBOOM(mt-vvadd)     1.000000        0.841389        0.823859        0.873316        0.907833
GigaBOOM(dhrystone)     1.000000        0.863810        0.823298        0.831009        0.843969
GigaBOOM(median)        1.000000        0.851271        0.810137        0.829453        0.855361
GigaBOOM(mt-vvadd)      1.000000        0.836701        0.800044        0.820091        0.833342
}{\dataTableETargetTuning}
% ===================================================

\pgfplotsset{
    width=0.86\linewidth,
    height=0.36\linewidth
}

\begin{tikzpicture}
\begin{axis}[
    width=\linewidth,
    height=0.22\linewidth, % 你想要的高度比例
    ymajorgrids=true,
    ybar,
    symbolic x coords={
        Adder(random),
        Multiplier(random),
        Rocket(dhrystone),
        SmallBOOM(dhrystone),
        SmallBOOM(median),
        SmallBOOM(mt-vvadd),
        GigaBOOM(dhrystone),
        GigaBOOM(median),
        GigaBOOM(mt-vvadd)
    },
    xlabel near ticks,
    xtick align=inside,
    xtick=data,
    xticklabels={
        \shortstack{Adder\\(random)}, 
        \shortstack{Multiplier\\(random)}, 
        \shortstack{Rocket\\(dhrystone)}, 
        \shortstack{SmallBOOM\\(dhrystone)}, 
        \shortstack{SmallBOOM\\(median)}, 
        \shortstack{SmallBOOM\\(mt-vvadd)}, 
        \shortstack{GigaBOOM\\(dhrystone)},
        \shortstack{GigaBOOM\\(median)},
        \shortstack{GigaBOOM\\(mt-vvadd)}
    },
    enlarge x limits=0.06,
    ylabel={Runtime},
    ylabel near ticks,
    ylabel shift=-5pt,
    ylabel style={align=center},  % 启用多行对齐
    % ytick align=outside,
    ytick={0.5, 0.75, 1.0, 1.25},
    bar width = 4pt,
    ymin=0.5,
    ymax=1.3,
    legend style={at={(0.5, 1.0)}, draw=none,anchor=south,legend columns=-1, font=\footnotesize},
    x tick label style={font=\scriptsize},
    y tick label style={font=\footnotesize},
    label style={font=\footnotesize}, % 调整轴标签的字体大小
    every node near coord/.append style={font=\tiny, color=black}
]
    % use TeX as calculator:
    \addplot +[ybar, fill=myred, draw=black, area legend] table [x=dataset,  y={e_target=2}] {\dataTableETargetTuning};
    \addplot +[      fill=myorange, draw=black, area legend] table [x=dataset,  y={e_target=4}] {\dataTableETargetTuning};
    \addplot +[      fill=mymiddleblue, draw=black, area legend] table [x=dataset,  y={e_target=8}] {\dataTableETargetTuning};
    \addplot +[      fill=mylight, draw=black, area legend] table [x=dataset,  y={e_target=16}] {\dataTableETargetTuning};
    \addplot +[      fill=mypink, draw=black, area legend] table [x=dataset,  y={e_target=32}] {\dataTableETargetTuning};

    \legend{$n_{\mathrm{evt,tgt}}=2$, $n_{\mathrm{evt,tgt}}=4$, $n_{\mathrm{evt,tgt}}=8$, $n_{\mathrm{evt,tgt}}=16$, $n_{\mathrm{evt,tgt}}=32$}
\end{axis}
\end{tikzpicture}
    \caption{Fused kernel runtime with different target event count per thread $n_{\mathrm{evt,tgt}}$. Normalized to $n_{\mathrm{evt,tgt}}=2$.}
    \label{fig:e_target_tuning}
\end{figure}
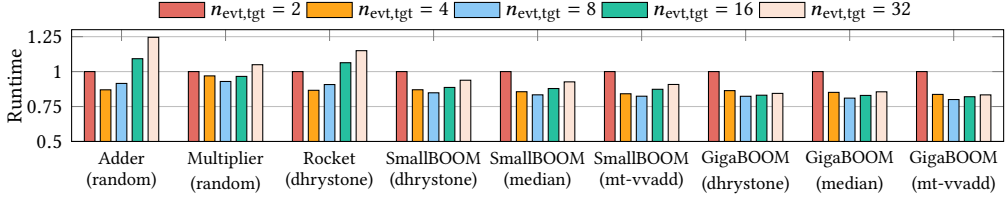

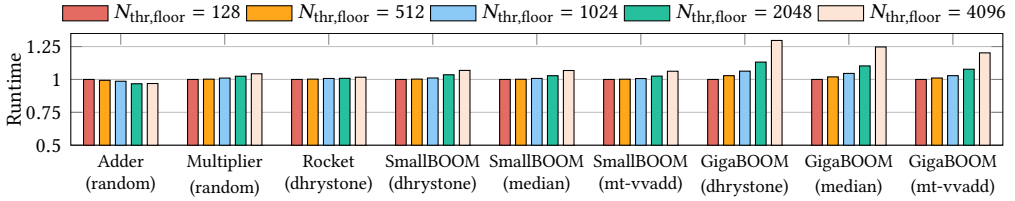
\begin{figure}[tb!]
    \centering
    %\definecolor{myred}{RGB}{255, 128, 128}
%\definecolor{mygreen}{RGB}{128, 255, 128}     %FFABA4q
%\definecolor{myblue}{RGB}{128, 80, 40}    %1D72DD
\definecolor{myorange}{RGB}{255,166,26}
\definecolor{myred}{RGB}{227,107,98}
\definecolor{myyellow}{RGB}{255,214,75}
\definecolor{mymiddleblue}{RGB}{139,201,246} 
\definecolor{mylight}{RGB}{38,192,159}  
\definecolor{mygreen}{RGB}{205,252,197}   
%FFABA4q
\definecolor{myblue}{RGB}{29,114,221}    %1D72DD
\definecolor{mybrown}{RGB}{120, 80, 40} 
\definecolor{mypink}{RGB}{252,228,215}   %FCE4D7

% ====== PGFPlots filecontents (auto-generated) ======
\pgfplotstableread[col sep=space]{
dataset                 para_floor=128  para_floor=512  para_floor=1024 para_floor=2048 para_floor=4096 para_floor=8192
Adder(random)           1.000000        0.993380        0.986401        0.967283        0.969119        0.983516
Multiplier(random)      1.000000        1.002187        1.010531        1.024545        1.043268        1.110108
Rocket(dhrystone)       1.000000        1.002410        1.007452        1.008545        1.016998        1.033295
SmallBOOM(dhrystone)    1.000000        1.002550        1.011115        1.035661        1.068836        1.151022
SmallBOOM(median)       1.000000        1.001413        1.007666        1.028426        1.068035        1.149941
SmallBOOM(mt-vvadd)     1.000000        1.001796        1.006765        1.025429        1.062451        1.139004
GigaBOOM(dhrystone)     1.000000        1.028754        1.063164        1.131977        1.296532        1.480793
GigaBOOM(median)        1.000000        1.019557        1.045904        1.102963        1.247076        1.441399
GigaBOOM(mt-vvadd)      1.000000        1.010902        1.028806        1.077679        1.202124        1.361398
}{\dataTableParallelismFloorTuning}
% ===================================================

\pgfplotsset{
    width=0.86\linewidth,
    height=0.36\linewidth
}

\begin{tikzpicture}
\begin{axis}[
    width=\linewidth,
    height=0.22\linewidth, % 你想要的高度比例
    ymajorgrids=true,
    ybar,
    symbolic x coords={
        Adder(random),
        Multiplier(random),
        Rocket(dhrystone),
        SmallBOOM(dhrystone),
        SmallBOOM(median),
        SmallBOOM(mt-vvadd),
        GigaBOOM(dhrystone),
        GigaBOOM(median),
        GigaBOOM(mt-vvadd)
    },
    xlabel near ticks,
    xtick align=inside,
    xtick=data,
    xticklabels={
        \shortstack{Adder\\(random)}, 
        \shortstack{Multiplier\\(random)}, 
        \shortstack{Rocket\\(dhrystone)}, 
        \shortstack{SmallBOOM\\(dhrystone)}, 
        \shortstack{SmallBOOM\\(median)}, 
        \shortstack{SmallBOOM\\(mt-vvadd)}, 
        \shortstack{GigaBOOM\\(dhrystone)},
        \shortstack{GigaBOOM\\(median)},
        \shortstack{GigaBOOM\\(mt-vvadd)}
    },
    enlarge x limits=0.06,
    ylabel={Runtime},
    ylabel near ticks,
    ylabel shift=-5pt,
    ylabel style={align=center},  % 启用多行对齐
    % ytick align=outside,
    ytick={0.5, 0.75, 1.0, 1.25},
    bar width = 4pt,
    ymin=0.5,
    ymax=1.35,
    legend style={at={(0.5, 1.0)}, draw=none, anchor=south, legend columns=-1, font=\footnotesize},
    x tick label style={font=\scriptsize},
    y tick label style={font=\footnotesize},
    label style={font=\footnotesize}, % 调整轴标签的字体大小
    every node near coord/.append style={font=\tiny, color=black}
]
    % use TeX as calculator:
    \addplot +[ybar, fill=myred, draw=black, area legend] table [x=dataset,  y={para_floor=128}] {\dataTableParallelismFloorTuning};
    \addplot +[      fill=myorange, draw=black, area legend] table [x=dataset,  y={para_floor=512}] {\dataTableParallelismFloorTuning};
    \addplot +[      fill=mymiddleblue, draw=black, area legend] table [x=dataset,  y={para_floor=1024}] {\dataTableParallelismFloorTuning};
    \addplot +[      fill=mylight, draw=black, area legend] table [x=dataset,  y={para_floor=2048}] {\dataTableParallelismFloorTuning};
    \addplot +[      fill=mypink, draw=black, area legend] table [x=dataset,  y={para_floor=4096}] {\dataTableParallelismFloorTuning};

    \legend{$N_{\mathrm{thr,floor}}=128$, $N_{\mathrm{thr,floor}}=512$, $N_{\mathrm{thr,floor}}=1024$, $N_{\mathrm{thr,floor}}=2048$, $N_{\mathrm{thr,floor}}=4096$}
\end{axis}
\end{tikzpicture}
    \caption{Fused kernel runtime with different parallelism floor $N_{\mathrm{thr,floor}}$. Normalized to $N_{\mathrm{thr,floor}}=128$.}
    \label{fig:parallelism_floor_tuning}
\end{figure}

We analyze event-density-aware partitioning parameters and use the observed trends to explain the selected defaults.
For $n_{\mathrm{evt,tgt}}$, we fix $N_{\mathrm{thr,floor}}=512$ and sweep values from 2 to 32 with exponential steps.
Intuitively, a smaller $n_{\mathrm{evt,tgt}}$ (finer granularity) would improve parallelism and thus reduce kernel execution time.
However, as shown in \Cref{fig:e_target_tuning}, experiments show a counterintuitive trend: an excessively small $n_{\mathrm{evt,tgt}}$ leads to a higher kernel execution time.
This is because, as given in \Cref{eq:total_work_of_gate_state_reconstruction}, the overhead of gate-state reconstruction scales with the number of threads.
A smaller $n_{\mathrm{evt,tgt}}$ results in more threads and therefore incurs higher reconstruction overhead.
This illustrates a fundamental trade-off between the degree of parallelism and the resulting overhead.
Our results indicate that $n_{\mathrm{evt,tgt}}=8$ achieves the optimal balance, minimizing runtime while maintaining sufficient parallelism.

We also analyze the parallelism floor $N_{\mathrm{thr,floor}}$ with $n_{\mathrm{evt,tgt}}=8$, as shown in \Cref{fig:parallelism_floor_tuning}.
Compared with $N_{\mathrm{thr,floor}}=128$, $N_{\mathrm{thr,floor}}=512$ increases geometric-mean runtime by only 0.7\%, while larger floors of 1024, 2048, and 4096 increase it by 1.8\%, 4.4\%, and 10.3\%, respectively.
This matches the reconstruction-overhead trade-off: larger floors launch more threads for sparse gates.
We therefore choose $N_{\mathrm{thr,floor}}=512$ because it is close to the best runtime with a higher minimum floor.

\section{Related Work} \label{sec:related_work}

% Prior work related to GPU-accelerated gate-level time-based power analysis includes academic power-analysis research, practical power-analysis tools, and GPU acceleration for electronic design automation (EDA).
Relevant prior work spans academic power analysis research, practical tools, and GPU acceleration for electronic design automation (EDA).
We first review academic power-analysis research, which can be broadly classified into ML-based and simulation-based approaches.

% \subsection{ML-Based Power Analysis}
% \textbf{ML-Based Power Analysis.}
In the context of ML-based methods, gate-level power analysis research primarily employs RTL switching activity data to train predictive models for estimating power characteristics at the gate level.
PRIMAL~\cite{dac_2019_primal} introduces an ML framework that encodes RTL register switching activities into learnable feature embeddings, thereby predicting gate-level cycle-accurate power waveform from RTL simulation traces.
% GRANNITE~\cite{dac_2020_grannite} presents a graph convolutional network (GCN) framework for average power estimation by propagating toggle rates through combinational logic.
% GRIPT~\cite{tcad_2025_gript} advances GRANNITE by replacing fixed-weight GCN with feature-aware attention mechanisms, enabling untrained cell prediction across technology nodes.
GRANNITE~\cite{dac_2020_grannite} estimates average power with a graph convolutional network (GCN) that propagates toggle rates through combinational logic, and GRIPT~\cite{tcad_2025_gript} improves it with feature-aware attention for untrained-cell prediction across technology nodes.
More recently, ATLAS~\cite{tcad_2026_atlas} and a GNN method~\cite{date_2026_cross_design_power} predict cross-design, cycle-level power from gate-level netlists.
Regarding the RTL level, MasterRTL~\cite{iccad_2023_master_rtl} proposes a framework using operator graph representations to facilitate pre-synthesis average power estimation.
APOLLO~\cite{micro_2021_apollo} presents an automated RTL power modeling framework that enables cycle-accurate power estimation with proxy selection and linear regression.
Overall, ML-based approaches typically \emph{estimate} power from learned design representations and thus neither guarantee \emph{per-gate, per-event} time-based correctness nor generate a \emph{complete} report covering all power components, peak power, and waveforms.
% Overall, ML-based approaches are typically formulated as \emph{estimation} problems that infer power metrics from RTL-level proxies, and they generally do not provide \emph{per-gate, per-event} time-based correctness guarantees required for precise power calculation. 
% In addition, many ML-based approaches focus on specific power targets (e.g., average power or cycle-level waveform prediction) rather than generating a comprehensive report that consistently covers leakage, internal, switching, glitch components together with peak power values and time-resolved power waveforms. 

% \subsection{Simulation-Based Power Analysis}
In the domain of simulation-based methods, Strober~\cite{isca_2016_strober} accelerates power estimation by sparsely sampling RTL snapshots through field-programmable gate array (FPGA) emulation and replaying them using commercial power analysis tools.
SMART-GPO~\cite{aspdac_2025_smart_gpo} employs a randomized sampling approach that samples only 1\% of cycles for efficient glitch power estimation.
These simulation-based approaches are more physically grounded than purely ML-based estimation, but they still introduce \emph{approximations} through sampling and limited temporal coverage. 
% As a result, they may miss infrequent yet important temporal behaviors and do not fully realize comprehensive \emph{per-gate, per-event} time-based power calculation across the entire execution. 

Practical tools also cover related flows:
OpenSTA~\cite{opensta} provides static timing analysis and supports \emph{averaged} power analysis, but it does not perform the gate-level time-based power computation targeted in this work.
Trace2Power~\cite{trace2power} \emph{exports switching-activity} for downstream power-analysis flows rather than computing a complete power report.
Commercial signoff tools such as Synopsys PrimeTime PX~\cite{synopsys_primetime} support gate-level time-based power analysis and therefore provide the closest baseline for validating complete power-computation results.

% In the context of average power, McPAT-Calib~\cite{tcad_2023_mcpat_calib} develops a machine learning (ML)  calibration framework for out-of-order RISC-V microarchitecture average power estimation at a 7nm technology node.

% Despite these efforts, the current studies concentrate solely on a subset of power analysis and fail to deliver a comprehensive power report, which should encompass various power components, including leakage, internal, switching, glitch, peak power values, and power waveform. 
% Additionally, existing studies primarily focus on power estimation rather than precise power calculation. 
% A comprehensive and accurate power report is crucial for diagnosing power-related issues, as any oversight or inaccuracy may lead to critical and undetected issues.
%Therefore, it is crucial to perform gate-level time-based power analysis to ensure the creation of a comprehensive and accurate power report.
% Notably, commercial tools such as PrimeTime PX~\cite{synopsys_primetime} support gate-level time-based power analysis on the CPU with multi-thread support, delivering detailed and accurate power reports.
% However, they still suffer from performance limitations when handling large-scale designs.

% \subsection{GPU-Accelerated EDA Tools}
% GPUs have exhibited remarkable computational capabilities and advantages across numerous domains of scientific computing, including electronic design automation (EDA).
Finally, GPU acceleration has been widely explored for various EDA tasks, such as logic simulation~\cite{todaes_2011_Chatterjee_gate_lev_sim_gpu, todaes_2011_deng_massive_gpu_logic_sim, iccad_2021_fdu_gate_lve_resim, dac_2022_gatspi}, placement~\cite{tcad_2021_dreamplace}, routing~\cite{iccad_2022_gpu_accu_rect_steiner_tree,tcad_2023_gamer}, and STA~\cite{iccad_2020_gpu_accu_sta, date_2024_heter_sta}.
Notably, GATSPI~\cite{dac_2022_gatspi} accelerates the delay-aware logic re-simulation step in a gate-level power-improvement flow and generates switching activity for downstream power analysis.
It addresses \emph{activity generation} rather than the gate-level time-based power analysis stage accelerated in this paper.
More recently, HeteroPower~\cite{iseda_2026_hetero_power} proposes a CPU-GPU heterogeneous engine for gate-level averaged power analysis that reuses GPU-resident STA data to accelerate power computation.
However, it targets \emph{averaged} power analysis rather than event-driven time-based power analysis.
% Although GPU acceleration has been investigated in these domains, the acceleration of gate-level time-based power analysis using GPUs remains underexplored.

% Overall, existing work covers different scopes: ML and simulation methods focus on estimation or sampled analysis; OpenSTA and Trace2Power provide averaged-power or activity-export functions; PrimeTime PX provides complete CPU-based gate-level time-based signoff analysis.
% GPU-accelerated EDA work targets activity generation or averaged-power analysis rather than complete event-driven time-based power analysis, motivating our GPU acceleration focus.
Overall, prior work either estimates or samples power behavior, provides averaged-power analysis or activity-export functionality, or supports complete gate-level time-based power analysis in CPU-based commercial flows, leaving this complete analysis flow underexplored on GPUs.

\section{Conclusions}\label{sec:conclusion_and_future}
In this paper, we propose, to the best of our knowledge, the first GPU-accelerated gate-level time-based power analysis framework. 
% We introduce an event-equalized chunking methodology. 
% This approach not only mitigates memory insufficiency issues but also facilitates pipelined CPU-GPU execution, effectively hiding host-side event preparation latency.
% Furthermore, we present our GPU-specific optimizations, including the spatial-temporal parallelization strategy and a state-indexed power lookup method to tackle the issue of inefficient power retrieval.
We introduce a state-indexed power lookup method to address inefficient state-dependent power retrieval.
We further propose an event-density-aware, per-gate cycle partitioning strategy that exploits variations in gate event density to allocate an appropriate number of GPU threads to each gate.
Finally, we present a kernel fusion strategy that computes dynamic and leakage power within a single kernel, thereby reducing redundant work in separate kernels.
Our framework achieves high accuracy and a 1.94$\times$ to 37.63$\times$ end-to-end speedup over multi-threaded PrimeTime PX.
% As there is, as far as we know, no open source time-based power analysis tool, 
% Our power analysis tool relies on the VCD files generated from logic simulation, which necessitates waiting for the simulation to complete before conducting power analysis. 
% For this reason, we plan to integrate our GPU-accelerated power analysis tool into the logic simulation process to streamline the power analysis workflow.
% In the future, we plan to adapt our framework into a multi-GPU version to achieve faster power analysis.

{
    \bibliographystyle{ACM-Reference-Format}
    \bibliography{ref/Top,ref/Power}
}

\end{document}